\newif\ifpreprint

\preprintfalse % Enable for double column reprint

\ifpreprint
\documentclass[journal=jctcce,manuscript=letter]{achemso}
\else
\documentclass[journal=jctcce,manuscript=letter,layout=twocolumn]{achemso}
\fi

\usepackage[T1]{fontenc} % Use modern font encodings

\usepackage{amsmath}
\usepackage{newtxtext,newtxmath}

\usepackage{graphicx}
\usepackage{dcolumn}
\usepackage{braket}
\usepackage{multirow}
\usepackage{threeparttable}
\usepackage{xspace}
\usepackage{verbatim}
\usepackage[version=4]{mhchem} % Formula subscripts using \ce{}
\usepackage{comment}
\usepackage{color,soul}

\usepackage{mathtools}
\usepackage[dvipsnames]{xcolor}
\usepackage{xspace}
\usepackage{ifthen}

\usepackage{qcircuit}

\usepackage{graphicx,longtable,dcolumn,mhchem}
\usepackage{rotating,color}
\usepackage{lscape}
\usepackage{amsmath}
\usepackage{dsfont}
\usepackage{soul}
\usepackage{physics}
\newcolumntype{d}{D{.}{.}{-1}}

\newcommand{\ie}{\textit{i.e}}

\newcommand{\Pop}{6-31+G(d)}
\newcommand{\DAVDZ}{d-\emph{aug}-cc-pVDZ}
\newcommand{\DAVTZ}{d-\emph{aug}-cc-pVTZ}
\newcommand{\DAVQZ}{d-\emph{aug}-cc-pVQZ}
\newcommand{\DAVFZ}{d-\emph{aug}-cc-pVFZ}
\newcommand{\AVDZ}{\emph{aug}-cc-pVDZ}
\newcommand{\ACVTZ}{\emph{aug}-cc-pCVTZ}
\newcommand{\AVTZ}{\emph{aug}-cc-pVTZ}
\newcommand{\AVQZ}{\emph{aug}-cc-pVQZ}
\newcommand{\AVFZ}{\emph{aug}-cc-pV5Z}

\usepackage[colorlinks = true,
            linkcolor = blue,
            urlcolor  = black,
            citecolor = blue,
            anchorcolor = black]{hyperref}

\definecolor{goodorange}{RGB}{225,125,0}
\definecolor{goodgreen}{RGB}{5,130,5}
\definecolor{goodred}{RGB}{220,50,25}
\definecolor{goodblue}{RGB}{30,144,255}

\newcommand{\note}[2]{
\ifthenelse{\equal{#1}{F}}{
\colorbox{goodorange}{\textcolor{white}{\footnotesize \fontfamily{phv}\selectfont #1}}
    \textcolor{goodorange}{{\footnotesize \fontfamily{phv}\selectfont #2}}\xspace
}{}
\ifthenelse{\equal{#1}{R}}{
\colorbox{goodred}{\textcolor{white}{\footnotesize \fontfamily{phv}\selectfont #1}}
    \textcolor{goodred}{{\footnotesize \fontfamily{phv}\selectfont #2}}\xspace
}{}
\ifthenelse{\equal{#1}{N}}{
\colorbox{goodgreen}{\textcolor{white}{\footnotesize \fontfamily{phv}\selectfont #1}}
    \textcolor{goodgreen}{{\footnotesize \fontfamily{phv}\selectfont #2}}\xspace
}{}
\ifthenelse{\equal{#1}{M}}{
\colorbox{goodblue}{\textcolor{white}{\footnotesize \fontfamily{phv}\selectfont #1}}
    \textcolor{goodblue}{{\footnotesize \fontfamily{phv}\selectfont #2}}\xspace
}{}
}

\usepackage{titlesec}

\usepackage[fontsize=11pt]{scrextend}
\titleformat{\section}
{\normalfont\sffamily\bfseries\color{Blue}}
{\thesection.}{0.25em}{\uppercase}

\titleformat{\subsection}
{\normalfont\sffamily\bfseries}
{\thesubsection}{0.25em}{}

\titleformat{\subsubsection}
{\normalfont\sffamily}
{\thesubsubsection}{0.25em}{}

\titleformat{\suppinfo}
{\normalfont\sffamily\bfseries}
{\thesubsection}{0.25em}{}

\titlespacing*{\section}{0pt}{0.5\baselineskip}{0.01\baselineskip}
\titlespacing*{\subsection}{0pt}{0.125\baselineskip}{0.01\baselineskip}
\titlespacing*{\subsubsection}{0pt}{0.125\baselineskip}{0.01\baselineskip}

\author{Amara Chrayteh}
	\affiliation[CEISAM, Nantes]{Universit\'e de Nantes, CNRS,  CEISAM UMR 6230, F-44000 Nantes, France}
\author{Aymeric Blondel}
	\affiliation[CEISAM, Nantes]{Universit\'e de Nantes, CNRS,  CEISAM UMR 6230, F-44000 Nantes, France}
\author{Pierre-Fran\c{c}ois Loos}
	\affiliation[LCPQ, Toulouse]{Laboratoire de Chimie et Physique Quantiques, Universit\'e de Toulouse, CNRS, UPS, France}
\author{Denis Jacquemin}
	\email{Denis.Jacquemin@univ-nantes.fr}
	\affiliation[CEISAM, Nantes]{Universit\'e de Nantes, CNRS,  CEISAM UMR 6230, F-44000 Nantes, France}
		
\let\oldmaketitle\maketitle
\let\maketitle\relax
     \title{A Mountaineering Strategy to Excited States: Highly-Accurate Oscillator Strengths and Dipole Moments of Small Molecules}
\date{\today}

\begin{tocentry}
	\centering
	\includegraphics[width=.7\textwidth]{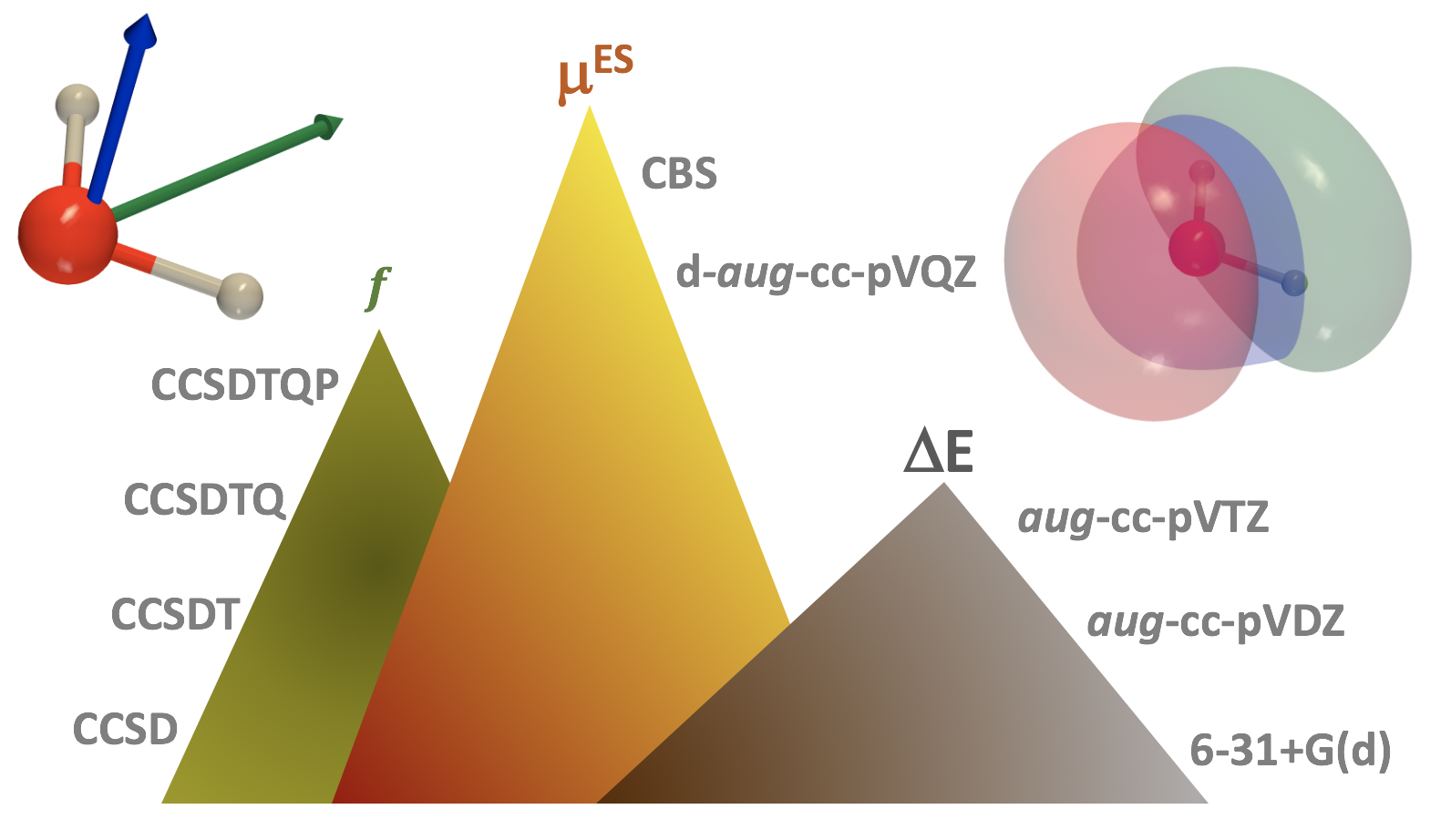}
\end{tocentry}

\begin{document}

\ifpreprint
\else
\twocolumn[
\begin{@twocolumnfalse}
\fi
\oldmaketitle

%%%%%%%%%%%%%%%%
%%% ABSTRACT %%%
%%%%%%%%%%%%%%%%
\begin{abstract}
This work presents a series of highly-accurate excited-state properties obtained using high-order coupled-cluster  (CC) calculations performed with a series of diffuse containing basis sets, as well as
extensive comparisons with experimental values. Indeed, we have computed both the main ground-to-excited transition property, the oscillator strength, as well as the ground- and excited-state 
dipole moments, considering {thirteen} small molecules (hydridoboron, hydrogen chloride, water, hydrogen sulfide, boron fluoride, carbon monoxide, dinitrogen, ethylene, formaldehyde, thioformaldehyde, nitroxyl, 
{fluorocarbene}, and silylidene). We systematically include corrections up to the quintuple (CCSDTQP) in the CC expansion and extrapolate to the complete basis set limit. When comparisons 
with experimental measurements are possible, that is, when a number of consistent experimental data can be found, theory typically provides values falling within the experimental error bar for 
the excited-state properties.  Besides completing our previous studies focussed on transition energies (\textit{J.~Chem.~Theory Comput.} \textbf{14} (2018) 4360--4379,  \textit{ibid.}~\textbf{15} 
(2019) 1939--1956, \textit{ibid.}~\textbf{16} (2020) 1711--1741, and \textit{ibid.}~\textbf{16} (2020) 3720--3736),  this work also provides ultra-accurate dipoles and oscillator strengths that could 
be employed for future theoretical benchmarks.
\end{abstract}

\ifpreprint
\else
\end{@twocolumnfalse}
]
\fi

\ifpreprint
\else
\small
\fi

\noindent

%%%%%%%%%%%%%%%%%%%%
%%% INTRODUCTION %%%
%%%%%%%%%%%%%%%%%%%%
\section{Introduction}

In the formidable quest aiming at reaching high accuracy in the modeling of electronically excited states (ESs), the primary focus has been set on vertical excitation energies, \cite{Loo20c} defined as
the difference in total energies between a given ES and its corresponding ground state (GS) at fixed geometry (typically the ground-state equilibrium geometry).  The main reasons for this choice are, on the one hand, the availability of many 
theoretical models for computing total ES energies, and, on the other hand, the fact that obtaining accurate transition energies is generally viewed as a prerequisite for further ES investigations. Along the years, more and
more accurate vertical transition energies have become available. Probably the most illustrative example is provided by the well-known Thiel set, \cite{Sch08,Sau09,Sil10c} encompassing 223 values obtained at the CC3 and CASPT2 levels,
as well as our recent efforts to obtain chemically-accurate vertical energies within $\pm$0.03 eV of the full configuration interaction (FCI) limit for more than 400 ES. \cite{Loo18a,Loo19c,Loo20a,Loo20d} Whilst such sets are obviously useful to 
benchmark lower-order methods, they remain nevertheless intrinsically limited by two factors. First, vertical transition energies remain inaccessible experimentally (in the vast majority of the cases), preventing direct comparisons with 
measurements.  To circumvent this issue, several groups have turned their attention to 0-0 energies, \cite{Fur02,Die04b,Hat05c,Goe09,Goe10a,Sen11b,Jac12d,Win13,Fan14b,Jac15b,Oru16,Sch17,Loo18b,Loo19a,Loo19b} defined as the 
differences between the ES and GS energies determined at their respective minimum and corrected for zero-point vibrational effects, because 0-0 energies allow straightforward theory-experiment comparisons. To compute 0-0 energies, 
one must however determine ES geometries and vibrations, which limits the number of methods that can be applied. This explains why compromise ``hybrid'' protocols employing different levels of theory for the structural and energetic 
parameters are popular in this particular field. \cite{Goe09,Jac15b,Oru16,Loo18b} Interestingly, the accuracy of the underlying geometries has been shown to be rather irrelevant, \cite{Sen11b,Loo19a} indicating that benchmarks of 
0-0 energies still mainly assess the quality of the (adiabatic) transition energies. Second, vertical transition energies do provide a very partial characterization of the ES as one typically needs to determine transition probabilities 
(oscillator strengths, $f$) as well as ES properties (such as structures and dipole moments, $\mu$) to attain a better grasp of the photophysics. On a more theoretical viewpoint, it is also known that a method can provide accurate 
transition energies while failing to deliver accurate ES properties, \cite{Bre18a,Taj19,Taj20a} indicating that benchmarks relying on transition energies as unique gauge can lead to incomplete and/or erroneous conclusions regarding the 
strengths and weaknesses of a given theoretical model.

In the above-defined framework, let us now summarize the efforts that have been made to define accurate reference ES properties for significant sets of compounds.  

First, for ES geometries, which are not our focus 
here, several ensembles of structures have been reported at CASPT2, \cite{Pag03} SAC-CI, \cite{Bou13,Bou14b} Mk-MRCCSD, \cite{Jag12}  QMC, \cite{Gua13} and CC2 \cite{Tun16} levels.  To the best of our knowledge, the most extensive 
dataset of accurate ES structural parameters remains the one defined by some of us during the last few years: it contains bond lengths, valence angles, and torsion angles determined at the CC3 and CASPT2 levels with 
\emph{aug}-cc-pVTZ for several dozens of small organic compounds.  \cite{Bud17,Jac18a,Bre18a,Loo19a,Fih19} In this framework, it is also worth mentioning the works of Szalay's group on the shape of potential energy surfaces, 
in which CCSDT references are defined, \cite{Taj18,Taj19,Taj20a} as well as the publications of Olivucci's group focussing on the topology of conical intersections, in which the results of high-level multi-reference 
calculations are available. \cite{Goz12,Tun15}  

Second, for transition properties, the most accurate reference values we are aware of (for a significant set of transitions) are: i) the CC3/TZVP oscillator strengths obtained by Kannar 
and Szalay \cite{Kan14} for the Thiel set (this work also includes 15 CCSDT/TZVP oscillator strengths) {that can be compared to CASPT2/TZVP values obtained earlier for the same set};\cite{Sch08} 
ii) the numerous CC3/\emph{aug}-cc-pVTZ oscillator strengths determined to complete our FCI energy calculations; \cite{Loo18a,Loo20a,Loo20d}  
and iii) the transition dipole moments computed at the CC3 or ADC(3) levels for 15 molecules by Robinson. \cite{Rob18}  There are also earlier sets of oscillator strengths computed at lower levels of theory, \emph{e.g.}, CCSD, \cite{Car10d} SAC-CI, \cite{Li07b} 
and CC2. \cite{Sil10b,Jac16b}  The typical error associated with these approaches (with respect to FCI) remains unknown at this stage.  To the best of our knowledge, a significant set of FCI-quality oscillator strengths has yet to be published. 

Third, theoretical datasets for accurate ES 
dipoles are apparently even scarcer. The most advanced sets we are aware of contain: i) CASPT2/TZVP, \cite{Sch08} and CC2/\emph{aug}-cc-pVTZ reference  values  \cite{Sil10b} determined for Thiel's set; ii) CC2/{\AVQZ} results for small-
and medium-sized molecules; \cite{Hel11}  and iii) CC2/{\AVTZ} data for ``real-life'' organic dyes. \cite{Jac16d}  Comparisons with experimental ES dipoles performed in Ref.~\citenum{Hel11} yield an error of ca.~0.2 Debye (D) for both ADC(2) and CC2, a 
value that can be viewed  as acceptable, but is nevertheless far from chemical accuracy and not on par with the precision provided by state-of-the-art approaches for GS dipole moments. \cite{Hai18} Of course, one can find specific works focussing 
on a small number of ES dipoles and using high level(s) of theory (see our Result Section for references). However, such specific studies preclude valuable statistical conclusions. 

Globally, these previous studies have 
demonstrated that both $f$ and $\mu^{\mathrm{ES}}$ are much more basis set sensitive than transition energies and geometries. Indeed, a specific challenge comes from intensity-borrowing effects that can vastly change the 
properties of two close-lying ES of the same symmetry while their energies remain almost unaffected. When a change of basis set slightly tunes the energy gap between two ESs, it might simultaneously drastically affect 
the properties. This highlights that properties are much harder to accurately estimate than energies, not only for implementation or computational reasons, but also due to more fundamental aspects.

Another important fact is that, in contrast to vertical transition energies, both oscillator strengths and ES dipole moments are accessible experimentally, so comparisons between theoretical and experimental measurements are, in principle, 
possible. However, as we pointed out previously for geometries, \cite{Bud17} such comparisons are generally far from straightforward. For the oscillator strengths, the two main measurement techniques are electron impact and optical 
spectroscopies. As nicely summarized elsewhere, \cite{Bor06a} the former typically requires extrapolations of the cross-sections measured at various momenta/angles (such extrapolations are not error-free), whereas the latter 
can be plagued by saturation and interaction effects (yielding to underestimation of the actual $f$ value). In any case, the so-called ``electronic'' or ``optical'' $f$, which is of interest here, is not directly measurable, and  post-processing of the 
experimental raw data (lifetimes or cross sections) is needed to access its ``experimental'' value. \cite{Nie80b,Bor06a}  The measurements of ES dipoles are also cumbersome. \cite{Sch08b} The typical strategy is to investigate 
solvatofluorochromism, that is, to measure the shift of the emission wavelength in a series of solvents of various polarities, and to fit the results with the Lippert-Mataga equation within an Onsager-like interaction model. Such approach is obviously 
not a direct gas-phase measurement and comes with significant assumptions, making the final error often too large to allow benchmarking. \cite{Lom87} Alternatively, one can also measure $\mu^{\mathrm{ES}}$ directly 
in gas phase by studying how  external electric fields tune the position and shape of the vibronic peaks (the so-called Stark effect). \cite{Sch08b,Lom87} Only the latter approach can be considered as sufficiently robust for reference purposes. 
Yet, such Stark effect measurements do typically provide an ``adiabatic'' version of $\mu^{\mathrm{ES}}$, that is, the ES dipole measured at the ES equilibrium geometry. If a non-trifling structural reorganization takes place after absorption, such as in 
formaldehyde, this means that the value of $\mu^{\mathrm{ES}}$ computed at the GS equilibrium structure has again no experimental counterpart. In addition, the direction of the ES dipole moment cannot be determined directly from the measurements
of the Stark effect.
 
For all these reasons, it is of interest to define a set of coherent near-FCI oscillator strengths and excited-state dipole moments. {Indeed, the oscillator strengths are directly related to the transition dipole moments, whereas the dipole
moments can be seen as a measure of the quality of the total (GS and ES) densities, so that these data go beyond the ``simple'' characterization of vertical transition energies.}  This contribution therefore aims at tackling this ambitious objective 
for a set of small molecules  similar to the one treated in our original \emph{Mountaineering} paper. \cite{Loo18a} To this end, we take advantage of the CC hierarchy, going from CCSD to CCSDTQP for both properties, in combination with 
increasingly large atomic basis sets including one or two sets of diffuse basis functions. Whilst such calculations provide ``definite answers'', they remain at the limit of today's computational capabilities and are achievable for compact molecules 
only, which stands as a clear limit of the present contribution. Nevertheless, providing highly-accurate numbers for properties directly related to the quality of the transition and the ES density is, we believe, useful. Very recently, Hait and Head-Gordon 
have used, in a density-functional theory (DFT) context, the GS dipole moment as a metric to estimate the quality of the GS density given by many exchange-correlation functionals. \cite{Hai18}  {As these authors nicely stated:  the dipole \emph{ ``is 
perhaps the simplest observable that captures errors in the underlying density, (\emph{... and is}) a relevant density derived quantity to examine for DFA testing and development''.}}  The extension to ESs (and time-dependent DFT) 
is obviously natural, but such task is difficult due to the lack of definite reference values.  In this framework, we also note some very recent efforts for computing ES first-order properties with the many-body expansion FCI (MBE-FCI) approach. \cite{Eri20}

%%%%%%%%%%%%%%%%%%%%
%%% METHODS %%%
%%%%%%%%%%%%%%%%%%%%
\section{Computational methods}

\subsection{Geometries and basis sets}

All our geometries are obtained at the CC3/{\AVTZ} \cite{Chr95b,Koc95} without using the frozen-core (FC) approximation (\ie, we correlate all the electrons). These geometries are given in the {Supporting Information (SI)}. Note that several
structures come from previous works, \cite{Loo18a,Loo20d} {and we used this level of theory here for consistency.} New optimizations have been achieved with the DALTON 2017 \cite{dalton} and CFOUR 2.1 \cite{cfour} 
codes applying default parameters in both cases.

As in our previous works, \cite{Loo18a,Loo19c,Loo20a,Loo20d} we consider the diffuse-containing Pople's  {\Pop} and Dunning's (d-)\emph{aug}-cc-pVXZ (X $=$ D, T, Q, and 5) atomic basis sets in all ES 
calculations.

In contrast with our previous studies, in which the complete basis set limit (CBS) could be obtained by a brute-force approach, $f$ and $\mu^{\mathrm{ES}}$ converge, in some cases, slower than the energy with respect to the basis set size.
Therefore, we have performed CBS extrapolation by applying the well-known Helgaker formula, \cite{Hel97b}
\begin{equation}
P_{\mathrm{CBS}} = \frac{P_{X-1}  \left( X-1\right)^3 - P_{X} X^3 }{\left( X-1\right)^3 - X^3},
\end{equation}
in which $X$ equals 2, 3, 4, \ldots for D, T, Q, \ldots in the Dunning series and $P$ is the property under investigation. Interestingly, this approach was successfully used for GS dipole moments. \cite{Hal99b} In practice, we
performed four extrapolations using both the CCSD and CCSDT results obtained with the pairs of largest singly- and doubly-augmented basis sets accessible {of at least triple-$\zeta$ size}, {i.e., i)} CCSD/{\AVQZ} and {\AVFZ};  
CCSDT/{\AVTZ} and {\AVQZ};  CCSD/{\DAVTZ} and {\DAVQZ}; and  CCSDT/{\DAVTZ} and {\DAVQZ} for molecules in Sections \ref{res-31}--\ref{res-35}; { ii) CCSD}/{\AVTZ} { and} {\AVQZ}; { and CCSD}/{\DAVTZ} { and} {\DAVQZ}
{ for molecules in Sections}  \ref{res-36}--\ref{res-39}. In such a way, we can estimate the extrapolation error, and provide error bars for the CBS values, {although such error bar is likely underestimated for the latter set
of compounds}. These CCSD {and/or} CCSDT CBS values are next used to  correct the properties obtained at higher levels (\emph{e.g.}, CCSDTQ) with a finite basis using the approach described below.

\subsection{Reference calculations}

We have chosen to use the MRCC (2017 and 2019) program, \cite{Kal20,mrcc} for performing our CC calculations, as this code allows to set up an arbitrary CC expansion order.  We therefore use the 
CCSD, \cite{Pur82,Scu87,Koc90b,Sta93,Sta93b} CCSDT, \cite{Nog87,Scu88,Kuc01,Kow01,Kow01b} CCSDTQ, \cite{Kuc91,Kal03,Kal04,Hir04} and CCSDTQP \cite{Kal03,Kal04,Hir04} hierarchy for energies, oscillator strengths, 
GS and ES dipoles. All these values have been obtained within the FC approximation.  {The interested reader
may find discussions about the impact of this approximation and the importance of core correlation functions for transition energies in some of our previous works.} \cite{Sue19,Loo19a} For CCSD, we performed 
several test calculations with GAUSSIAN 16,  \cite{Gaussian16} Q-CHEM 5.2, \cite{Kry13}  DALTON 2017, \cite{dalton} {and $e^T$ 1.0} \cite{eT} and we could not detect any significant discrepancy with respect to the MRCC 
results. At this stage, it is important to stress that all these calculations rely on the so-called linear-response (LR) formalism, \cite{Koc90,Chr98d,Kal04} so that while the same transition energies would be obtained with the equation-of-motion (EOM) 
approach, the properties would be different. However, it is known that the two formalisms become equivalent when the CC wave function becomes exact. As we strive here to be as close as possible from the FCI limit, 
we trust that our  theoretical best estimates (TBEs)  are not significantly affected by the selection of the LR implementation.   {At the CCSD level, we also provide a comparison between the EOM and LR oscillator strength values obtained with 
various codes in Table S1 in the SI. The differences found between the EOM and LR formalisms are very small.} 
The interested reader may {also} find extensive comparisons of oscillator strengths determined within the two formalisms elsewhere. \cite{Car09b}
We also note that all our oscillator strengths are given in the length gauge, the most commonly applied gauge, but again this choice is likely irrelevant when one is targeting 
near-exact values. The interested reader can find comparisons between CC oscillator strengths determined with the length, velocity, and mixed length-velocity gauges on small compounds elsewhere.\cite{Paw04} As expected 
the impact of the gauge was found to decrease when increasing the order of the CC expansion, especially when triples are included. Finally, we report in the Tables below the so-called \emph{orbital-relaxed} dipoles, 
which are more accurate than the so-called  \emph{orbital-unrelaxed} dipoles in which the impact of the external field on the orbitals is neglected. Details on various approaches and their implementations 
for correlated first-order properties can be found elsewhere. \cite{Tru88,Chr95,Chr98d}

Beyond the basis set extrapolation discussed above, we also define TBEs in the following. To this end, we use an incremental strategy for the transition 
energies and dipoles, \emph{e.g.},  at the {\AVTZ} level,
\begin{equation}
\begin{split}
	P(\mathrm{TBE}) 
	& = P(\mathrm{Low/AVTZ}) 
	\\
	& +  P(\mathrm{High/AVDZ}) - P(\mathrm{Low/AVDZ}),
\end{split}
\end{equation}
where Low and High denote, \emph{e.g.}, CCSDTQ and CCSDTQP (see footnotes in the Tables for specific details).  For oscillator strengths, we applied the corresponding multiplicative approach, \emph{e.g.}, 
\begin{equation}
	f(\mathrm{TBE}) = f(\mathrm{Low/AVTZ}) \frac{ f(\mathrm{High/AVDZ})}{f(\mathrm{Low/AVDZ})}.
\end{equation}
{Such incremental strategy that used a double-$\zeta$ result to estimate ``Q'' or ``P'' effects is commonly employed in the CC literature.}\cite{Kal04,Bal06,Kam06b,Wat12,Fel14,Fra19}

%%%%%%%%%%%%%%%%%%%%
%%% RESULTS %%%
%%%%%%%%%%%%%%%%%%%%
\section{Results and Discussion}
%%% TABLE I %%%
\begin{table*}[t]
\scriptsize
\caption{Ground-state dipole moment $\mu^{\mathrm{GS}}$, vertical transition energies $\Delta E_{\mathrm{vert}}$, oscillator strengths $f$, and excited-state dipole moments, $\mu_{\mathrm{vert}}^{\mathrm{ES}}$ and $\mu_{\mathrm{adia}}^{\mathrm{ES}}$, determined for BH (GS and ES geometries) and HCl (GS geometry). 
Transition energies are in eV and dipoles in D. } 
\label{Table-1}
%\vspace{-0.3 cm}
\begin{tabular}{cc|ccccc|cccc}
\hline 
		&				&\multicolumn{5}{c}{BH} & \multicolumn{4}{c}{HCl}  \\
		&				& $^1\Sigma^+$ & \multicolumn{4}{c}{$^1\Pi$ (Val)}			& $^1\Sigma^+$& \multicolumn{3}{c}{$^1\Pi$  (Val)}	\\
Basis 	& Method			&$\mu^{\mathrm{GS}}$	& $\Delta E_{\mathrm{vert}}$& $f$	& $\mu_{\mathrm{vert}}^{\mathrm{ES}}$		&$\mu_{\mathrm{adia}}^{\mathrm{ES}}$ 
						&$\mu^{\mathrm{GS}}$& $\Delta E_{\mathrm{vert}}$	&$f$		& $\mu_{\mathrm{vert}}^{\mathrm{ES}}$						\\
\hline
{\AVDZ}	&CCSD			&1.389		&2.970	&0.051	&0.530	&0.513				&1.147		&7.862		&0.066	&-2.773\\
		&CCSDT			&1.371		&2.946	&0.049	&0.543	&0.527				&1.131		&7.815		&0.065	&-2.745\\
		&CCSDTQ		&1.370		&2.947	&0.049	&0.545	&0.528				&1.130		&7.822		&0.065	&-2.728\\
		&CCSDTQP		&1.370		&2.947	&0.049	&0.545	&0.528				&1.130		&7.823		&0.065	&-2.727\\
{\AVTZ}	&CCSD			&1.433		&2.928	&0.050	&0.550	&0.534				&1.097		&7.906		&0.056	&-2.526\\
		&CCSDT			&1.410		&2.900	&0.048	&0.558	&0.541				&1.085		&7.834		&0.055	&-2.515\\
		&CCSDTQ		&1.409		&2.901	&0.048	&0.559	&0.542				&1.084		&7.837		&0.055	&-2.502\\
{\AVQZ}	&CCSD			&1.440		&2.918	&0.050	&0.555	&0.538				&1.111		&7.954		&0.051	&-2.410\\
		&CCSDT			&1.416		&2.890	&0.048	&0.561	&0.544				&1.098		&7.880		&0.050	&-2.410\\
{\AVFZ}	&CCSD			&1.443		&2.915	&0.050	&0.556	&0.540				&1.109		&7.961		&0.048	&-2.336\\
		&				&			&		&		&        	&					&			&			&		&	\\
{\DAVDZ}	&CCSD			&1.388		&2.969	&0.051	&0.519	&0.503				&1.135		&7.836		&0.064	&-2.697\\
		&CCSDT			&1.370		&2.945	&0.049	&0.533	&0.516				&1.119		&7.787		&0.063	&-2.670\\
{\DAVTZ}	&CCSD			&1.432		&2.927	&0.049	&0.550	&0.534				&1.096		&7.894		&0.055	&-2.491\\
		&CCSDT			&1.409		&2.900	&0.048	&0.558	&0.541				&1.084		&7.822		&0.054	&-2.480\\
{\DAVQZ}	&CCSD			&1.440		&2.918	&0.050	&0.555	&0.538				&1.111		&7.949		&0.050	&-2.405\\
		&CCSDT			&1.416		&2.890	&0.048	&0.561	&0.544				&1.098		&7.876		&0.050	&-2.406\\
		&				&			&		&		&		&					&			&			&		&	\\
{\AVTZ}	& TBE$^a$		&1.409		&2.901	&0.048	&0.559	&0.542				&1.084		&7.837		&0.055	&-2.501\\
CBS		& TBE$^b$		&1.42$\pm$0.00&2.88$\pm$0.01&0.048$\pm$0.001&0.57$\pm$0.01&0.55$\pm$0.01	&1.11$\pm$0.01	&7.91$\pm$0.01&0.046$\pm$0.001&	 -2.32$\pm$0.01	\\
		&				&			&			&		&				&		&	\\
Lit.		& Th.			&1.425$^c$	&2.944$^d$&	&		&	0.57$^e$				&1.106$^f$	&7.94$^g$	&0.081$^g$		&\\
		&	Exp.			&1.27$\pm$0.21$^h$&		&0.044$^i$	& &0.58$\pm$0.04$^h$	&			&8.05$^j$		&0.051$^j$		&\\	
		&				&			&			&		&			&			&			&			&0.042$\pm$0.004$^k$	&\\
		&				&			&			&		&			&			&			&			&0.052$\pm$0.006$^l$	&\\
\hline																			
\end{tabular}
%\vspace{-0.3 cm}
\begin{flushleft}
$^a${Computed using CCSDTQ/{\AVTZ} values and CCSDTQP/{\AVDZ} corrections. Note that CCSDTQ is equivalent to FCI for BH;}
$^b${See Computational Methods section;}
$^c${Average between the MR-ACPF/\emph{aug}-cc-pCV7Z(i) values obtained for $r=1.220$ and 1.225 \AA\ in Ref.~\citenum{Kop15};}
$^d${FCI/\emph{aug}-cc-pVDZ value of Ref.~\citenum{Koc95};}
$^e${CC2/{\AVQZ} result from Ref.~\citenum{Hel11};}
$^f${CCSD(T)/CBS value from Ref.~\citenum{Hai18};}
$^g${CISDTQ/\emph{aug}-cc-pCVQZ value from Ref.~\citenum{Eng12}. A $f$ value of 0.071 is also reported in Table II of the same work;}
$^h${Stark (emission) measurements of Ref.~\citenum{Tho69};}
$^i${$f_{00}$ obtained from laser-induced fluorescence in Ref.~\citenum{Dou89}. There is no major contributions from other bands according to this work (see references therein for previous experimental values). A slightly
older experiment (Ref.~\citenum{Duf78}) reports a smaller estimate of 0.045$\pm$0.02;}
$^j${Absorption values from Ref.~\citenum{Che02} ($\Delta E$ corresponds to the maximum of absorption);}
$^k${EELS value from  Ref.~\citenum{Li06c};}
$^l${HR-EELS value from Ref.~\citenum{Xu19}.}
\end{flushleft}
\end{table*}

Below, we discuss individual molecules going up on size progressively. Concerning literature references, we do not intend to provide an exhaustive list of all previous works for each system considered here, but rather to 
highlight the studies and comparisons that we have found valuable for the present work.

\subsection{BH and HCl}
\label{res-31}

Let us start by a tiny compound, BH. Our results are listed in Table \ref{Table-1} and although the size of this molecule seems ridiculous   (only 4 valence electrons), some valuable conclusions can be obtained. 
We note that  $\mu^{\mathrm{GS}}$ is slightly too large with CCSD (irrespective of the basis set) but the CC convergence is fast and CCSDT is obviously sufficient. Our TBE/CBS of 1.42 D for $\mu^{\mathrm{GS}}$ is equivalent to a recent 
ultra-accurate estimate, \cite{Kop15} and also falls within the error bar of the only available experimental value we are aware of: 1.27$\pm$0.21 D.  \cite{Tho69}  Such large error bar is explainable: the experiment relied on an 
analysis of the emission spectrum. \cite{Tho69}  It is quite obvious that the theoretical estimate is more trustworthy in this specific case, indicating that previous error analyses based on the 1.27 D value likely 
significantly exaggerated the ADC(2) and CC2 overestimations, \cite{Hel11} but underestimated the QMC error \cite{Lu03} for $\mu^{\mathrm{GS}}$. For the lowest transition energy in BH, our values fit with 
previous FCI calculations, \cite{Koc95} and one again notices rather quick convergence of $\Delta E_{\mathrm{vert}}$ with respect to both basis set size and CC order.  For the oscillator strength, there is also an astonishing 
stability of the values, as the considered ES is well separated from higher-lying ones of the same spatial symmetry. Our TBE is close to the most recent measurement (of $f_{00}$) we could find. \cite{Dou89} For 
$\mu^{\mathrm{ES}}$, one notes a very small decrease of the amplitude between the GS and ES geometries, which is a logical consequence of the tiny geometrical relaxation (+0.018 \AA), and our TBE is within the 
rather small experimental error bar (see bottom of Table \ref{Table-1}). {Finally, as can be seen in Table S2, these results are not affected by the FC approximation, e.g., the difference between CCSDTQ(FC)/}{\AVTZ}
{and CCSDTQ(Full)/}{\ACVTZ} {is 0.001 for $f$ and 0.002 D for the dipole moments. }

For HCl, the GS dipole moment does not cause any specific challenge and our TBE is equivalent to the one recently reported by the Head-Gordon group.\cite{Hai18} As expected, \cite{Loo18a} CCSDTQ provides
converged $\Delta E_{\mathrm{vert}}$ and this holds for both $f$ and $\mu^{\mathrm{ES}}$. When increasing the size of the basis set, one sees a significant decrease of the oscillator strength (ca.~-25\%\ from {\AVDZ} to {\AVFZ}) and of the ES
dipole (-16\%\ for the same basis pair), whereas the transition energy varies by 1\%\ only. Nevertheless, the CBS extrapolations are stable for all investigated properties.  For HCl, it is difficult to obtain a very accurate experimental $f$ value, 
in part due to the mixing of the $^1\Pi$ and $^3\Pi$ states. \cite{Che02} There is therefore a broad range of measured values (see bottom of Table  \ref{Table-1} as well as Table 2 in Ref.~\citenum{Xu19}), and our TBE/CBS of 
0.046 is compatible with the two most recent experiments. In contrast, we could not find any experimental  $\mu^{\mathrm{ES}}$ estimate, which is a logical consequence of the dissociative character of the lowest singlet ES of HCl.

\subsection{H$_2$O and H$_2$S}

%%% TABLE H2O %%%
\begin{table*}[htp]
\scriptsize
\caption{Ground-state dipole moment $\mu^{\mathrm{GS}}$, vertical transition energies $\Delta E_{\mathrm{vert}}$, oscillator strengths $f$, and excited state dipole moments $\mu_{\mathrm{vert}}^{\mathrm{ES}}$ determined for H$_2$O (GS geometry). See caption of Table \ref{Table-1} for details.} 
\label{Table-2}
\vspace{-0.3 cm}
\begin{tabular}{cc|ccccccccc}
\hline 
		&			& $^1A_1$	&	\multicolumn{3}{c}{$^1B_1 (\mathrm{Ryd}, n\rightarrow 3s)$}	&	\multicolumn{2}{c}{$^1A_2 (\mathrm{Ryd}, n\rightarrow 3p)$}	&	\multicolumn{3}{c}{$^1A_1 (\mathrm{Ryd}, n\rightarrow 3s)$}	\\
Basis 	& Method		&$\mu^{\mathrm{GS}}$	&$\Delta E_{\mathrm{vert}}$	&$f$		& $\mu_{\mathrm{vert}}^{\mathrm{ES}}$		&	$\Delta E_{\mathrm{vert}}$		&$\mu_{\mathrm{vert}}^{\mathrm{ES}}$	
&$\Delta E_{\mathrm{vert}}$	&$f$			&$\mu_{\mathrm{vert}}^{\mathrm{ES}}$		\\
\hline
{\AVDZ}	&CCSD			&1.870		&7.447		&0.057	&-1.404		&9.213		&-0.936	&9.861	&0.103	&-1.095 \\
		&CCSDT			&1.849		&7.497		&0.058	&-1.420		&9.279		&-0.978	&9.903	&0.104	&-1.149 \\
		&CCSDTQ		&1.848		&7.529		&0.058	&-1.415		&9.313		&-0.974	&9.937	&0.105	&-1.143 \\
		&CCSDTQP		&1.848		&7.522		&0.058	&-1.414		&9.318		&-0.972	&9.941	&0.105	&-1.141 \\
{\AVTZ}	&CCSD			&1.864		&7.597		&0.053	&-1.549		&9.361		&-1.056	&9.957	&0.098	&-1.163 \\
		&CCSDT			&1.842		&7.591		&0.054	&-1.565		&9.368		&-1.110	&9.949	&0.100	&-1.221 \\
		&CCSDTQ		&1.840		&7.620		&0.054	&-1.559		&9.401		&-1.107	&9.981	&0.100	&-1.215 \\
{\AVQZ}	&CCSD			&1.873		&7.660		&0.052	&-1.655		&9.422		&-1.237	&10.004	&0.095	&-1.226 \\
		&CCSDT			&1.850		&7.637		&0.053	&-1.667		&9.410		&-1.294	&9.980	&0.097	&-1.286 \\
{\AVFZ}	&CCSD			&1.876		&7.683		&0.051	&-1.726		&9.444		&-1.309	&10.010	&0.089	&-1.229 \\
		&				&			&			&		&			&			&		&		&		&		\\
{\DAVDZ}	&CCSD			&1.861		&7.429		&0.052	&-1.759		&9.179		&-1.658	&9.731	&0.050	&-0.884 \\
		&CCSDT			&1.841		&7.479		&0.053	&-1.754		&9.244		&-1.703	&9.792	&0.057	&-1.021 \\
{\DAVTZ}	&CCSD			&1.865		&7.592		&0.051	&-1.764		&9.348		&-1.626	&9.869	&0.057	&-1.154 \\
		&CCSDT			&1.843		&7.586		&0.052	&-1.767		&9.353		&-1.681	&9.872	&0.062	&-1.232 \\
{\DAVQZ}	&CCSD			&1.874		&7.659		&0.051	&-1.765		&9.416		&-1.622	&9.932	&0.058	&-1.228 \\
		&CCSDT			&1.851		&7.636		&0.052	&-1.770		&9.403		&-1.678	&9.917	&0.063	&-1.295 \\
		&				&			&			&		&			&			&\\
{\AVTZ}	& TBE$^a$		&1.840		&7.614		&0.054	&-1.558		&9.405		&-1.106	&9.985	&0.100	&-1.213 \\
CBS		& TBE$^b$		&1.86$\pm$0.01&7.71$\pm$0.02& 0.052$\pm$0.001&-1.77$\pm$0.04&9.49$\pm$0.02&-1.67$\pm$0.01&9.99$\pm$0.01&0.062$\pm$0.002&-1.30$\pm$0.03\\
		&				&			&			&			&			&		&		\\
Lit.		& Th.			&1.853$^c$	&7.66$^c$	&0.054$^d$	&-1.787$^c$	& 9.42$^c$  &-1.682$^c$ &9.97$^e$ & 0.100$^d$\\
		&				&			&7.70$^e$	&0.049$^f$	&			&9.47$^e$ & 	\\
		&				&			&7.71$^f$		&		&			&				& &9.92$^g$ & 0.055$^g$	&\\
		& Exp.			&1.850$^h$	&7.41$^i$		&0.046$\pm$0.007$^j$		&			&9.20$^i$		& 	&9.67$^i$	 & 0.051$^k$\\
\hline																			
\end{tabular}
\vspace{-0.3 cm}
\begin{flushleft}
$^{a,b}${See corresponding footnotes in Table \ref{Table-1};}
$^c${CASPT2/{\DAVQZ} values from Ref.~\citenum{Pal08};} 
$^d${LR-CC3/{\AVTZ} value from Ref.~\citenum{Loo18a}; }
$^e${Basis set corrected exFCI/{\AVQZ} values from Ref.~\citenum{Loo18a}; }
$^f${2FVCAS/MR-CI/CBS values ($f$ in length gauge) from Ref.~\citenum{Bor06a};}
$^g${2FVCAS/MR-CI/d-\emph{aug}-cc-pV5Z values ($f$ in length gauge) from Ref.~\citenum{Bor06b};}
$^h${Average of the three (very close) experimental values reported in Table 1 of  Ref.~\citenum{Joh80};}
$^i${Energy loss experiment from Ref.~\citenum{Ral13};}
$^j${Electron impact from Ref.~\citenum{Tho07};}
$^k${Electron impact from Ref.~\citenum{Mot05} integrated by Borges in Ref.~\citenum{Bor06b}.}
\end{flushleft}
\end{table*}

The results obtained for water and hydrogen sulfide are listed in Tables \ref{Table-2} and \ref{Table-3}, respectively. The CCSDT/{\AVTZ} estimate of the ground-state dipole of water is already within 0.01 D of both previous CASPT2\cite{Pal08}
and experimental values.\cite{Joh80} For all transition energies, as discussed in our earlier work, \cite{Loo18a} the calculations are converged with CCSDTQ, but one needs a rather large basis set (especially in terms of diffuse functions) to be chemically accurate, which
is quite usual for Rydberg transitions in small compounds. Nevertheless, CCSDT/{\AVTZ} delivers $\Delta E_{\mathrm{vert}}$ with an error of 1--3\%\ only as compared to the most accurate estimates (see bottom of Table \ref{Table-2}) for all three 
transitions. For the lowest $B_1$ excitation, the oscillator strength $f$ varies rather mildly with the selected level of theory and basis set, although one notices a general decreasing trend when improving the method. Our TBE of 0.052$\pm$0.001 falls
in the error bar of a recent experiment, \cite{Tho07} and is only slightly larger than the previous most accurate TBE we are aware of. \cite{Bor06a}  More exhaustive lists of additional theoretical and experimental values can be 
found in Table IV of Ref.~\citenum{Tho07} and Table 6 of Ref.~\citenum{Bor06a} for the lowest transition. The reported $f$ values in these tables are in the range 0.041--0.060.  The magnitude of $\mu^{\mathrm{ES}}$ for this $B_1$ ES increases 
significantly with the basis set size, and a difference of -0.21 D exists between our {\AVTZ} and CBS TBEs.  The latter compares well with an earlier CASPT2 value.\cite{Pal08} This trend is even exacerbated for the  $\mu^{\mathrm{ES}}$ value associated with the $A_2$ state
that changes by -0.56 D from  {\AVTZ} to CBS, whereas no significant changes can be noticed between CCSDT and CCSDTQ. For the lowest $A_1$ ES, there is a significant mixing with a close-lying state of the same symmetry, 
and the addition of a second set of diffuse is mandatory to obtain reasonable estimates of the oscillator strength (which is much too large with {\AVTZ}). Our TBE/CBS $f$ value of 0.062 could still be slightly too large, but the experimental
values range from 0.041 to 0.073 (see Table 3 in Ref.~\citenum{Bor06b}). In contrast to what we found for the two lower-lying ES, $\mu^{\mathrm{ES}}$ for the $A_1$ ES is not very sensitive to the basis set size, with a difference of 
-0.09 D only between the  {\AVTZ} and CBS TBEs.  We can already conclude, from the data of Table \ref{Table-2}, that the basis set required to reach accurate estimates not only differs from one ES to another, but might
also be very different, for a given ES, in oscillator strengths and ES dipoles. On a brighter note, the improvements brought by the Q and P excitations are rather limited, meaning that CCSDT seems already sufficient (for water at least).

%%% TABLE H2S %%%
\begin{table*}[htp]
\scriptsize
\caption{Ground-state dipole moment $\mu^{\mathrm{GS}}$, vertical transition energies $\Delta E_{\mathrm{vert}}$, oscillator strengths $f$, and excited state dipole moments $\mu^{\mathrm{ES}}$ determined for H$_2$S (GS geometry). See caption of Table \ref{Table-1} for details.} 
\label{Table-3}
\vspace{-0.3 cm}
\begin{tabular}{cc|cccccc}
\hline 
		&				& $^1A_1$ 	&	\multicolumn{2}{c}{$^1A_2 (\mathrm{Ryd}, n\rightarrow 4p)$}	&	\multicolumn{3}{c}{$^1B_1 (\mathrm{Ryd}, n\rightarrow 4s)$}	\\
Basis 	& Method			&$\mu^{\mathrm{GS}}$			&	$\Delta E_{\mathrm{vert}}$		& $\mu^{\mathrm{ES}}$		&$\Delta E_{\mathrm{vert}}$		&$f$			&$\mu^{\mathrm{ES}}$	\\
\hline
{\AVDZ}	&CCSD			&1.031			&6.343			&0.113		&6.141		&0.068		&-1.983\\
		&CCSDT			&1.016			&6.286			&0.131		&6.098		&0.067		&-1.946\\
		&CCSDTQ		&1.015			&6.286			&0.137		&6.103		&0.067		&-1.934\\
		&CCSDTQP		&1.015			&6.286			&0.137		&6.103		&0.067		&-1.933\\
{\AVTZ}	&CCSD			&0.990			&6.246			&0.503		&6.295		&0.064		&-1.893\\
		&CCSDT			&0.978			&6.185			&0.496		&6.237		&0.063		&-1.875\\
		&CCSDTQ		&0.977			&6.181			&0.498		&6.238		&0.063		&-1.866\\
{\AVQZ}	&CCSD			&1.001			&6.212			&0.650		&6.349		&0.062		&-1.822\\
		&CCSDT			&0.990			&6.153			&0.636		&6.288		&0.061		&-1.815\\
{\AVFZ}	&CCSD			&0.998			&6.177			&0.691		&6.368		&0.061		&-1.794\\
		&				&				&				&			&			&			&		\\
{\DAVDZ}	&CCSD			&1.017			&6.297			&0.445		&6.130		&0.065		&-1.811\\
		&CCSDT			&1.002			&6.241			&0.458		&6.086		&0.065		&-1.776\\
{\DAVTZ}	&CCSD			&0.989			&6.228			&0.597		&6.292		&0.062		&-1.774\\
		&CCSDT			&0.977			&6.167			&0.587		&6.234		&0.062		&-1.761\\
{\DAVQZ}	&CCSD			&1.001			&6.206			&0.668		&6.347		&0.061		&-1.758\\
		&CCSDT			&0.989			&6.147			&0.653		&6.286		&0.061		&-1.755\\
		&				&				&				&			&			&			&\\
{\AVTZ}	& TBE$^a$		&0.977			&6.181			&0.498		&6.238		&0.063		&-1.865\\
CBS		& TBE$^b$		&0.99$\pm$0.01	&6.10$\pm$0.03	&0.72$\pm$0.01&6.33$\pm$0.01&0.060$\pm$0.001&-1.74$\pm$0.02	\\
		&				&				&				&			&			&		&		\\
Lit.		&	Th.			&0.989$^c$		&6.12$^c$;6.10$^d$;6.10$^e$&0.653$^c$	&6.27$^c$;6.29$^d$;6.33$^e$	&0.063$^f$&-1.733$^c$\\
		&	Exp.			&0.974$\pm$0.005$^g$&				&			&6.326$^h$	&0.0542$^i$;0.0547$^j$\\
\hline																			
\end{tabular}
\vspace{-0.3 cm}
\begin{flushleft}
$^{a,b}${See corresponding footnotes in Table \ref{Table-1};}
$^c${CASPT2/{\DAVQZ} values from Ref.~\citenum{Pal08};} 
$^d${Basis set corrected exFCI/{\AVQZ} values from Ref.~\citenum{Loo18a}; }
$^e${SS-RASPT2/ANO-RCC-VTZP+diffuse values from Ref.\citenum{Ert20};  }
$^f${LR-CC3/{\AVTZ} value from Ref.~\citenum{Loo18a}; }
$^g${From Ref.~\citenum{Hui65};}
$^h${From Ref.~\citenum{Mas79} (see Table 9 of this work);}
$^i${From Ref.~\citenum{Lee87}, obtained by integrating the experimental absorption spectrum in the 5.2--7.7 eV region:}
$^j${From Ref.~\citenum{Fen99}, obtained by integrating the experimental absorption spectrum in the 5.2--7.7 eV region.}
\end{flushleft}
\end{table*}

As can be seen in Table \ref{Table-3}, the ground-state dipole moment of hydrogen sulfide remains almost unchanged when increasing the CC order or the size of the basis set: all estimates fall
in a tight window: 1.00$\pm$0.03 D. Our TBE/CBS of 0.99 D is very close from the experimental value of 0.974$\pm$0.005 D, although we use a theoretical geometry. For the lowest $^1A_2$
transition, the excitation energies are almost converged with CCSDT, but a rather large basis set is required, like in water. Our TBE/CBS of 6.10$\pm$0.03 eV agrees with 
previous high-order estimates.  \cite{Pal08,Loo18a,Ert20}  To the best of our knowledge there are no accurate experimental estimates for this dark state. While the methodological effects remain firmly
under control for the ES dipole, CCSDT being again sufficient, the basis set effects are huge: at the CCSDT level, we have $\mu^{\mathrm{ES}} = 0.13$ D with \emph{aug}-cc-pVDZ but more than four times larger (0.65 D)
with d-\emph{aug}-cc-pVQZ. Our TBE/CBS of 0.72$\pm$0.02 D is slightly larger than the best previous estimate we have found (0.65 D). \cite{Pal08} The transition energy to the second
ES, of $B_1$ symmetry, is also basis set sensitive, although in that case larger bases yield larger (and not smaller) transition energies, as shown in Table \ref{Table-3}. Our TBE/CBS of 6.33$\pm$0.01 eV is close
to previous estimates, \cite{Pal08,Loo18a,Ert20} and also consistent with measurements, \cite{Mas79} although we recall that such comparisons should be made with care. The dipole moment of the 
$B_1$ state is very large and relatively insensitive to the basis set as compared to its $A_2$ counterpart. Our TBE/CBS of -1.74$\pm$0.02 D is very close to an earlier CASPT2 estimate. \cite{Pal08}
Finally, the computed oscillator strength is rather methodologically insensitive and our best estimate  of 0.060 is within 0.010 of the two most recent measurements we could found, \cite{Lee87,Fen99}
and compares favorably with two (rather old) theoretical estimates: 0.081 \cite{Per97d} and 0.075. \cite{Rau84} 

\subsection{BF}

For this diatomic, the lowest $\Pi$ ES behaves rather nicely, and one notices in Table \ref{Table-4} that we could reach very stable estimates for all investigated properties, CCSDT/{\AVTZ} delivering
already sufficiently accurate values. It is noteworthy, that for any given basis set, CCSD underestimates $\mu^{\mathrm{GS}}$ ($\mu^{\mathrm{ES}}$) by ca.~4\%\ (25\%), highlighting the
difficulty posed by ES for ``simple'' methods. The experimental value of $\mu^{\mathrm{GS}}$ was measured to be 0.5$\pm$0.2 D, \cite{Lov71} a value that was suggested to be too low more than five 
scores ago. \cite{Hon93} Indeed, the present TBE/CBS of 0.80 D is significantly above the upper limit of the experimental error bar, but in good agreement with an earlier MRCI+Q value (0.84 D)
obtained with a very large basis set. \cite{Mag13}  Our best estimate for $\Delta E_{\mathrm{vert}}$, 6.39 eV, is slightly larger than a rather old CIPSI value of 6.329 eV, \cite{Mer97} whereas 
unfortunately, Ref.~\citenum{Mag13} lists adiabatic energies only. For the oscillator strength, the TBE of Table \ref{Table-4} is larger than previous experimental \cite{Hub79} and theoretical \cite{Hon93} 
data, but the uniformity of our estimates gives confidence in their quality. Eventually, we predict a significant drop in polarity when going from the GS to the ES, with a $\mu^{\mathrm{ES}}$ value of ca.~0.27 D.
This value is much larger than the MRCI+Q $\mu^{\mathrm{ES}}$ reported by Magoulas and coworkers (0.01 D), \cite{Mag13} but the latter is obtained on the ES rather than the GS geometry.%...

%%% TABLE BF %%%
\begin{table}[htp]
\scriptsize
\caption{Ground-state dipole moment $\mu^{\mathrm{GS}}$, vertical transition energies $\Delta E_{\mathrm{vert}}$ , oscillator strengths $f$, and excited state dipole moments $\mu^{\mathrm{ES}}$ determined for BF (GS geometry). See caption of Table \ref{Table-1} for details.} 
\label{Table-4}
\vspace{-0.3 cm}
\begin{tabular}{cc|cccc}
\hline 
		&				& $^1A_1$ 	&	\multicolumn{3}{c}{$^1\Pi (\mathrm{Val}, \sigma \rightarrow \pi^\star)$}	\\
Basis 	& Method			&$\mu^{\mathrm{GS}}$		&$\Delta E_{\mathrm{vert}}$		&$f$			&$\mu^{\mathrm{ES}}$	\\
\hline
{\AVDZ}	&CCSD			&0.832			&6.534	&0.479	&0.240\\
		&CCSDT			&0.861			&6.491	&0.475	&0.311\\
		&CCSDTQ		&0.861			&6.486	&0.474	&0.316\\
		&CCSDTQP		&0.860			&6.485	&0.474	&0.316\\
{\AVTZ}	&CCSD			&0.794			&6.464	&0.475	&0.222\\
		&CCSDT			&0.824			&6.423	&0.469	&0.293\\
		&CCSDTQ		&0.824			&6.417	&0.468	&0.300\\
{\AVQZ}	&CCSD			&0.783			&6.449	&0.475	&0.207\\
		&CCSDT			&0.812			&6.411	&0.468	&0.279\\
{\AVFZ}	&CCSD			&0.782			&6.443	&0.475	&0.202\\
		&				&				&		&		&	\\
{\DAVDZ}	&CCSD			&0.822			&6.521	&0.478	&0.262\\
		&CCSDT			&0.850			&6.477	&0.474	&0.330\\
{\DAVTZ}	&CCSD			&0.794			&6.459	&0.476	&0.237\\
		&CCSDT			&0.824			&6.419	&0.469	&0.308\\
{\DAVQZ}	&CCSD			&0.784			&6.448	&0.475	&0.213\\
		&CCSDT			&0.813			&6.409	&0.468	&0.285\\
		&				&				&		&		&	\\
{\AVTZ}	& TBE$^a$		&0.824			&6.417	&0.468	&0.299\\
CBS		& TBE$^b$		&0.80$\pm$0.01	&6.39$\pm$0.01&0.467$\pm$0.001&0.27$\pm$0.01\\
		&				&				&		&		&	\\
Lit.		&	Th.			&0.84$^c$		&6.329$^d$&0.30$^e$& 0.01$^c$\\
		&	Exp.			&0.5$\pm$0.2$^f$	&		&0.40$^g$\\
\hline	
\end{tabular}
\vspace{-0.3 cm}
\begin{flushleft}
$^{a,b}${See corresponding footnotes in Table \ref{Table-1};}
$^c${MRCI+Q/\emph{aug}-cc-pV6Z values from Ref.~\citenum{Mag13}, the ES dipole was obtained on the corresponding ES geometry;} 
$^d${CIPSI/(11s7p4d3f)/[6s4p4d3f] estimate from Ref.~\citenum{Mer97};}
$^e${Sum of the MRCI $f$ value determined for GS $\nu''=0$, from Table 5 of Ref.~\citenum{Hon93};}
$^f${From Ref.~\citenum{Lov71};}
$^g${$f_{00}$ value from Ref.~\citenum{Hub79}.}
\end{flushleft}
\end{table}

\subsection{CO}

%%% TABLE CO %%%
\begin{table*}[htp!]
\scriptsize
\caption{Ground-state dipole moment $\mu^{\mathrm{GS}}$, vertical transition energies $\Delta E_{\mathrm{vert}}$, oscillator strengths $f$, and excited state dipole moments, $\mu_{\mathrm{vert}}^{\mathrm{ES}}$ and $\mu_{\mathrm{adia}}^{\mathrm{ES}}$, determined for various transitions in CO at its GS geometry.
The transition energies are in eV and the dipoles in D. } 
\label{Table-5}
\vspace{-0.3 cm}
\begin{tabular}{cc|ccccccccc}
\hline 
		&				& $^1\Sigma^+$ & \multicolumn{4}{c}{$^1\Pi (\mathrm{Val}, n \rightarrow \pi^\star)$}	& \multicolumn{2}{c}{$^1\Sigma^- (\mathrm{Val}, \pi \rightarrow \pi^\star)$}	 & \multicolumn{2}{c}{$^1\Delta (\mathrm{Val}, \pi \rightarrow \pi^\star)$}\\
Basis 	& Method			&$\mu^{\mathrm{GS}}$	& $\Delta E_{\mathrm{vert}}$& $f$	& $\mu_{\mathrm{vert}}^{\mathrm{ES}}$ 	&$\mu_{\mathrm{adia}}^{\mathrm{ES}}$ 
						& $\Delta E_{\mathrm{vert}}$	& $\mu_{\mathrm{vert}}^{\mathrm{ES}}$
						& $\Delta E_{\mathrm{vert}}$	& $\mu_{\mathrm{vert}}^{\mathrm{ES}}$						\\
\hline
{\AVDZ}	&CCSD			&0.078	&8.671	&0.167	&-0.157&-0.438		&10.096	&1.574	&10.210	&1.405	\\
		&CCSDT			&0.121	&8.574	&0.173	&-0.079&-0.331		&10.062	&1.597	&10.178	&1.432	\\
		&CCSDTQ		&0.130	&8.563	&0.174	&-0.072&-0.317		&10.057	&1.613	&10.169	&1.446	\\
		&CCSDTQP		&0.132	&8.561	&0.175	&-0.069&-0.312		&10.057	&1.615	&10.168	&1.448	\\
{\AVTZ}	&CCSD			&0.051	&8.587	&0.161	&-0.227&-0.478		&9.986	&1.577	&10.123	&1.413	\\
		&CCSDT			&0.104	&8.492	&0.164	&-0.137&-0.358		&9.940	&1.595	&10.076	&1.432	\\
		&CCSDTQ		&0.113	&8.480	&0.166	&-0.129&-0.344		&9.932	&1.612	&10.066	&1.449	\\
{\AVQZ}	&CCSD			&0.039	&8.574	&0.160	&-0.264&-0.511		&9.992	&1.567	&10.127	&1.402	\\
		&CCSDT			&0.094	&8.480	&0.163	&-0.169&-0.385		&9.940	&1.586	&10.073	&1.421	\\
{\AVFZ}	&CCSD			&0.037	&8.571	&0.160	&-0.280&-0.523		&		&		&10.130	&1.398	\\
		&				&		&		&		&	    &			&		&		&		&		\\
{\DAVDZ}	&CCSD			&0.085	&8.663	&0.167	&-0.149&-0.442		&10.087	&1.561	&10.199	&1.386	\\
		&CCSDT			&0.128	&8.565	&0.173	&-0.069&-0.334		&10.053	&1.584	&10.167	&1.413	\\
{\DAVTZ}	&CCSD			&0.053	&8.582	&0.160	&-0.232&-0.483		&9.983	&1.568	&10.120	&1.401	\\
		&CCSDT			&0.106	&8.487	&0.164	&-0.141&-0.364		&9.937	&1.586	&10.073	&1.421	\\
{\DAVQZ}	&CCSD			&0.040	&8.572	&0.160	&-0.269&-0.513		&9.992	&1.563	&10.126	&1.397	\\
		&CCSDT			&0.094	&8.478	&0.163	&-0.173&-0.389		&9.940	&1.581	&10.073	&1.416	\\
		&				&		&		&		&		&		&		&		&		&		\\
{\AVTZ}	& TBE$^a$		&0.115	&8.478	&0.166	&-0.126	&-0.339	&9.932	&1.614	&10.065	 & 1.450	 \\
CBS		& TBE$^b$		&0.097$\pm$0.002&	8.46$\pm$0.01	& 0.165$\pm$0.001 & -0.19$\pm$0.01 &-0.39$\pm$0.01	&9.94$\pm$0.01 &1.60$\pm$0.01 &10.07$\pm$0.01 &1.43$\pm$0.01	\\
		&				&		&		&		&\\
Lit.		& Th.			&0.1172$^c$&8.541$^d$&0.121$^e$	&-0.135$^f$	&&10.045$^d$	&		&10.182$^d$			\\
		&				&0.091$^f$&8.48$^g$& 0.168$^g$	&	&-0.05$^h$;-0.19$^i$&9.98$^g$	&		&10.10$^g$			\\%Loo18a	
		&	Exp.			&0.122$^j$	&8.51$^k$	&0.181$^l$	&&-0.15$\pm$0.05$^m$		&9.88$^k$	&		&10.23$^k$\\
		&				&			&			&0.194$^n$&	&-0.335$\pm$0.013$^o$\\
		&				&\\
\hline 
		&				& \multicolumn{3}{c}{$^1\Sigma^+ (\mathrm{Ryd})$}	 & \multicolumn{3}{c}{$^1\Sigma^+ (\mathrm{Ryd})$}	 & \multicolumn{3}{c}{$^1\Pi (\mathrm{Ryd})$}	 \\
Basis 	& Method			&$\Delta E_{\mathrm{vert}}$& $f$	& $\mu_{\mathrm{vert}}^{\mathrm{ES}}$	&$\Delta E_{\mathrm{vert}}$& $f$	& $\mu_{\mathrm{vert}}^{\mathrm{ES}}$	&
						$\Delta E_{\mathrm{vert}}$& $f$	& $\mu_{\mathrm{vert}}^{\mathrm{ES}}$							\\
\hline
{\AVDZ}	&CCSD			&11.171	&0.003	&-4.134	&11.710	&0.248	&6.323	&11.973	&0.132	&0.192	\\
		&CCSDT			&10.944	&0.001	&-4.519	&11.518	&0.240	&6.699	&11.767	&0.124	&0.149	\\
		&CCSDTQ		&10.926	&0.000	&-4.572	&11.510	&0.238	&6.768	&11.758	&0.122	&0.143	\\
		&CCSDTQP		&10.919	&0.000	&-4.592	&11.506	&0.238	&6.792	&11.753	&0.121	&0.141	\\
{\AVTZ}	&CCSD			&11.222	&0.008	&-3.376	&11.751	&0.208	&5.600	&11.960	&0.115	&0.195	\\
		&CCSDT			&10.987	&0.004	&-3.894	&11.540	&0.203	&6.094	&11.737	&0.110	&0.138	\\
		&CCSDTQ		&10.963	&0.003	&-3.994	&11.523	&0.202	&6.205	&11.720	&0.108     	& 0.126        \\
{\AVQZ}	&CCSD			&11.190	&0.010	&-2.792	&11.733	&0.183	&5.097	&11.916	&0.103	&0.084	\\
		&CCSDT			&10.954	&0.006	&-3.350	&11.514	&0.180	&5.619	&11.687	&0.098	&0.007	\\
{\AVFZ}	&CCSD			&11.133	&0.012	&-2.077	&11.691	&0.160	&4.614	&11.851	&0.091	&-0.158	\\
		&				&		&		&		&		&		&		&		&		&		\\
{\DAVDZ}	&CCSD			&10.795	&0.009	&-1.462	&11.393	&0.134	&3.498	&11.535	&0.067	&-1.957	\\
		&CCSDT			&10.569	&0.006	&-1.981	&11.175	&0.129	&3.927	&11.313	&0.061	&-2.146	\\
{\DAVTZ}	&CCSD			&10.960	&0.010	&-1.337	&11.567	&0.134	&3.814	&11.700	&0.074	&-1.449	\\
		&CCSDT			&10.726	&0.007	&-1.855	&11.340	&0.130	&4.216	&11.468	&0.069	&-1.648	\\
{\DAVQZ}	&CCSD			&11.010	&0.011	&-1.298	&11.617	&0.134	&3.946	&11.749	&0.075	&-1.233	\\
		&CCSDT			&10.774	&0.007	&-1.810	&11.388	&0.130	&4.337	&11.516	&0.070	&-1.437	\\
		&				&\\
{\AVTZ}	& TBE$^a$		&10.956	&0.003	&-4.014	&11.519	&0.202	&6.229	&11.715	&0.107	&0.124\\
CBS		& TBE$^b$		&10.84$\pm$0.06	& 0.005$\pm$0.001 & -2.49$\pm$0.59 &11.43$\pm$0.04	& 0.146$\pm$0.016 & 4.94$\pm$0.47
						&11.58$\pm$0.05	& 0.078$\pm$0.009 & -0.70$\pm$0.60\\
		&				&\\
Lit.		& Th.			&10.983$^d$	&0.029$^e$	&-2.79$^i$ &			&0.133$^e$	& 5.34$^i$		&	\\
		&				&10.80$^g$	&0.003$^g$	&		&11.42$^g$		&0.200$^g$	&		&11.55$^g$	&0.106$^g$	&\\
		&	Exp.			&10.78$^k$	&0.009$^l$	&-1.60$\pm$0.15$^m$		&11.40$^k$		&0.121$^l$	&4.52$\pm$0.35$^m$		&11.53$^k$ & 0.074$^l$ &\\
		&				&		&			&	-1.95$\pm$0.03$^o$	&	&0.136$^n$		&4.50$\pm$0.07$^o$		&		&0.074$^n$\\
\hline																			
\end{tabular}
\vspace{-0.3 cm}
\begin{flushleft}
$^{a,b}${See corresponding footnotes in Table \ref{Table-1};}
$^c${CCSD(T)/CBS value from Ref.~\citenum{Hai18};} 
$^d${CCSDT/PVTZ+ values from Ref.~\citenum{Kuc01};}
$^e${SOPPA (with ``all corrections'') from Ref.~\citenum{Nie80b};}
$^f${FCI/cc-pVDZ from Ref.~\citenum{Coe13b};}
$^g${From Ref.~\citenum{Loo18a}: the values od $\Delta E_{\mathrm{vert}}$ are basis set corrected exFCI/{\AVQZ} values whereas $f$ are  LR-CC3/{\AVTZ} values. Note that a factor of two linked to the degeneracy was incorrectly omitted in this earlier work
for the oscillator strengths of the two $\Pi$ transitions;   }
$^h${CC2/{AVQZ} result from Ref.~\citenum{Hel11};}
$^i${From Ref.~\citenum{Coo87};}
$^j${From Ref.~\citenum{Hub79};}
$^k${Vertical values estimated in Ref.~\citenum{Nie80b} on the basis of the experimental spectroscopic data of Ref.~\citenum{Hub79};}
$^l${Dipole (e,e) spectroscopy of Ref.~\citenum{Cha93b}. For the higher ES, we give the contributions given for the two vibrationnal quanta;}
$^m${Stark measurements of the 0-0 bands from Ref.~\citenum{Fis76}. Note that the value for the lowest ES is considered as an upper limit in this work. The sign of the dipole is assumed to be consistent with theory;}
$^n${Dipole ($\gamma$,$\gamma$) measurements from Ref.~\citenum{Kan15}, summing over the different vibrational contributions;}
$^o${Two-photon laser induced fluorescence spectroscopy measurement of Stark effect from Ref.~\citenum{Dra93}. The sign of the dipole is assumed to be consistent with theory.}
\end{flushleft}
\end{table*}
%%% TABLE CO %%%

For C=O, we considered six different excited states of various nature, and our results are listed in Table \ref{Table-5}.  For the three lowest transitions of valence character, one notices that CCSDT is again sufficient, 
with very small corrections brought by the Q and P excitations, e.g., the Q-induced changes attain ca.~$\pm$0.01--0.02 D only for $\mu^{\mathrm{ES}}$.  For these three transitions, the basis set effects are also firmly 
under control for all properties, and, both $f$ and $\mu^{\mathrm{ES}}$ are not significantly affected by the second set of diffuse orbitals and  {\AVQZ} is likely sufficiently large to obtain accurate estimates. 
As a consequence, it is rather straightforward to get stable CBS extrapolations.  The moderate impact brought by the quadruples seems to pertain for the three higher-lying Rydberg ESs 
of carbon monoxide, except for the dipole moment of the second $^1\Sigma^+$ transition for which $\mu^{\mathrm{ES}}$ increases by +0.111 D when going from CCSDT/{\AVTZ} to CCSDTQ/{\AVTZ}. As expected, much 
larger basis sets are needed to obtain converged properties for the Rydberg ESs than for their valence counterparts.  Indeed, when going from {\AVTZ} to {\DAVQZ}, the CCSDT ES dipoles of the three lowest Rydberg states drastically change from -3.89 to -1.81 D, 
from +6.09 to +4.34 D, and from +0.14 to -1.44 D, respectively.  In the same time, the CCSDT value of $f$ for the two significantly dipole-allowed transitions decrease by -36\%\ when 
considering the same basis set pair. Obviously such behavior makes the error bar obtained for our TBE/CBS estimates non-negligible (see below).  In contrast, and as already noted above, the transition energies of these 
Rydberg ESs can be effectively estimated with high accuracy.

When comparing with previously reported experimental or theoretical estimates, one should keep in mind that the properties of carbon monoxide are strongly affected by the bond length, \cite{Coo87} and that we used a 
theoretical geometry with a slightly too elongated double bond (2.142 bohr \emph{versus} 2.132 bohr experimentally).  Our TBE/CBS for $\mu^{\mathrm{GS}}$ is 0.10 D, which is slightly too low as compared to both the 
experimental measurement \cite{Hub79} and the theoretical calculations performed at the experimental geometry. \cite{Scu91,Hai18} For the lowest and well-studied $\Pi$ ES, the present 8.46$\pm$0.01 eV $\Delta E_{\mathrm{vert}}$ 
value is close to our recent FCI estimate (8.48 eV)\cite{Loo18a} as well as to the experimental value (8.51 eV) deduced with the help of the measured spectroscopic constants and the reconstruction of the potential energy surfaces. \cite{Nie80} 
The computed value of $f = 0.165$ is in very reasonable agreement with the estimates obtained by dipole (e,e) and ($\gamma$,$\gamma$) spectroscopies: 0.181 \cite{Cha93b} and 0.194, \cite{Kan15} respectively. We refer the 
interested readers to Tables 3 and 4 of Ref.~\citenum{Cha93b}, Table 6 of Ref.~\citenum{Car78}, and Table II of Ref.~\citenum{Roc98} for more complete lists of experimental and theoretical estimates of the oscillator strength. There is a quite significant 
elongation of the double bond in the lowest ES. As a consequence the value of $\mu^{\mathrm{ES}}$ determined on the GS (-0.19$\pm$0.01 D) and ES (-0.39$\pm$0.01 D) structures do differ significantly.  The two Stark effect measurements 
we are aware of yield  $\mu^{\mathrm{ES}} = -0.15\pm0.05$ D (analysis of the 0-0 band) \cite{Fis76} and $\mu^{\mathrm{ES}} = -0.34\pm0.01$ D (two-photon LIF), \cite{Dra93} which are also somehow inconsistent, yet of the same order of magnitude
as our GS and ES values, respectively. For the two other valence transitions, of $\Sigma^-$ and $\Delta$ symmetries, the TBE/CBS for $\Delta E_{\mathrm{vert}}$ show small errors (1--2 \%) as compared to the ``experimental'' values. \cite{Nie80b} 
We could not find any experimental estimate of the ES dipoles for these states, likely because they are dark, and we believe that the data listed in Table \ref{Table-5} stand as the most accurate ES dipole values proposed to date. They indicate that the dipole 
moments of these two ESs are parallel to the one of the GS but have much larger amplitudes. For the three Rydberg transitions considered, our best estimates of the transition energies are fully compatible with the measurements. 
The lowest Rydberg ES has a very low oscillator strength, a result consistent with previous theoretical\cite{Nie80b,Loo18a} and experimental \cite{Cha93b} investigations. More interestingly, its dipole moment is large and negative. The two
available measurements return dipoles of -1.60$\pm$0.15 \cite{Fis76} and -1.95$\pm$0.03 D, \cite{Dra93} whereas the strong basis set effects bring uncertainty to our TBE/CBS (-2.49$\pm$0.59 D). The lower bound (-1.90 D) 
fits the latest experimental value quite well. We have also computed the ES dipole of this state at its equilibrium geometry, but the changes are very limited ({see Table S4 in the SI}).  The second Rydberg $\Sigma^+$ transition is much more 
dipole allowed than the first, and our TBE/CBS of 0.146$\pm$0.016 for the oscillator strength is consistent with the most recent measurement (0.136). \cite{Kan15} The dipole moment of this ES is large, positive, and rather unaffected by 
structural relaxation effects ({Table S4 in the SI}), our extrapolated value of 4.94$\pm$0.47 D agreeing well with the two experimental estimates of 4.52$\pm$0.35\cite{Fis76} and 4.50$\pm$0.07 D. \cite{Dra93} Eventually, for the highest ES  
of CO considered herein, the extrapolated $f$ value perfectly fits the measurements. \cite{Cha93b,Kan15} We provide, as far as we know, the first estimate of its dipole moment, which is rather small and negative (see bottom of 
Table \ref{Table-5}).

\subsection{N$_2$}
\label{res-35}

%%% TABLE N2 %%%
\begin{table*}[htp]
\scriptsize
\caption{Vertical transition energies $\Delta E_{\mathrm{vert}}$ and oscillator strengths $f$ determined for N$_2$ (GS geometry). See caption of Table \ref{Table-1} for details.} 
\label{Table-6}
\vspace{-0.3 cm}
\begin{tabular}{cc|cccccccc}
\hline 

		&				&	\multicolumn{2}{c}{$^1\Pi_u (\mathrm{Ryd})$} &	\multicolumn{2}{c}{$^1\Sigma_u^+ (\mathrm{Ryd})$} &	\multicolumn{2}{c}{$^1\Pi_u (\mathrm{Ryd})$} &	\multicolumn{2}{c}{$^1\Pi_u (\mathrm{Ryd})$} \\
Basis 	& Method			&$\Delta E_{\mathrm{vert}}$	&$f$		&$\Delta E_{\mathrm{vert}}$	&$f$&$\Delta E_{\mathrm{vert}}$	&$f$&$\Delta E_{\mathrm{vert}}$	&$f$\\
\hline
{\AVDZ}	&CCSD			&13.451	&0.531	&13.250	&0.311	&13.765	&0.014	&14.497	&0.148 	\\
		&CCSDT			&13.174	&0.469	&13.131	&0.334	&13.591	&0.020	&14.228	&0.163 	\\
		&CCSDTQ		&13.131	&0.458	&13.109	&0.337	&13.560	&0.027	&14.221	&0.164 	\\
		&CCSDTQP		&13.127	&0.457	&13.107	&0.338	&13.558	&0.028	&14.216	&0.164 	\\
{\AVTZ}	&CCSD			&13.422	&0.439	&13.264	&0.263	&13.674	&0.053	&14.307	&0.136 	\\
		&CCSDT			&13.140	&0.435	&13.118	&0.281	&13.455	&0.008	&14.034	&0.148 	\\
		&CCSDTQ		&13.095	&0.424	&13.090	&0.285	&13.419	&0.014	&14.014	&0.149 	\\
{\AVQZ}	&CCSD			&13.354	&0.357	&13.242	&0.242	&13.638	&0.113	&14.216	&0.134 	\\
		&CCSDT			&13.108	&0.422	&13.088	&0.258	&13.372	&0.000	&13.935	&0.145 	\\
{\AVFZ}	&CCSD			&13.235	&0.266	&13.195	&0.219	&13.621	&0.181	&14.100	&0.126 	\\
		&CCSDT			&13.037	&0.356	&13.039	&		&13.306	&0.044	&13.816	&0.138 	\\
		&				&		&		&		&		&		&		&		&		\\
{\DAVDZ}	&CCSD			&12.784	&0.170	&12.822	&0.172	&13.640	&0.101	&13.537	&0.234	\\
		&CCSDT			&12.669	&0.198	&12.712	&0.183	&13.262	&0.256	&13.300	&0.014	\\
{\DAVTZ}	&CCSD			&12.978	&0.178	&13.026	&0.181	&13.599	&0.314	&13.682	&0.007	\\
		&CCSDT			&12.827	&0.224	&12.885	&0.193	&13.257	&0.168	&13.396	&0.073	\\
{\DAVQZ}	&CCSD			&13.039	&0.181	&13.090	&0.184	&13.601	&0.276	&13.730	&0.043	\\
		&CCSDT			&12.876	&0.234	&12.939	&0.196	&13.256	&0.460	&13.443	&0.082	\\
		&				&				&		&		&	\\
{\AVTZ}	& TBE$^a$		&13.090	&0.423	&13.088	&0.286	&13.417	&0.015	&14.009	&0.148\\
CBS		& TBE$^b$		&12.83$\pm$0.08 & 0.22$\pm$0.06$^c$	&12.96$\pm$0.01	&0.207$\pm$0.005&	13.27$\pm$0.07 &	$^d$	& 13.57$\pm$0.12 & 0.10$\pm$0.03$^c$\\	
		&				&				&		&		&	\\
Lit.		&	Th.			&12.73$^e$	&0.458$^e$	&12.95$^e$	&0.296$^e$	&13.27$^e$	&0.000$^e$	&\\
		&				&			&0.091$^f$;0.063$^g$& & 0.65$^f$;0.221$^g$ & & 0.32$^f$;0.091$^g$ & &0.083$^g$ \\
		&	Exp.			&12.78$^h$	&0.243$^i$&12.96$^h$	&0.279$^i$&13.10$^h$	&0.145$^i$	& & 0.080$^i$\\
		&				&12.90$^j$	&		&12.98$^j$	&0.223$^k$	&13.24$^j$	&		&13.63$^j$	\\	
\hline	
\end{tabular}
\vspace{-0.3 cm}
\begin{flushleft}
$^{a,b}${See corresponding footnotes in Table \ref{Table-1};}
$^c${The extrapolation is very challenging due to the mixing, tentative values;}
$^d${Too unstable to report any reasonable CBS estimate;}
$^e${From Ref.~\citenum{Loo18a}: energies are basis set corrected exFCI/{\AVQZ} values and the oscillator strengths are LR-CC3/{\AVTZ} values. Note that a factor of two linked to the degeneracy was incorrectly omitted in this earlier work for the $\Pi_u$ transitions; }
$^f${RPA values in length gauge from Ref.~\citenum{Odd85};}
$^g${SOPA values from Ref.~\citenum{Neu04};}
$^h${Experimental vertical values given in Ref.~\citenum{Odd85} deduced from spectroscopic data of Ref.~\citenum{Hub79};}
$^i${Integrated intensities from electron scattering of Ref.~\citenum{Cha93}.}
$^j${Experimental vertical values given in Ref.~\citenum{Ben90} deduced from spectroscopic data of Ref.~\citenum{Hub79};}
$^k${Integrated electron impact induced emission intensities of Ref.~\citenum{Aje89}. }
\end{flushleft}
\end{table*}

The understanding of the nature of the ES in the highly symmetric N$_2$ molecule is certainly challenging from both a experimental and a theoretical point of view. \cite{Odd85,Ben90,Neu04}  Of particular relevance for the present work 
is the nature and relative ordering of the three lowest $^1\Pi_u$ states. Here, we have classified them following the nature of the underlying MOs,\cite{zzz-mountain} and characterized all of them as Rydberg, although  
alternative yet reasonable assignments can be found in the literature (\emph{e.g.}, see Tables I and II in Ref.~\citenum{Neu04}). Our results are given in Table \ref{Table-6}. The convergence with respect to the CC excitation order is rather quick for both energies
and oscillator strengths, with nevertheless non-negligible contributions from the quadruples for the transition energies, a likely consequence of the presence of a triple bond.  If basis set effects follow the expected trends
for the energies of these high-lying ES (\emph{i.e.}, a decrease of $\Delta E_{\mathrm{vert}}$ when increasing the basis set size, and a significant impact of the second set of diffuse functions), the changes in the oscillator strength of the $^1\Pi_u$  states
when enlarging the basis set are dramatic. As an illustration, the oscillator strength of the second  $^1\Pi_u$  transition is 0.000 at the CCSDT/{\AVQZ} level but 0.460 at  the CCSDT/{\DAVQZ} level. If the extreme sensitivity of N$_2$'s oscillator strengths to 
the computational setup is known for years (\emph{e.g.}, see Table 6 in Ref.~\citenum{Odd85}), it remains a very striking example of the state mixing nightmare. As a consequence, while we could obtain solid TBE/{\AVTZ} values in
all cases, the extrapolation to CBS of the $f$ values of the $^1\Pi_u$ transitions is clearly problematic. In comparison, while the basis set effects are far from being negligible for the $^1\Sigma_u^+$ ES, it is definitely possible to establish robust CBS $f$ values for this transition.

For the vertical excitation energies, our TBE/CBS of 12.83, 13.27, and 13.57 eV for the three lowest $^1\Pi_u$ transitions and of 12.96 eV for the lowest $^1\Sigma_u^+$ transition are all in reasonable agreement with current 
state-of-the-art values. Indeed, as can be seen from the bottom of Table \ref{Table-6}, these values are close to the experimental vertical data deduced elsewhere, \cite{Ben90} as well as to our recent FCI/CBS estimates. \cite{Loo18a}
When turning to the oscillator strengths, there is quite a diversity in the experimentally-measured values (see Table 8 of Ref.~\citenum{Cha93} for integrated values, and Table 1 of Ref.~\citenum{Liu16c} for individual vibronic contributions from various earlier works). 
For the $^1\Sigma_u^+$ transition, several estimates are available and they show a wide range on both the theoretical and experimental sides (see Table 3 of Ref.~\citenum{Lav04}), and our TBE/CBS $f$ value of 0.207 is reasonably in line with 
the most recent electron impact value that we have found (0.223), \cite{Aje89} yet significantly smaller than a value obtained by electron scattering (0.279). \cite{Cha93}  For the first and third $^1\Pi_u$ transitions, our estimates of the 
oscillator strength come with large error bars (0.22$\pm$0.06 and 0.10$\pm$0.03, respectively), which nevertheless cover the most recent experimental values (0.243 and 0.080), \cite{Cha93} an outcome that we found satisfying. As stated 
above, no reasonable estimate could be obtained for the oscillator strengths of the remaining transition. It is likely that high-level multi-reference calculations would be welcome in this specific case.

\subsection{Ethylene}
\label{res-36}

The ESs of this model $\pi$-conjugated hydrocarbon have puzzled chemists for years, in particular the relative ordering and nature of the two lowest singlet ESs  considered here (Table \ref{Table-7}). 
As can be seen, the oscillator strength of the lowest Rydberg state of $B_{3u}$ symmetry is much too large with 6-31+G(d) but rapidly converges with the Dunning series, the addition of a second set of diffuse functions 
playing no significant role. In fact the $f$ values for this transition converges faster when climbing the methodology ladder than the $\Delta E_{\mathrm{vert}}$ values. For the valence $\pi \rightarrow \pi^\star$ $B_{1u}$ 
transition, CCSD seems to slightly overestimate the oscillator strength and the convergence with basis set size is nearly reached with \emph{aug}-cc-pVTZ. Our best estimates are {7.42 eV} ($f$=0.076) and 7.90 eV ($f$={0.338}),
for the Rydberg and valence transitions, respectively. 

%%% TABLE ETHY %%%
\begin{table}[H]
\scriptsize
\caption{Vertical transition energies $\Delta E_{\mathrm{vert}}$ and oscillator strengths $f$ of ethylene (GS geometry). See caption of Table \ref{Table-1} for details.} 
\label{Table-7}
\vspace{-0.3 cm}
\begin{tabular}{cc|cccc}
\hline 
		&				&	\multicolumn{2}{c}{$^1B_{3u} (\mathrm{Ryd})$} &\multicolumn{2}{c}{$^1B_{1u} (\mathrm{Val})$}  \\
Basis 	& Method			&$\Delta E_{\mathrm{vert}}$	&$f$		&$\Delta E_{\mathrm{vert}}$	&$f$		\\
\hline
{\Pop}	&CCSD			&7.814	&0.152	&8.275	&0.380	\\
		&CCSDT			&7.725	&0.151	&8.152	&0.365	\\
		&CCSDTQ		&7.722	&0.150	&8.137	&0.364	\\
		&CCSDTQP		&7.722	&0.150	&8.135	&0.364	\\
{\AVDZ}	&CCSD			&7.323	&0.080	&8.035	&0.365	\\
		&CCSDT			&7.294	&0.080	&7.944	&0.352	\\
		&CCSDTQ		&7.303	&0.080	&7.932	&0.351	\\
{\AVTZ}	&CCSD			&7.417	&0.078	&8.020	&0.362	\\
		&CCSDT			&7.365	&0.078	&7.918	&0.346	\\
{\AVQZ}	&CCSD			&7.451	&0.078	&8.023	&0.360	\\
		&				&		&		&		&		\\
{\DAVDZ}	&CCSD			&7.301	&0.078	&8.008	&0.345\\
		&CCSDT			&7.273	&0.078	&7.920	&0.336\\
{\DAVTZ}	&CCSD			&7.409	&0.077	&8.009	&0.353\\
		&CCSDT			&7.357	&0.077	&7.908	&0.339\\
{\DAVQZ}	&CCSD			&{7.446}&{0.077}&{8.017}&{0.355}\\		
		&				&		&		&		&	\\
{\AVTZ}	& TBE$^a$		&7.374	&0.078	&7.905	&0.345	\\
CBS		& TBE$^b$		&7.42$\pm$0.02&0.076$\pm$0.001	&7.90$\pm$0.01	&{0.338}$\pm$0.005	\\
		&				&		&		&		&	\\
Lit.		&	Th.			&7.45$^c$&0.069$^c$&8.00$^c$&0.333$^c$	\\
		&				&7.43$^d$&0.078$^d$&7.92$^d$&0.348$^d$ \\
		&	Exp.			&7.11$^e$&$\sim$0.04$^f$	&7.60$^e$ &0.34$^g$\\
\hline	
\end{tabular}
\vspace{-0.3 cm}
\begin{flushleft}
$^a${CCSDT/{\AVTZ} values corrected for Q effects using {\AVDZ} and for P effects using {\Pop};}
$^b${See Computational section;}
$^c${From Ref.~\citenum{Fel14}: best composite theory for energies, icCAS(12/15)-CI/VDZ+ for the oscillator strengths of the Rydberg transition and icINO(12/16)-CI/VDZ+ extrapolated to FCI for the oscillator strengths of the valence transition;}
$^d${FCI/CBS (for transition energies) and CC3/{\AVTZ} (for $f$) values from Ref.~\citenum{Loo18a}; }
$^e${Experimental values collected in Ref.~\citenum{Rob85b} (see the discussions in Refs.~\citenum{Sch08}, \citenum{Fel14}, and  \citenum{Ser93});}
$^f${From Ref.~\citenum{Mer69} (see text);}
$^g${Vacuum absorption from Ref.~\citenum{Zel53}.}
\end{flushleft}
\end{table}

On the theoretical side, the most advanced theoretical study of ethylene's ES likely remains the 2014 investigation of Feller \textit{et al}. \cite{Fel14} With their best composite theory, these authors reported transition energies of 7.45 eV 
and 8.00 eV with respective $f$ values of 0.069 and 0.333 (see footnotes of Table  \ref{Table-7} for details). One can also compare to the values of Ref.~\citenum{Loo18a}: FCI/CBS estimates of 7.43 and 7.92 eV 
with CC3/AVTZ $f$ values of 0.078 and 0.346, respectively. Older works report oscillator strengths of 0.389 for the valence state at the CCSDT/TZVP level, \cite{Kan14} and of 0.078 and 0.358 with CCSD. \cite{Wat96b} The 
experimental measurements of the oscillator strength do not allow to attribute values to individual transitions due to strong overlapping. \cite{Coo95} The generally used experimental reference values are 0.34
or 0.29 for the valence transition,  both estimates being obtained from measurements of the vacuum absorption spectrum performed in 1953 \cite{Zel53}  and 1955, \cite{Ham55}  respectively. However, a more recent dipole (e,e) spectroscopy 
study suggests that the originally measured oscillator strengths in the 7.4--8.0 eV regions are probably too low by ca.~10--15\%. \cite{Coo95} Our TBE of {0.338} is therefore reasonably in line with the current experimental knowledge. 
For the Rydberg ES, the only experimental estimate  we are aware of has been reported in 1969 as a \emph{``total $f$ perhaps about 0.04''}. \cite{Mer69} Given the consistency of all theoretical estimates, it seems 
rather reasonable to state that our current TBE of 0.076 is significantly more trustworthy. In ethylene, in contrast to N$_2$, theory has clearly the edge because the considered transitions have different spatial symmetries.

\subsection{Formaldehyde and thioformaldehyde}

The ESs of formaldehyde have been extensively studied before with almost all possible theoretical approaches, \cite{For92b,Had93,Hea94,Hea95,Gwa95,Wib98,Wib02,Hir04,Pea08,Sch08,She09b,Car10,Li11,Hel11,Lea12,Hoy16,Kan17,Loo18a}
and we have considered here two valence and three Rydberg states. Our results are collected in Table \ref{Table-8} for these fives ESs.  The hallmark $n \rightarrow \pi^\star$ transition behaves nicely from a theoretical point of view, and the convergences
of both $\Delta E_{\mathrm{vert}}$ and $\mu^{\mathrm{ES}}$ with respect to the basis set size are very quick, with a negligible effect of the second set of diffuse functions.  Likewise, the corrections brought by the
Q and P terms in the CC expansion are rather negligible, and CCSDT/\emph{aug}-cc-pVTZ values are trustworthy. The higher-lying $\pi \rightarrow \pi^\star$ valence ES is significantly more challenging as, on the one hand, the second set of diffuse functions
significantly increases the transition energy (by ca.~+0.2 eV), decreases the oscillator strength (by roughly 10\%), and greatly amplifies the ES dipole (by a factor of 2 or 3), whereas, on the other hand, going from CCSDT to CCSDTQ yields a non-negligible 
drop of the computed dipole. Extrapolation to the CBS limit is therefore uneasy for the latter property. The oscillator strengths of the three Rydberg ESs are all relatively small, but their absolute and relative amplitudes are fairly independent on the
selected level of theory and basis sets, though double augmentation induces a small decrease of the magnitude of the oscillator strength.  The value of $\mu^{\mathrm{ES}}$ for the lowest-lying Rydberg ES cannot be adequately described with the selected Pople basis set,
but are easy to extrapolate using Dunning's series. For the second (third) Rydberg ES, all tested approaches agree on the rather small (moderate) amplitude for  $\mu^{\mathrm{ES}}$, but the basis set effects are quite drastic. For instance,
considering the higher-lying Rydberg ES, the CCSDT   $\mu^{\mathrm{ES}}$ value is -0.06 D with {\Pop}, -2.19 D with {\AVDZ}, and -0.37 D with {\DAVTZ}. Clearly it is challenging to get a definitive CBS estimate.

%%% TABLE Forma %%% INVERT GEOM in SI
\begin{table*}[htp]
\scriptsize
\caption{Ground-state dipole moment $\mu^{\mathrm{GS}}$, vertical transition energies $\Delta E_{\mathrm{vert}}$, oscillator strengths $f$, and excited state dipole moments, $\mu_{\mathrm{vert}}^{\mathrm{ES}}$ and $\mu_{\mathrm{adia}}^{\mathrm{ES}}$, determined of formaldehyde. See caption of Table \ref{Table-1} for details.}
%Note that CCSDTQP was beyond reach on the less symmetric ES geometry of the lowest transition.} %DJ: to remove ?
\label{Table-8}
\vspace{-0.3 cm}
\begin{tabular}{cc|cccccccccc}
\hline 
		&			& $^1A_1$	&	\multicolumn{3}{c}{$^1A_2 (\mathrm{Val}, n \rightarrow \pi^\star)$}	&	\multicolumn{4}{c}{$^1B_2 (\mathrm{Ryd}, n \rightarrow 3s)$}		\\
Basis 	& Method		&$\mu^{\mathrm{GS}}$	&$\Delta E_{\mathrm{vert}}$	&$\mu_{\mathrm{vert}}^{\mathrm{ES}}$	&$\mu_{\mathrm{adia}}^{\mathrm{ES}}$
					&$\Delta E_{\mathrm{vert}}$	& $f$&$\mu_{\mathrm{vert}}^{\mathrm{ES}}$	&$\mu_{\mathrm{adia}}^{\mathrm{ES}}$	\\
\hline
{\Pop}	&CCSD			&2.584	&4.031	&1.710	&1.870	&7.238	&0.017	&	-0.901&-0.479	\\
		&CCSDT			&2.529	&4.012	&1.649	&1.790	&7.232	&0.021	&	-1.459&-0.990	\\
		&CCSDTQ		&2.518	&4.022	&1.629	&1.751	&7.279	&0.020	&	-1.364&-0.906	\\
		&CCSDTQP		&2.517	&4.023	&1.627	&{1.746}&7.287	&0.020	&	-1.342&-0.885	\\
{\AVDZ}	&CCSD			&2.427	&4.020	&1.397	&1.602	&7.043	&0.018	&	-2.078&-1.634	\\
		&CCSDT			&2.368	&3.986	&1.337	&1.524	&7.040	&0.020	&	-2.457&-1.976	\\
		&CCSDTQ		&2.356	&3.997	&1.319	&1.486	&7.091	&0.020	&	-2.403&-1.932	\\
{\AVTZ}	&CCSD			&2.457	&4.013	&1.416	&1.620	&7.231	&0.018	&	-1.929&-1.476	\\
		&CCSDT			&2.389	&3.954	&1.346	&1.534	&7.165	&0.020	&	-2.379&-1.873	\\
{\AVQZ}	&CCSD			&2.475	&4.024 	&1.436	&1.635	&7.296	&0.018	&	-1.861&-1.406	\\
		&				&          	&         	&          	&         	&              &               &                &             \\	
{\DAVDZ}	&CCSD			&2.417	&4.012	&1.386	&1.596	&7.027	&0.016	&	-1.839&-1.410		\\
		&CCSDT			&2.359	&3.978	&1.327	&1.518	&7.024	&0.019	&	-2.282&-1.803		\\
{\DAVTZ}	&CCSD			&2.454	&4.011	&1.418	&1.619	&7.224	&0.017	&	-1.751&-1.309		\\
		&CCSDT			&2.387	&3.952	&1.347	&1.533	&7.158	&0.020	&	-2.248&-1.743		\\
{\DAVQZ}	&{CCSD}		&{2.475}&{4.023}&{1.439}&{1.636}&{7.192}&{0.018}&{-1.747}&{-1.299}	\\
		&				&		&		&		&		&		&		&		  &		&\\
{\AVTZ}	& TBE$^a$		&2.375	&3.966	&1.325	&1.491	&7.225	&0.020	&	-2.302&-1.809	\\
CBS		& TBE$^b$	&{2.41}$\pm$0.01&{3.99}$\pm${0.01}&{1.36$\pm$0.01}& {1.52}$\pm$0.01&{7.34$\pm$0.01}& 0.020$\pm$0.001&{-2.15$\pm$0.03}&{-1.66$\pm$0.03}\\
		&				&		&		&		&		\\
Lit.		&	Th.			&2.393$^c$&3.98$^d$&1.33$^e$&1.48$^e$	&7.12$^d$&0.018$^d$;0.025$^e$&-2.52$^e$	&-3.45$^f$\\
		&				&2.33$^e$&3.97$^g$	&		&1.73$^h$	&7.30$^g$ & 0.021$^g$;0.018$^i$& 	&\\
		&				&{2.44}$^j$&{3.98}$^j$&{1.46}$^j$&\\
		&				&\\
		&	Exp.			&2.332$^k$&4.07$^l$&	&	1.53$\pm$0.11$^m$	&7.11$^k$&0.028$^n$ & &-0.33$\pm$0.16$^o$\\%
		&				&		&		&	&1.47$^p$	&				&0.041$^q$\\
\hline 
		&			&	\multicolumn{3}{c}{$^1B_2 (\mathrm{Ryd}, n\rightarrow 3p)$}&\multicolumn{3}{c}{$^1A_1  (\mathrm{Ryd}, n\rightarrow 3p)$}	 &\multicolumn{3}{c}{$^1A_1  (\mathrm{Val}, \pi\rightarrow \pi^\star)$}\\
Basis 	& Method		&$\Delta E_{\mathrm{vert}}$	& $f$&$\mu_{\mathrm{vert}}^{\mathrm{ES}}$	&$\Delta E_{\mathrm{vert}}$	& $f$&$\mu_{\mathrm{vert}}^{\mathrm{ES}}$	
&$\Delta E_{\mathrm{vert}}$& $f$&$\mu_{\mathrm{vert}}^{\mathrm{ES}}$\\
\hline
{\Pop}	&CCSD			&7.994	&0.042	&0.199	&8.282	&0.060	&0.082	&10.042	&0.183	&2.057\\
		&CCSDT			&8.007	&0.040	&0.625	&8.295	&0.058	&-0.062	&9.829	&0.163	&1.737\\
		&CCSDTQ		&8.045	&0.040	&0.504	&8.341	&0.058	&-0.076	&9.779	&0.159	&1.670\\
		&CCSDTQP		&8.051	&0.041	&0.483	&8.350	&0.058	&-0.074	&9.776	&0.158	&1.663 \\
{\AVDZ}	&CCSD			&7.993	&0.044	&0.414	&8.052	&0.058	&-2.058	&9.752	&0.157	&1.470\\
		&CCSDT			&8.002	&0.042	&0.631	&8.068	&0.057	&-2.190	&9.588	&0.147	&1.356\\
		&CCSDTQ		&8.045	&0.042	&0.557	&8.119	&0.057	&-2.202	&9.544	&0.143	&1.292\\
{\AVTZ}	&CCSD			&8.120	&0.040	&0.319	&8.210	&0.054	&-1.438	&9.670	&0.139	&1.503\\
		&CCSDT			&8.070	&0.038	&0.549	&8.164	&0.052	&-1.617	&9.488	&0.131	&1.372\\
{\AVQZ}	&CCSD			&8.153	&0.039	&0.292	&8.267	&0.053	&-1.145	&9.647	&0.125	&1.388\\
		&				&		&		&		&              &               &           \\  	
{\DAVDZ}	&CCSD			&7.834	&0.036	&0.062	&7.962	&0.049	&-0.019	&9.885	&0.126	&3.672\\	
		&CCSDT			&7.846	&0.035	&0.280	&7.983	&0.049	&-0.301	&9.748	&0.119	&3.553\\
{\DAVTZ}	&CCSD			&8.026	&0.035	&-0.104	&8.169	&0.050	&-0.044	&9.862	&0.124	&3.783\\	
		&CCSDT			&7.978	&0.033	&0.117	&8.126	&0.049	&-0.369	&9.719	&0.108	&4.347\\
{\DAVQZ}	&{CCSD}		&{8.095}&{0.035}&{-0.115}&{8.240}&{0.050}&{-0.055}&{9.869}&{0.123}&{3.918}\\
		&				&		&		&		&		&		&		\\
{\AVTZ}	& TBE$^a$		&8.119	&0.039	&0.454	&8.224	&0.052	&-1.628	&9.441	&0.127		&1.301	\\
CBS		& TBE$^b$		&{8.16$\pm$0.02}&0.035$\pm$0.002&{0.21$\pm$0.20}&8.28$\pm$0.04&0.050$\pm${0.001}&{-0.69$\pm$0.43}&{9.52$\pm$0.12}&{0.107$\pm$0.002}&{2.46$\pm$1.36}\\
		&				&\\
Lit.		&	Th.			&7.94$^d$&0.040$^d$;0.041$^e$&0.85$^e$		&8.16$^d$&0.043$^d$;0.058$^e$&-2.16$^e$	&9.83$^d$	&0.100$^d$	&\\
		&				&8.14$^g$&0.037$^g$;0.035$^i$&	&8.27$^g$	& 0.052$^g$;0.050$^i$&			&9.26$^g$& 0.135$^g$;0.093$^i$\\
%		&				&		&		&		&		&		&		&{9.31}$^j$&	{0.451}$^j$&{2.68}$^j$\\
		&				&\\
		&	Exp.			&7.97$^l$	&0.017$^n$	&	&8.14$^l$	&0.032$^n$	\\
		&				&			&0.018$^q$	&	&			&0.061$^q$	\\
\hline
\end{tabular}
\vspace{-0.3 cm}
\begin{flushleft}
$^{a,b}${See corresponding footnotes in Table \ref{Table-7};}
$^c${CCSD(T)/CBS value from Ref.~\citenum{Hai18};} 
$^d${MR-AQCC-LRT calculations from Ref.~\citenum{Mul01};} 
$^e${CCSDT/{\AVDZ} results from Ref.~\citenum{Hir04};}
$^f${MRDCI value from Ref.~\citenum{Hac94};}
$^g${exFCI/{\AVTZ} transition energies corrected for basis set effects up to {\DAVFZ} and LR-CC3/{\AVTZ} for $f$ from Ref.~\citenum{Loo18a};}
$^h${CC2/{\AVQZ} figure from Ref.~\citenum{Hel11};}
$^i${EOM-CCSD from Ref.~\citenum{Gom10b};}
$^j${{CASPT2/TZVP values from Ref.} \citenum{Sch08};}
$^k${Electric resonance spectroscopy from Ref.~\citenum{Fab77};}
$^l${Various experimental sources collected in Ref.~\citenum{Rob85b};}
$^m${Stark effect measurement on lineshapes from Ref.~\citenum{Han68b};}
$^n${EELS values from Ref.~\citenum{Wei71};}
$^o${Values measured from polarized electrochromism reported in Refs. \citenum{Cau78,Cau79};}
$^p${Stark effect from quantum beat spectroscopy from Ref.~\citenum{Vac89};  }
$^q${Dipole (e,e) spectroscopy from Ref.~\citenum{Coo96}.}
\end{flushleft}
\end{table*}

%%% TABLE ThioF %%% INVERT GEOM in SI
\begin{table*}[htp]
\scriptsize
\caption{Ground-state dipole moment $\mu^{\mathrm{GS}}$, vertical transition energies $\Delta E_{\mathrm{vert}}$, oscillator strengths $f$, and excited state dipole moments, $\mu_{\mathrm{vert}}^{\mathrm{ES}}$ and $\mu_{\mathrm{adia}}^{\mathrm{ES}}$, determined of thioformaldehyde. See caption of Table \ref{Table-1} for details.} 
\label{Table-9}
\vspace{-0.3 cm}
\begin{tabular}{cc|ccccccccccc}
\hline 
		&			& $^1A_1$	&	\multicolumn{3}{c}{$^1A_2 (\mathrm{Val}, n\rightarrow \pi^\star)$}	&	\multicolumn{4}{c}{$^1B_2 (\mathrm{Ryd}, n\rightarrow 4s)$}	&	\multicolumn{3}{c}{$^1A_1 (\mathrm{Val}, \pi\rightarrow \pi^\star)$}	\\
Basis 	& Method		&$\mu^{\mathrm{GS}}$	&$\Delta E_{\mathrm{vert}}$	&$\mu_{\mathrm{vert}}^{\mathrm{ES}}$	&$\mu_{\mathrm{adia}}^{\mathrm{ES}}$
					&$\Delta E_{\mathrm{vert}}$	& $f$&$\mu_{\mathrm{vert}}^{\mathrm{ES}}$	&$\mu_{\mathrm{adia}}^{\mathrm{ES}}$	&$\Delta E_{\mathrm{vert}}$& $f$&$\mu_{\mathrm{vert}}^{\mathrm{ES}}$\\
\hline
{\Pop}	&CCSD			&1.747	&2.302	&0.933	&0.968	&5.937	&0.019	&-3.378	&-3.070	&6.961	&0.261	&1.964\\
		&CCSDT			&1.733	&2.244	&0.948	&0.990	&5.875	&0.018	&-3.570	&-3.257	&6.790	&0.223	&1.745\\
		&CCSDTQ		&1.720	&2.246	&0.919	&0.947	&5.890	&0.019	&-3.563	&-3.253	&6.713	&0.191	&1.334\\
		&CCSDTQP		&1.719	&2.247	&0.917	&0.943	&5.893	&0.019	&-3.562	&-3.252	&6.708	&0.189	&1.305\\
{\AVDZ}	&CCSD			&1.742	&2.325	&0.851	&0.873	&5.841	&0.012	&-3.938	&-3.693	&6.749	&0.251	&1.875\\
		&CCSDT			&1.716	&2.253	&0.870	&0.898	&5.796	&0.011	&-4.140	&-3.885	&6.597	&0.182	&1.753\\
		&CCSDTQ		&1.704	&2.255	&0.848	&0.864	&5.817	&0.011	&-4.134	&-3.885	&6.512	&0.152	&1.264\\
{\AVTZ}	&CCSD			&1.737	&2.291	&0.848	&0.870	&5.970	&0.014	&-3.374	&-3.149	&6.633	&0.206	&2.379\\
		&CCSDT			&1.706	&2.207	&0.865	&0.890	&5.900	&0.012	&-3.665	&-3.422	&6.467	&0.163	&1.698\\
{\AVQZ}	&CCSD			&1.759	&2.296	&0.869	&0.890	&6.018	&0.014	&-3.162	&-2.944	&6.607	&0.200	&2.385\\
		&				&     		&       	&         	&         	&        	&		&       	 &         	&              &               &              \\
{\DAVDZ}	&CCSD			&1.735	&2.324	&0.858	&0.877	&5.804	&0.015	&-3.258	&-3.055	&6.678	&0.165	&-0.038\\
		&CCSDT			&1.709	&2.252	&0.875	&0.901	&5.761	&0.014	&-3.495	&-3.281	&6.577	&0.191	&1.749\\
{\DAVTZ}	&CCSD			&1.737	&2.291	&0.851	&0.872	&5.958	&0.015	&-3.050	&-2.847	&6.627	&0.206	&2.376\\
		&CCSDT			&1.707	&2.207	&0.867	&0.892	&5.888	&0.013	&-3.379	&-3.154	&6.463	&0.162	&2.193\\
{\DAVQZ}	&{CCSD}		&{1.759}&{2.296}&{0.871}&{0.892}&{6.012}&{0.014}&{-2.999}&{-2.794}&{6.604}&{0.199}&{2.383}\\
		&				&		&		&		&		&		&		&		&		&\\
{\AVTZ}	& TBE$^a$		&1.694	&2.210	&0.840	&0.851	&5.923	&0.013	&-3.658	&-3.421	&6.377 	&0.135	&1.179\\
CBS		& TBE$^b$		&{1.73$\pm$0.01}&{2.22}$\pm$0.01&{0.088$\pm$0.01}&{0.089$\pm$0.01}&{6.01$\pm$0.01}&0.013$\pm$0.001&{-3.27$\pm$0.03}&{-3.05$\pm$0.02}&{6.33$\pm$0.01} & {0.13$\pm$0.01} & {1.20$\pm$0.01}	\\
		&				&		&		&		&		\\
Lit.		&	Th.			&1.700$^c$&2.20$^d$&		&0.96$^e$	&5.99$^d$&0.012$^f$	&		&		&6.34$^g$&0.178$^f$\\
		&				&1.72$^e$&\\
		&				&		&\\
		&	Exp.			&1.649$^h$&2.033$^i$&		&0.850$\pm$0.002$^j$		&5.841$^i$&		&	&-2.2$\pm$0.3$^k$	&6.60$^i$\\
		&				&1.647$^l$&			&	&0.815$\pm$0.020$^m$		&		&\\
\hline
\end{tabular}
\vspace{-0.3 cm}
\begin{flushleft}
$^{a,b}${See corresponding footnotes in Table \ref{Table-7};}
$^c${CCSD(T)/\emph{aug}-cc-pVQZ from Ref.~\citenum{Ben07b};}
$^d${exFCI/{\AVTZ} transition energies corrected for basis set effects up to {\DAVQZ} from Ref.~\citenum{Loo18a};}
$^e${CC2/\emph{aug}-cc-pVQZ from Ref.~\citenum{Hel11};}
$^f${LR-CC3/{\AVTZ} $f$ from Ref.~\citenum{Loo18a};}
$^g${CCSDTQ/{\AVDZ} transition energy corrected for basis set effects up to {\DAVQZ} from Ref.~\citenum{Loo18a};}
$^h${Molecular beam electric resonance value from Ref.~\citenum{Fab77};}
$^i${0-0 energies listed in Table 13 of Ref.~\citenum{Clo83};}
$^j${Microwave-optical double resonance measurements of Stark effect from Ref.~\citenum{Suz85};}
$^k${Stark effect on the absorption spectrum from Ref.~\citenum{Goe81};}
$^l${Stark effect measurement on the microwave spectra from Ref.~\citenum{Joh71};}
$^m${Intermodulated fluorescence of Stark effect from Ref.~\citenum{Fun96}.}
\end{flushleft}
\end{table*}

A very accurate measurement of  $\mu^{\mathrm{GS}}$ for formaldehyde is available at 2.3321$\pm$0.0005 D (molecular beam electric resonance spectroscopy), \cite{Fab77} and our theoretical TBE of {2.41}$\pm$0.01 D
seems slightly too large, but is in very good agreement with previous CCSD(T)/CBS (2.393 D) \cite{Hai18} and  CCSD(T)/{\AVQZ} (2.382 D) estimates. \cite{Ben07b} As early as 2004, Hirata proposed a CCSDT/{\AVDZ} value of
$\mu^{\mathrm{GS}}$ at 2.33 D, \cite{Hir04} right on the experimental spot, but an experimental geometry was used and the orbital relaxation effects neglected, which might have induced a very slight error compensation. For the hallmark 
lowest $n \rightarrow \pi^\star$ transition, one can find several experimental estimates of $\mu^{\mathrm{ES}}$: 1.48$\pm$0.07, \cite{Fre64} 1.56$\pm$0.07, \cite{Fre66} 1.4$\pm$0.1 D, \cite{Bri68b} 1.53$\pm$0.11D, \cite{Han68b} and 1.47 D. \cite{Vac89} 
Somehow surprisingly, the most recent value obtained by Stark quantum beat spectroscopy has hardly been considered as reference in theoretical works than the \emph{maximal} measured value of 1.56 D.  
On the theory side, one can highlight two significant earlier contributions {(on the ES geometry)}: 1.48 D (CCSDT/{\AVDZ}) \cite{Hir04} and 1.73 D (CC2/{\AVQZ}). \cite{Hel11} We somehow reconcile these earlier results 
by using both large basis sets and high CC levels {and considering both geometries}, leading to a TBE/CBS of {1.52}$\pm$0.01 D {for the adiabatic value}, right at the center of the experimental cloud. It is noteworthy that the geometrical relaxation 
induces a non-negligible increase of the magnitude of the dipole moment for the $A_2$ ($^1A''$) ES. We indeed found a TBE/CBS value of {1.36}$\pm$0.02 D for the GS geometry.  {For this $\mu_{\mathrm{vert}}^{\mathrm{ES}}$,
earlier estimates include the 1.46 D (CASPT2/TZVP)}\cite{Sch08} {and 1.38 D (CC2/\emph{aug}-cc-pVTZ).}\cite{Sil10b}  For the lowest $B_2$ Rydberg transition, we are aware of one experiment only (polarized electrochromism), leading to an ES 
dipole of -0.33$\pm$0.16 D. \cite{Cau78,Cau79}  While theory does confirm the sign change, it returns a much larger amplitude for the dipole with {-2.15$\pm$0.03} D (GS structure) or {-1.66$\pm$0.03} D (ES geometry).  
At the CCSDT/{\AVDZ} level, Hirata reported -2.52 D (vertical), \cite{Hir04} likely the best previous estimate. This significant discrepancy between theory and experiment was previously attributed to an (experimental) mixing 
between the two lowest-lying $B_2$ transitions. \cite{Hac94} It seems reasonable to state that theory has the edge in this case. For the higher-lying ES, no Stark effect measurement are available, and our values are very likely
more accurate than the previous ones reported at the CCSDT level but with a rather small basis set. \cite{Hir04} Nevertheless, the CBS extrapolation is uneasy and large error bars are obtained for all these high-lying ESs.  
The oscillator strengths of the three lowest Rydberg transitions, $B_2 (n \rightarrow 3s)$, $B_2 (n \rightarrow 3p)$, and $A_1 (n \rightarrow 3s)$ have been respectively measured as 0.038, 0.017$\pm$0.02, and 0.038$\pm$ 0.04 
(absorption spectroscopy), \cite{Men71}  0.028, 0.017, and 0.032  (EELS), \cite{Wei71} 0.032, 0.019, and 0.036 (absorption), \cite{Sut86} and 0.041, 0.028, and 0.061 [dipole (e,e) spectroscopy]. \cite{Coo96} Although the orders 
of magnitude are consistent with the present calculations, the theoretical values do not follow the same ranking as they yield $f$ values of 0.020, 0.035, and 0.050. Such discrepancy has been attributed by other 
groups to the difficulty of assigning individual vibronic bands to a specific electronic transition in the experimental spectra. \cite{Mat01c,Hir04} Interestingly, our current values are agreeing very well with previous theoretical estimates, 
that returned 0.018, 0.040, and 0.043 (MR-AQCC-LRT), \cite{Mul01} 0.025, 0.041, and 0.058 (CCSDT), \cite{Hir04} 0.018, 0.035, and 0.050 (EOM-CCSD), \cite{Gom10b} and 0.021, 0.037, and 0.052 (CC3). \cite{Loo18a} For the 
brighter $\pi \rightarrow \pi^\star$ transition, we are not aware of experimental $f$ values, but theoretical values reported in previous works are of the order of 0.1: 0.100, \cite{Mul01} 0.093, \cite{Gom10b} and 0.135, \cite{Loo18a} and our 
current TBE of {0.107$\pm$0.002} lies in the middle of these earlier data. {In the original Thiel benchmark, the next $A_1$ ES with a larger $f$ was actually considered.}\cite{Sch08}

In thioformaldehyde (Table \ref{Table-9}), one notes relatively stable $\Delta E_{\mathrm{vert}}$ and $\mu^{\mathrm{ES}}$ for the lowest dipole-forbidden $A_2$ transition: the convergence is rather fast with respect to both CC expansion 
and basis set size, so that we can safely report accurate TBE/CBS for both the GS and ES structures. The change of ES dipole between the two geometries is limited as well, contrasting with formaldehyde. This is because there is no puckering effect in
thioformaldehyde's  $A_2$ ES: the true minimum belongs to the  $C_{2v}$ point group. \cite{Jen82b,Dun91,Bud17} For the first Rydberg transition ($B_2$) the impact of the basis set size is logically more pronounced with, \emph{e.g.}, a +0.76 D 
change between the CCSDT/{\AVTZ} and CCSDT/{\DAVTZ} dipoles, making the TBE/CBS extrapolation uncertainty larger than for the $A_2$ ES. The difference between the values of $\mu^{\mathrm{ES}}$ determined at the GS and ES equilibrium geometries 
are also larger for the Rydberg excitation than for the lowest transition, despite the planarity of all geometries. When selecting Dunning's basis sets, the weak oscillator strength of the $B_2$ transition always falls in a rather tight window (0.011--0.015). 
 The valence $A_1 (\pi \rightarrow \pi^\star)$ transition is clearly no cakewalk: not only the enlargement of the basis set yields significant changes of $\mu^{\mathrm{ES}}$  (\emph{e.g.}, +0.49 D from CCSDT/{\AVTZ} to CCSDT/{\DAVTZ), 
but, in addition, the impact of the quadruples in the CC expansion becomes significant: the Q term induces a drop of  $\mu^{\mathrm{ES}}$ by ca.~-0.40 D and a decrease of  $f$ by ca.~15\%.  For this transition CCSD is clearly insufficient to
obtain accurate ES properties. A chemical understanding of the underlying reasons for this large Q effect in the $A_1$ ES of thioformaldehyde would likely require an in-depth analysis of the various densities determined at various levels of theory, 
which is beyond our scope.

The value of $\mu^{\mathrm{GS}}$ in thioformaldehyde was measured very accurately: 1.6491 $\pm$ 0.0004 D, \cite{Fab77} and our TBE of {1.73$\pm$0.01} D is slightly higher. Same comment applies to a previous CCSD(T)/\emph{aug}-cc-pVQZ value of
(1.700 D) \cite{Ben07b} and a CC2/\emph{aug}-cc-pVQZ value of 1.72 D. \cite{Hel11} Like in formaldehyde,  one can find a series of Stark measurements relying on various spectroscopic techniques for the lowest 
$^1A_2$ ES. Quite a range of magnitudes have been reported for  $\mu^{\mathrm{ES}}$: 0.79$\pm$0.04 D (absorption spectroscopy), \cite{Dix78} 0.838$\pm$0.008 D (laser-induced fluorescence excitation), \cite{Dix83} 
0.850$\pm$0.002 D (microwave-optical double resonance), \cite{Suz85} and 0.815$\pm$0.020 D (intermodulated fluorescence). \cite{Fun96} Obviously, the error bars of these measurements are not overlapping, 
but the latter work warns that values between 0.77 and 0.93 D can be obtained. \cite{Fun96} Our TBEs of {0.88$\pm$0.01} D (GS geometry) and  {0.89$\pm$0.01} D (ES geometry) are therefore obviously compatible 
with the experimental measures. The only previous wave function-based TBE we are aware of is the CC2/\emph{aug}-cc-pVQZ value of 0.96 D estimated by Hellweg (ES geometry), \cite{Hel11} that appears approximately 0.10 D too large. 
For the second ES (of Rydberg nature), we know only one measurement of the Stark effect (on the absorption spectrum) that led to a  $\mu^{\mathrm{ES}}$ value of -2.2 $\pm$ 0.3 D.\cite{Goe81} Theory clearly 
confirms the flip of the dipole as compared to the GS, but our TBEs are significantly larger than this experimental value, irrespective of the considered geometry: {-3.27} (GS geometry) and 
{-3.05} D (ES geometry). Given the significant basis set dependence of  $\mu^{\mathrm{ES}}$  of this state, one clearly needs to be cautious but it is nevertheless likely that the experimental  value of -2.2 D is  too low.
To the best of our knowledge, there is no previous published value of  $\mu^{\mathrm{ES}}$ for the $A_1$ ES. Concerning the oscillator strengths, the previous TBEs are likely our CC3/{\AVTZ} values: 0.012 ($B_1$) and 
0.178 ($A_1$), \cite{Loo18a} which are consistent with the new values listed in Table \ref{Table-9}. On the ``experimental side'', an estimate of 0.38 was proposed for the valence transition, \cite{Jud78} but it is based on 
a empirical ratio of 10 compared to an earlier estimate of the oscillator strength for the corresponding Rydberg ES of formaldehyde. \cite{Men71} We trust that our current TBEs are significantly more accurate.

\subsection{Nitroxyl and fluorocarbene}

Table \ref{Table-10} provides the dipole moments and transition energies of the lowest ES of HNO.  Although this transition is not strictly forbidden by symmetry, all methods return very low $f$ values ($<$ 0.001). 
Thus, we have not bothered reporting the values of the oscillator strength.  

\begin{table}[H]
\scriptsize
\caption{Ground-state dipole moment $\mu^{\mathrm{GS}}$, vertical transition energies $\Delta E_{\mathrm{vert}}$, and excited state dipole moments, 
$\mu_{\mathrm{vert}}^{\mathrm{ES}}$ and $\mu_{\mathrm{adia}}^{\mathrm{ES}}$, determined of nitroxyl. We report the norm of the dipoles. See caption of Table \ref{Table-1} for details.} 
\label{Table-10}
\vspace{-0.3 cm}
\begin{tabular}{cc|cccc}
\hline 
		&			& $^1A'$	&	\multicolumn{3}{c}{$^1A'' (\mathrm{Val}, n\rightarrow \pi^\star)$}		\\
Basis 	& Method		&$\mu^{\mathrm{GS}}$	&$\Delta E_{\mathrm{vert}}$	&$\mu_{\mathrm{vert}}^{\mathrm{ES}}$	&$\mu_{\mathrm{adia}}^{\mathrm{ES}}$\\
\hline
{\Pop}	&CCSD			&1.902	&1.802	&1.982	&2.111	\\
		&CCSDT			&1.876	&1.797	&1.948	&2.076	\\
		&CCSDTQ		&1.869	&1.799	&1.938	&2.063	\\
		&CCSDTQP		&1.868     	&1.800	&1.927	&2.062	\\
{\AVDZ}	&CCSD			&1.701	&1.779	&1.719	&1.840	\\
		&CCSDT			&1.667	&1.767	&1.681	&1.799 	\\
		&CCSDTQ		&1.658	&1.770	&1.670	&1.785	\\
{\AVTZ}	&CCSD			&1.722	&1.756	&1.727	&1.850	\\
		&CCSDT			&1.683	&1.737	&1.688	&1.807	\\
{\AVQZ}	&CCSD			&1.735	&1.753	&1.735	&1.859	\\
		&				&		&		&		&	\\
{\DAVDZ}	&CCSD			&1.695	&1.778	&1.709	&1.831	\\
		&CCSDT			&1.661	&1.766	&1.671	&1.790	\\
{\DAVTZ}	&CCSD			&1.720	&1.755	&1.724	&1.847	\\
		&CCSDT			&1.681	&1.737	&1.685	&1.804	\\
{\DAVQZ}	&CCSD			&{1.735}&{1.753}&{1.679}	&{1.795}	\\
		&				&		&\\
{\AVTZ}	& TBE$^a$		&1.674	&1.740	&1.676	&1.791\\
CBS		& TBE$^b$		&1.69$\pm$0.01&1.73$\pm$0.01&1.69$\pm$0.01&1.80$\pm$0.01\\
		&				&		&\\
Lit.		&	Th.			&1.654$^c$&1.74$^d$\\
		&				&		&\\
		&	Exp.			&1.67$\pm$0.03$^e$&1.63$^f$	&	&	1.69$\pm$0.01$^g$\\
		&				&1.62$\pm$0.02$^h$&\\
\hline
\end{tabular}
\vspace{-0.3 cm}
\begin{flushleft}
$^{a,b}${See corresponding footnotes in Table \ref{Table-7};}
$^c${CCSD(T)/CBS value from Ref.~\citenum{Hai18};}
$^d${Unpublished exFCI/{\AVTZ} value (from our groups);}
$^e${From microwave spectroscopy (Ref.~\citenum{Sai73});}
$^f${0-0 energy from Ref.~\citenum{Dix80} and references therein.}
$^g${Mircowave optical double resonance value from Ref.~\citenum{Tak85} }
$^h${From Stark effects measurements (Ref.~\citenum{Joh77}).}
\end{flushleft}
\end{table}

As can be seen in Table \ref{Table-10}, the convergences with respect to the CC excitation order and basis size are rather fast: quadruples tune the dipole par $\sim$-0.01 D only, and basis set 
extension beyond triple-$\zeta$ is unnecessary. In other words CCSDT/{\AVTZ} provides very accurate estimates and the CBS extrapolations come with small error bars.  One notes that the geometrical relaxation
of the ES increases the predicted dipole by ca.~+0.11 D for all methods.  Our TBE/CBS for $\mu^{\mathrm{GS}}$, 1.69$\pm$0.01 D, is slightly above the Hai and Head-Gordon value (1.654 D), but the two available
experiments also show discrepancies larger than the reported uncertainties (see bottom of Table \ref{Table-10}). For the vertical transition energy, our TBE/{\AVTZ} is the same as the result of a CIPSI calculation performed
with the same basis set, and logically exceeds the experimental 0-0 energy.  For the ES dipole we are aware of two experiments, \cite{Dix80,Tak85} but the former  investigated $\mu_a$ (one of the two dipole components) 
only. The most recent experiment yields a total  $\mu^{\mathrm{ES}}$ of 1.69$\pm$0.01 D, \cite{Tak85} which indicates a very slight increase as compared to the GS dipole.  Our $\mu_{\mathrm{vert}}^{\mathrm{ES}}$
({1.69$\pm$0.01 D}) and  $\mu_{\mathrm{adia}}^{\mathrm{ES}}$ {(1.80$\pm$0.01 D}) values apparently slightly undershoots and overestimates the measured \emph{change} of dipole moment. Again, the final call
on the relative accuracy of theory and experiment is hard to make.

\begin{table}[H]
\scriptsize
\caption{Ground-state dipole moment $\mu^{\mathrm{GS}}$, vertical transition energies $\Delta E_{\mathrm{vert}}$, oscillator strengths $f$, and excited state dipole moments, 
$\mu_{\mathrm{vert}}^{\mathrm{ES}}$ and $\mu_{\mathrm{adia}}^{\mathrm{ES}}$, determined of fluorocarbene. We report the norm of the dipoles. See caption of Table \ref{Table-1} for details.} 
\label{Table-11}
\vspace{-0.3 cm}
\begin{tabular}{cc|cccc}
\hline 
		&			& $^1A'$	&	\multicolumn{3}{c}{$^1A''$}		\\
Basis 	& Method		&$\mu^{\mathrm{GS}}$	&$\Delta E_{\mathrm{vert}}$	&$f$ &$\mu_{\mathrm{vert}}^{\mathrm{ES}}$	\\
\hline
{\Pop}	&CCSD			&{1.572}&{2.581}	&{0.009}	&{1.316}  	\\  
		&CCSDT			&{1.552}&{2.573}	&{0.009}	&{1.287} 	\\  
		&CCSDTQ		&{1.549}&{2.577}	&{0.009}	&{1.282} 	\\  
		&CCSDTQP		&{1.549}&{2.578}	&{0.009}	&{1.282} 	\\  
{\AVDZ}	&CCSD			&{1.451}&{2.541}	&{0.007}	&{0.991} 	\\  
		&CCSDT			&{1.430}&{2.529}	&{0.006}	&{0.970} 	\\  
		&CCSDTQ		&{1.428}&{2.534}	&{0.006}	&{0.965} 	\\  
{\AVTZ}	&CCSD			&{1.465}&{2.507}	&{0.006}	&{0.991} 	\\  
		&CCSDT			&{1.441}&{2.493}	&{0.006}	&{0.969} 	\\  
{\AVQZ}	&CCSD			&{1.468}&{2.500}	&{0.006}	&{0.994} 	\\  
		&				&		&			&			&			\\  
{\DAVDZ}	&CCSD			&{1.445}&{2.536}	&{0.006}	&{0.978} 	\\  
		&CCSDT			&{1.425}&{2.524}	&{0.006}	&{0.958} 	\\  
{\DAVTZ}	&CCSD			&{1.461}&{2.505}	&{0.006}	&{0.987} 	\\  
		&CCSDT			&{1.437}&{2.491}	&{0.006}	&{0.965} 	\\  
{\DAVQZ}	&CCSD			&{1.468}&{2.500}	&{0.006}	&{0.994}	\\

		&				&		&\\
{\AVTZ}	& TBE$^a$		&{1.438	}&{2.498}	&{0.006}	&{0.964}	\\
CBS		& TBE$^b$		&{1.44$\pm$0.01}&{2.48$\pm$0.01}&{0.006$\pm$0.001}&{0.97$\pm$0.01}\\
		&				&		&\\
Lit.		&	Th.			&{1.426$^c$}&{2.49$^d$}&{0.006$^e$}&\\
		&				&		&\\
		&	Exp.			&{1.403$^f$}&{2.14$^g$}\\
		&				&		&\\
\hline
\end{tabular}
\vspace{-0.3 cm}
\begin{flushleft}
$^{a,b}${See corresponding footnotes in Table \ref{Table-7};}
$^c${CISD value from Ref.~\citenum{Scu86};}
$^d${exFCI/{\AVTZ} transition energies corrected for basis set effects up to {\AVFZ} from Ref.~\citenum{Loo20d};}
$^e${LR-CC3/{\AVTZ} $f$ from Ref.~\citenum{Loo20d};}
$^f${From Stark effects measurements (Ref.~\citenum{Joh77});}
$^g${Experiment 0-0 energy from Ref. ~\citenum{Kak81}.}
\end{flushleft}
\end{table}

{Table} \ref{Table-11} {provides the dipole moments, oscillator strength, and transition energies for the smallest halocarbene, HCF, a system isoelectronic to the previous one. Although we note
a small oscillation of the GS dipole and transition energies going from CCSD to CCSDT and CCSDTQ, it is obvious that the convergence with respect to the CC order is fast. Likewise, basis set effects are moderate in
the Dunning series, whereas the use of Pople's basis set yields grossly overestimated oscillator strengths, and ES dipole moments. In short reaching accurate values is not problematic. Our TBE/CBS for}
$\mu^{\mathrm{GS}}$, {1.44$\pm$0.01 D, is very close to an earlier CISD estimate, (1.43 D)} \cite{Scu86} {and both are slightly larger than the most recent measurement we are aware of (1.40 D)}.\cite{Wag00}
{As for nitroxyl, our TBE for the vertical transition energy is equivalent to the result of a recent CIPSI calculation, and both are logically larger than the experimental 0-0 energy. The small oscillator
strength determined here is also the same as our} CC3/{\AVTZ} {value. We could not find previous estimates of the ES dipole in the literature, and our calculations yield a decrease of ca.~50}\%\
{as compared to the ground state value, which contrasts with the very similar values obtained for the two states of HNO.}

\subsection{Silylidene}
\label{res-39}

\begin{table*}[htp]
\scriptsize
\caption{Ground-state dipole moment $\mu^{\mathrm{GS}}$, vertical transition energies $\Delta E_{\mathrm{vert}}$, oscillator strengths $f$, and excited state dipole moments $\mu_{\mathrm{vert}}^{\mathrm{ES}}$ determined of silylidene (GS geometry). See caption of Table \ref{Table-1} for details.} 
\label{Table-12}
\vspace{-0.3 cm}
\begin{tabular}{cc|cccccc}
\hline 
		&			& $^1A_1$	&	\multicolumn{2}{c}{$^1A_2 (\mathrm{Ryd}$}	&	\multicolumn{3}{c}{$^1B_2 (\mathrm{Ryd}$)}	\\
Basis 	& Method		&$\mu^{\mathrm{GS}}$	&$\Delta E_{\mathrm{vert}}$	&$\mu_{\mathrm{vert}}^{\mathrm{ES}}$		&$\Delta E_{\mathrm{vert}}$	& $f$&$\mu_{\mathrm{vert}}^{\mathrm{ES}}$	\\
\hline

{\Pop}	&CCSD			&0.091		&2.254	&-1.845	&3.966	&0.045	&-0.080\\
		&CCSDT			&0.028		&2.107	&-1.891	&3.874	&0.042	&-0.237\\
		&CCSDTQ		&0.019		&2.101	&-1.909	&3.876	&0.042	&-0.256\\
		&CCSDTQP		&0.018		&2.101	&-1.911	&3.877	&0.042	&-0.259\\
{\AVDZ}	&CCSD			&0.181		&2.289	&-1.836	&3.875	&0.036	&0.162\\
		&CCSDT			&0.115		&2.146	&-1.889	&3.795	&0.034	&0.005\\
		&CCSDTQ		&0.105		&2.140	&-1.905	&3.798	&0.034	&-0.012\\
{\AVTZ}	&CCSD			&0.235		&2.286	&-1.851	&3.877	&0.036	&0.161\\
		&CCSDT			&0.153		&2.128	&-1.905	&3.779	&0.033	&-0.018\\
{\AVQZ}	&CCSD			&0.249		&2.301	&-1.848	&3.891	&0.037	&0.186\\
		&				&			&		&		&		&		&		\\
{\DAVDZ}	&CCSD			&0.187		&2.290	&-1.831	&3.876	&0.036	&0.164\\
		&CCSDT			&0.121		&2.146	&-1.885	&3.796	&0.034	&0.007\\
{\DAVTZ}	&CCSD			&0.236		&2.287	&-1.848	&3.877	&0.036	&0.166\\
		&CCSDT			&0.155		&2.130	&-1.902	&3.779	&0.033	&-0.014\\
{\DAVQZ}	&{CCSD}		&{0.249}	&{2.301}&{-1.847}&{3.890}&{0.037}&{0.186}\\
		&				&			&		&		&		&		&\\
{\AVTZ}	& TBE$^a$		&0.142		&2.122	&-1.924	&3.783	&0.034	&-0.039\\
CBS		& TBE$^b$		&0.16$\pm$0.01&{2.15$\pm$0.00}&{-1.92}$\pm$0.01&{3.81$\pm$0.01}&0.034$\pm$0.01&{0.00$\pm$0.04}\\
		&				&			&		&		&		&		&\\
Lit.		&	Th.			&0.144$^c$	&2.12$^d$&		&3.80$^d$&0.033$^e$&		\\
		&				&			&		&		&		&		&\\
		&	Exp.			&			&1.88$^f$&		&3.63$^g$&		&	\\
\hline
\end{tabular}
\vspace{-0.3 cm}
\begin{flushleft}
$^{a,b}${See corresponding footnotes in Table \ref{Table-7};}
$^c${CCSD(T)/cc-pVCQZ from Ref.~\citenum{Lu12};}
$^d${exFCI/{\AVTZ} transition energies corrected for basis set effects up to {\AVFZ} from Ref.~\citenum{Loo20d};}
$^e${LR-CC3/{\AVTZ} $f$ from Ref.~\citenum{Loo20d};}
$^f${0-0 energy from Ref.~\citenum{Smi03};}
$^g${0-0 energy from Ref.~\citenum{Har97b}.}
\end{flushleft}
\end{table*}

Let us finish our tour by considering silylidene, H$_2$C=Si, a small original molecule presenting two well-defined low-lying Rydberg ES. \cite{Hil97,Har97b,Smi03} Our results
are collected in Table \ref{Table-12}.  The values of $\mu^{\mathrm{GS}}$ are small in magnitude for all methods and one notes that CCSD significantly overestimates the
dipole whereas one needs at least a triple-$\zeta$ basis set to obtain reasonable data. Our TBE of 0.16 D for the GS dipole  is close to the only previous theoretical value
we found. \cite{Lu12} There is, to the best of our knowledge, no experimental measurement available. The values of $\Delta E_{\mathrm{vert}}$ of the two lowest transitions are insensitive
to the addition of quadruples in the CC series and they converge quite well with respect to the basis set size.  The current TBEs are equal to the ones we obtained earlier applying
a different strategy to reach the FCI limit, \cite{Loo20a} and they remain slightly larger than the experimental 0-0 energies. \cite{Har97b,Smi03} For the two ESs, we
disclose here the two first estimates of the  dipole moments.  For the lowest excitation, the dipole clearly flips direction as compared to the GS, which contrasts with
thioformaldehyde, and also becomes much larger than the GS dipole with a trustworthy TBE of {-1.92}$\pm$0.01 D. For the second ES,  all methods predict relatively
small $\mu_{\mathrm{vert}}^{\mathrm{ES}}$ values, with the CCSD and CCSDT signs sometimes in disagreement for a given basis set. Although the addition of a second set of
diffuse orbitals has a quite small effect, the convergence with the size of the basis is quite slow, and we can only state that the final dipole should be almost null, though its
sign remains unknown. Finally, one notes that 6-31+G(d) provides too large oscillator strengths for the second ES, but that the stability is otherwise remarkable. One can likely
be confident in the proposed TBE value (0.034).

\section{Conclusions}

In this contribution, we have considered {30} singlet excited states in {thirteen} small molecules and strived to obtain oscillator strengths and dipole moments as accurate as possible. To this end,
we have performed a series of CC calculations going from (LR) CCSD to CCSDTQP using a large panel of basis sets containing one or two sets of diffuse functions.  In all cases, we have
obtained FCI/{\AVTZ} quality properties, as well as estimates of the corresponding FCI/CBS values, the latter coming with quite large uncertainties in some cases. While FCI results do obviously
yield rather definitive answers, we treated only small molecules here with computationally expensive methods, so that the transferability of this strategy to larger compounds is indeed limited.
Regarding the CC expansion, we found that the correction brought by the P term is always negligible, whereas the impact of Q is often rather small, although some exceptions to the latter statement 
have been observed for the considered set. 
For instance, a reduction of the oscillator strength and ES dipole by -15\%\ and -0.40 D, respectively, is observed for the valence $\pi \rightarrow \pi^\star$ transition of thioformaldehyde.  More problematic
is the convergence of the computed properties while increasing the size of the atomic basis set: this convergence can go from very smooth to erratic. For some states, huge differences 
between the results obtained with simply- and doubly-augmented basis sets are indeed found. The oscillator strengths determined for the three close-lying $\Pi_u$ excited states of dinitrogen being a typical
example of this problem caused by state mixing.  All in all, when choices have to be made, it seems a better option to use CCSDT with a very large basis set rather than CCSDTQ with a
smaller one when one wishes to perform comparisons with experiment, {which as  explained in the Introduction is always a challenging task}.  We have also found several examples herein in which one 
property, e.g., the oscillator strength, is rather independent from the selected basis set, whereas another, e.g., the ES dipole, is not. One must therefore carefully check the basis set convergence 
and dependence for all considered states and properties.

Despite these challenges, it is certainly noteworthy that for the vast majority of the properties studied here, we could not only establish the most accurate theoretical estimates available to date,
but also obtain values that are compatible with the experimental {knowledge} when these are accessible, which is not always the case even for the small molecules considered here. Theory sometimes
deliver smaller error bars than the corresponding experimental data.   This is certainly the case for the smallest compounds treated here, for which very large basis sets could be employed. 
It should be stressed that the measurements of both $f$ and $\mu^{\mathrm{ES}}$ are difficult, so that depending on the experimental techniques, apparently incompatible results are quite commonly 
reported and that the role of theory is likely critical. At this stage, it might be important to recall that we did not used any experimental input, as even our geometries are theoretically 
determined. The present effort is thus truly \emph{ab initio}. However, it only provides an idealized picture as we did not aimed at modeling vibronic effects.

 As we expected at the beginning of the study, getting the right answer for the right reason in the context of ES properties is certainly more challenging than for the corresponding energies. It is therefore 
 our hope that the reference values reported here will be useful benchmarks and will stimulate further studies in both the theoretical and experimental communities.

%%%%%%%%%%%%%%%%%%%%%%%%
%%% ACKNOWLEDGEMENTS %%%
%%%%%%%%%%%%%%%%%%%%%%%%
\section*{Acknowledgements}
AC and DJ are indebted to the LumoMat program for support in the framework of the Fluo-34 grant. PFL thanks the \textit{Centre National de la Recherche Scientifique} for funding. 
This research used resources of (i) the GENCI-TGCC (Grant No.~2019-A0060801738);  (ii) CALMIP under allocation 2020-18005 (Toulouse); (iii) CCIPL (\emph{Centre de Calcul Intensif des Pays de Loire}); 
(iv) a local Troy cluster and (v) HPC resources from ArronaxPlus  (grant ANR-11-EQPX-0004 funded by the French National Agency for Research). 

%%%%%%%%%%%%%%%%%
%%% SUPP INFO %%%
%%%%%%%%%%%%%%%%%
%\begin{suppinfo}
\section*{Supporting Information Available}
Extra data and Cartesian coordinates. The Supporting Information is available free of charge at https://pubs.acs.org/doi/10.1021/{doi}.
%\end{suppinfo}

%%%%%%%%%%%%%%%%%%%%
%%% BIBLIOGRAPHY %%%
%%%%%%%%%%%%%%%%%%%%
\bibliography{biblio-new}

\providecommand{\latin}[1]{#1}
\makeatletter
\providecommand{\doi}
  {\begingroup\let\do\@makeother\dospecials
  \catcode`\{=1 \catcode`\}=2\doi@aux}
\providecommand{\doi@aux}[1]{\endgroup\texttt{#1}}
\makeatother
\providecommand*\mcitethebibliography{\thebibliography}
\csname @ifundefined\endcsname{endmcitethebibliography}
  {\let\endmcitethebibliography\endthebibliography}{}
\begin{mcitethebibliography}{194}
\providecommand*\natexlab[1]{#1}
\providecommand*\mciteSetBstSublistMode[1]{}
\providecommand*\mciteSetBstMaxWidthForm[2]{}
\providecommand*\mciteBstWouldAddEndPuncttrue
  {\def\EndOfBibitem{\unskip.}}
\providecommand*\mciteBstWouldAddEndPunctfalse
  {\let\EndOfBibitem\relax}
\providecommand*\mciteSetBstMidEndSepPunct[3]{}
\providecommand*\mciteSetBstSublistLabelBeginEnd[3]{}
\providecommand*\EndOfBibitem{}
\mciteSetBstSublistMode{f}
\mciteSetBstMaxWidthForm{subitem}{(\alph{mcitesubitemcount})}
\mciteSetBstSublistLabelBeginEnd
  {\mcitemaxwidthsubitemform\space}
  {\relax}
  {\relax}

\bibitem[Loos \latin{et~al.}(2020)Loos, Scemama, and Jacquemin]{Loo20c}
Loos,~P.~F.; Scemama,~A.; Jacquemin,~D. The Quest for Highly-Accurate
  Excitation Energies: A Computational Perspective. \emph{J. Phys. Chem. Lett.}
  \textbf{2020}, \emph{11}, 2374--2383\relax
\mciteBstWouldAddEndPuncttrue
\mciteSetBstMidEndSepPunct{\mcitedefaultmidpunct}
{\mcitedefaultendpunct}{\mcitedefaultseppunct}\relax
\EndOfBibitem
\bibitem[Schreiber \latin{et~al.}(2008)Schreiber, Silva-Junior, Sauer, and
  Thiel]{Sch08}
Schreiber,~M.; Silva-Junior,~M.~R.; Sauer,~S. P.~A.; Thiel,~W. Benchmarks for
  Electronically Excited States: CASPT2, CC2, CCSD and CC3. \emph{J. Chem.
  Phys.} \textbf{2008}, \emph{128}, 134110\relax
\mciteBstWouldAddEndPuncttrue
\mciteSetBstMidEndSepPunct{\mcitedefaultmidpunct}
{\mcitedefaultendpunct}{\mcitedefaultseppunct}\relax
\EndOfBibitem
\bibitem[Sauer \latin{et~al.}(2009)Sauer, Schreiber, Silva-Junior, and
  Thiel]{Sau09}
Sauer,~S. P.~A.; Schreiber,~M.; Silva-Junior,~M.~R.; Thiel,~W. Benchmarks for
  Electronically Excited States: A Comparison of Noniterative and Iterative
  Triples Corrections in Linear Response Coupled Cluster Methods: CCSDR(3)
  \emph{versus} CC3. \emph{J. Chem. Theory Comput.} \textbf{2009}, \emph{5},
  555--564\relax
\mciteBstWouldAddEndPuncttrue
\mciteSetBstMidEndSepPunct{\mcitedefaultmidpunct}
{\mcitedefaultendpunct}{\mcitedefaultseppunct}\relax
\EndOfBibitem
\bibitem[Silva-Junior \latin{et~al.}(2010)Silva-Junior, Schreiber, Sauer, and
  Thiel]{Sil10c}
Silva-Junior,~M.~R.; Schreiber,~M.; Sauer,~S. P.~A.; Thiel,~W. Benchmarks of
  Electronically Excited States: Basis Set Effecs Benchmarks of Electronically
  Excited States: Basis Set Effects on CASPT2 Results. \emph{J. Chem. Phys.}
  \textbf{2010}, \emph{133}, 174318\relax
\mciteBstWouldAddEndPuncttrue
\mciteSetBstMidEndSepPunct{\mcitedefaultmidpunct}
{\mcitedefaultendpunct}{\mcitedefaultseppunct}\relax
\EndOfBibitem
\bibitem[Loos \latin{et~al.}(2018)Loos, Scemama, Blondel, Garniron, Caffarel,
  and Jacquemin]{Loo18a}
Loos,~P.-F.; Scemama,~A.; Blondel,~A.; Garniron,~Y.; Caffarel,~M.;
  Jacquemin,~D. A Mountaineering Strategy to Excited States: Highly-Accurate
  Reference Energies and Benchmarks. \emph{J. Chem. Theory Comput.}
  \textbf{2018}, \emph{14}, 4360--4379\relax
\mciteBstWouldAddEndPuncttrue
\mciteSetBstMidEndSepPunct{\mcitedefaultmidpunct}
{\mcitedefaultendpunct}{\mcitedefaultseppunct}\relax
\EndOfBibitem
\bibitem[Loos \latin{et~al.}(2019)Loos, Boggio-Pasqua, Scemama, Caffarel, and
  Jacquemin]{Loo19c}
Loos,~P.-F.; Boggio-Pasqua,~M.; Scemama,~A.; Caffarel,~M.; Jacquemin,~D.
  Reference Energies for Double Excitations. \emph{J. Chem. Theory Comput.}
  \textbf{2019}, \emph{15}, 1939--1956\relax
\mciteBstWouldAddEndPuncttrue
\mciteSetBstMidEndSepPunct{\mcitedefaultmidpunct}
{\mcitedefaultendpunct}{\mcitedefaultseppunct}\relax
\EndOfBibitem
\bibitem[Loos \latin{et~al.}(2020)Loos, Lipparini, Boggio-Pasqua, Scemama, and
  Jacquemin]{Loo20a}
Loos,~P.-F.; Lipparini,~F.; Boggio-Pasqua,~M.; Scemama,~A.; Jacquemin,~D. A
  Mountaineering Strategy to Excited States: Highly-Accurate Energies and
  Benchmarks for Medium Size Molecules. \emph{J. Chem. Theory Comput.}
  \textbf{2020}, \emph{16}, 1711--1741\relax
\mciteBstWouldAddEndPuncttrue
\mciteSetBstMidEndSepPunct{\mcitedefaultmidpunct}
{\mcitedefaultendpunct}{\mcitedefaultseppunct}\relax
\EndOfBibitem
\bibitem[Loos \latin{et~al.}(2020)Loos, Scemama, Boggio-Pasqua, and
  Jacquemin]{Loo20d}
Loos,~P.-F.; Scemama,~A.; Boggio-Pasqua,~M.; Jacquemin,~D. A Mountaineering
  Strategy to Excited States: Highly-Accurate Energies and Benchmarks for
  Exotic Molecules and Radicals. \emph{J. Chem. Theory Comput.} \textbf{2020},
  \emph{16}, 3720--3736\relax
\mciteBstWouldAddEndPuncttrue
\mciteSetBstMidEndSepPunct{\mcitedefaultmidpunct}
{\mcitedefaultendpunct}{\mcitedefaultseppunct}\relax
\EndOfBibitem
\bibitem[Furche and Ahlrichs(2002)Furche, and Ahlrichs]{Fur02}
Furche,~F.; Ahlrichs,~R. Adiabatic Time-Dependent Density Functional Methods
  for Excited States Properties. \emph{J. Chem. Phys.} \textbf{2002},
  \emph{117}, 7433--7447\relax
\mciteBstWouldAddEndPuncttrue
\mciteSetBstMidEndSepPunct{\mcitedefaultmidpunct}
{\mcitedefaultendpunct}{\mcitedefaultseppunct}\relax
\EndOfBibitem
\bibitem[Dierksen and Grimme(2004)Dierksen, and Grimme]{Die04b}
Dierksen,~M.; Grimme,~S. The Vibronic Structure of Electronic Absorption
  Spectra of Large Molecules: A Time-Dependent Density Functional Study on the
  Influence of \emph{Exact} Hartree-Fock Exchange. \emph{J. Phys. Chem. A}
  \textbf{2004}, \emph{108}, 10225--10237\relax
\mciteBstWouldAddEndPuncttrue
\mciteSetBstMidEndSepPunct{\mcitedefaultmidpunct}
{\mcitedefaultendpunct}{\mcitedefaultseppunct}\relax
\EndOfBibitem
\bibitem[H{\"a}ttig(2005)]{Hat05c}
H{\"a}ttig,~C. In \emph{Response Theory and Molecular Properties (A Tribute to
  Jan Linderberg and Poul J{\o}rgensen)}; Jensen,~H.~A., Ed.; Advances in
  Quantum Chemistry; Academic Press, 2005; Vol.~50; pp 37--60\relax
\mciteBstWouldAddEndPuncttrue
\mciteSetBstMidEndSepPunct{\mcitedefaultmidpunct}
{\mcitedefaultendpunct}{\mcitedefaultseppunct}\relax
\EndOfBibitem
\bibitem[Goerigk \latin{et~al.}(2009)Goerigk, Moellmann, and Grimme]{Goe09}
Goerigk,~L.; Moellmann,~J.; Grimme,~S. Computation of Accurate Excitation
  Energies for Large Organic Molecules with Double-Hybrid Density Functionals.
  \emph{Phys. Chem. Chem. Phys.} \textbf{2009}, \emph{11}, 4611--4620\relax
\mciteBstWouldAddEndPuncttrue
\mciteSetBstMidEndSepPunct{\mcitedefaultmidpunct}
{\mcitedefaultendpunct}{\mcitedefaultseppunct}\relax
\EndOfBibitem
\bibitem[Goerigk and Grimme(2010)Goerigk, and Grimme]{Goe10a}
Goerigk,~L.; Grimme,~S. Assessment of TD-DFT Methods and of Various Spin Scaled
  CIS$_n$D and CC2 Versions for the Treatment of Low-Lying Valence Excitations
  of Large Organic Dyes. \emph{J. Chem. Phys.} \textbf{2010}, \emph{132},
  184103\relax
\mciteBstWouldAddEndPuncttrue
\mciteSetBstMidEndSepPunct{\mcitedefaultmidpunct}
{\mcitedefaultendpunct}{\mcitedefaultseppunct}\relax
\EndOfBibitem
\bibitem[Send \latin{et~al.}(2011)Send, K{\"u}hn, and Furche]{Sen11b}
Send,~R.; K{\"u}hn,~M.; Furche,~F. Assessing Excited State Methods by Adiabatic
  Excitation Energies. \emph{J. Chem. Theory Comput.} \textbf{2011}, \emph{7},
  2376--2386\relax
\mciteBstWouldAddEndPuncttrue
\mciteSetBstMidEndSepPunct{\mcitedefaultmidpunct}
{\mcitedefaultendpunct}{\mcitedefaultseppunct}\relax
\EndOfBibitem
\bibitem[Jacquemin \latin{et~al.}(2012)Jacquemin, Planchat, Adamo, and
  Mennucci]{Jac12d}
Jacquemin,~D.; Planchat,~A.; Adamo,~C.; Mennucci,~B. A TD-DFT Assessment of
  Functionals for Optical 0-0 Transitions in Solvated Dyes. \emph{J. Chem.
  Theory Comput.} \textbf{2012}, \emph{8}, 2359--2372\relax
\mciteBstWouldAddEndPuncttrue
\mciteSetBstMidEndSepPunct{\mcitedefaultmidpunct}
{\mcitedefaultendpunct}{\mcitedefaultseppunct}\relax
\EndOfBibitem
\bibitem[Winter \latin{et~al.}(2013)Winter, Graf, Leutwyler, and
  H{\"a}ttig]{Win13}
Winter,~N. O.~C.; Graf,~N.~K.; Leutwyler,~S.; H{\"a}ttig,~C. Benchmarks for
  0--0 Transitions of Aromatic Organic Molecules: DFT/B3LYP{,} ADC(2){,} CC2{,}
  SOS-CC2 and SCS-CC2 Compared to High-resolution Gas-Phase Data. \emph{Phys.
  Chem. Chem. Phys.} \textbf{2013}, \emph{15}, 6623--6630\relax
\mciteBstWouldAddEndPuncttrue
\mciteSetBstMidEndSepPunct{\mcitedefaultmidpunct}
{\mcitedefaultendpunct}{\mcitedefaultseppunct}\relax
\EndOfBibitem
\bibitem[Fang \latin{et~al.}(2014)Fang, Oruganti, and Durbeej]{Fan14b}
Fang,~C.; Oruganti,~B.; Durbeej,~B. How Method-Dependent Are Calculated
  Differences Between Vertical, Adiabatic and 0-0 Excitation Energies? \emph{J.
  Phys. Chem. A} \textbf{2014}, \emph{118}, 4157--4171\relax
\mciteBstWouldAddEndPuncttrue
\mciteSetBstMidEndSepPunct{\mcitedefaultmidpunct}
{\mcitedefaultendpunct}{\mcitedefaultseppunct}\relax
\EndOfBibitem
\bibitem[Jacquemin \latin{et~al.}(2015)Jacquemin, Duchemin, and Blase]{Jac15b}
Jacquemin,~D.; Duchemin,~I.; Blase,~X. 0--0 Energies Using Hybrid Schemes:
  Benchmarks of TD-DFT, CIS(D), ADC(2), CC2, and BSE/GW formalisms for 80
  Real-Life Compounds. \emph{J. Chem. Theory Comput.} \textbf{2015}, \emph{11},
  5340--5359\relax
\mciteBstWouldAddEndPuncttrue
\mciteSetBstMidEndSepPunct{\mcitedefaultmidpunct}
{\mcitedefaultendpunct}{\mcitedefaultseppunct}\relax
\EndOfBibitem
\bibitem[Oruganti \latin{et~al.}(2016)Oruganti, Fang, and Durbeej]{Oru16}
Oruganti,~B.; Fang,~C.; Durbeej,~B. Assessment of a Composite CC2/DFT Procedure
  for Calculating 0--0 Excitation Energies of Organic Molecules. \emph{Mol.
  Phys.} \textbf{2016}, \emph{114}, 3448--3463\relax
\mciteBstWouldAddEndPuncttrue
\mciteSetBstMidEndSepPunct{\mcitedefaultmidpunct}
{\mcitedefaultendpunct}{\mcitedefaultseppunct}\relax
\EndOfBibitem
\bibitem[Schwabe and Goerigk(2017)Schwabe, and Goerigk]{Sch17}
Schwabe,~T.; Goerigk,~L. Time-Dependent Double-Hybrid Density Functionals with
  Spin-Component and Spin-Opposite Scaling. \emph{J. Chem. Theory Comput.}
  \textbf{2017}, \emph{13}, 4307--4323\relax
\mciteBstWouldAddEndPuncttrue
\mciteSetBstMidEndSepPunct{\mcitedefaultmidpunct}
{\mcitedefaultendpunct}{\mcitedefaultseppunct}\relax
\EndOfBibitem
\bibitem[Loos \latin{et~al.}(2018)Loos, Galland, and Jacquemin]{Loo18b}
Loos,~P.-F.; Galland,~N.; Jacquemin,~D. Theoretical 0--0 Energies with Chemical
  Accuracy. \emph{J. Phys. Chem. Lett.} \textbf{2018}, \emph{9},
  4646--4651\relax
\mciteBstWouldAddEndPuncttrue
\mciteSetBstMidEndSepPunct{\mcitedefaultmidpunct}
{\mcitedefaultendpunct}{\mcitedefaultseppunct}\relax
\EndOfBibitem
\bibitem[Loos and Jacquemin(2019)Loos, and Jacquemin]{Loo19a}
Loos,~P.-F.; Jacquemin,~D. Chemically Accurate 0-0 Energies with
  not-so-Accurate Excited State Geometries. \emph{J. Chem. Theory Comput.}
  \textbf{2019}, \emph{15}, 2481--2491\relax
\mciteBstWouldAddEndPuncttrue
\mciteSetBstMidEndSepPunct{\mcitedefaultmidpunct}
{\mcitedefaultendpunct}{\mcitedefaultseppunct}\relax
\EndOfBibitem
\bibitem[Loos and Jacquemin(2019)Loos, and Jacquemin]{Loo19b}
Loos,~P.-F.; Jacquemin,~D. Evaluating 0-0 Energies with Theoretical Tools: a
  Short Review. \emph{ChemPhotoChem} \textbf{2019}, \emph{3}, 684--696\relax
\mciteBstWouldAddEndPuncttrue
\mciteSetBstMidEndSepPunct{\mcitedefaultmidpunct}
{\mcitedefaultendpunct}{\mcitedefaultseppunct}\relax
\EndOfBibitem
\bibitem[Bremond \latin{et~al.}(2018)Bremond, Savarese, Adamo, and
  Jacquemin]{Bre18a}
Bremond,~E.; Savarese,~M.; Adamo,~C.; Jacquemin,~D. Accuracy of TD-DFT
  Geometries: a Fresh Look. \emph{J. Chem. Theory Comput.} \textbf{2018},
  \emph{14}, 3715--3727\relax
\mciteBstWouldAddEndPuncttrue
\mciteSetBstMidEndSepPunct{\mcitedefaultmidpunct}
{\mcitedefaultendpunct}{\mcitedefaultseppunct}\relax
\EndOfBibitem
\bibitem[Tajti and Szalay(2019)Tajti, and Szalay]{Taj19}
Tajti,~A.; Szalay,~P.~G. Accuracy of Spin-Component-Scaled CC2 Excitation
  Energies and Potential Energy Surfaces. \emph{J. Chem. Theory Comput.}
  \textbf{2019}, \emph{15}, 5523--5531\relax
\mciteBstWouldAddEndPuncttrue
\mciteSetBstMidEndSepPunct{\mcitedefaultmidpunct}
{\mcitedefaultendpunct}{\mcitedefaultseppunct}\relax
\EndOfBibitem
\bibitem[Tajti \latin{et~al.}(2020)Tajti, Tulip{\'a}n, and Szalay]{Taj20a}
Tajti,~A.; Tulip{\'a}n,~L.; Szalay,~P.~G. Accuracy of Spin-Component Scaled
  ADC(2) Excitation Energies and Potential Energy Surfaces. \emph{J. Chem.
  Theory Comput.} \textbf{2020}, \emph{16}, 468--474\relax
\mciteBstWouldAddEndPuncttrue
\mciteSetBstMidEndSepPunct{\mcitedefaultmidpunct}
{\mcitedefaultendpunct}{\mcitedefaultseppunct}\relax
\EndOfBibitem
\bibitem[Page and Olivucci(2003)Page, and Olivucci]{Pag03}
Page,~C.~S.; Olivucci,~M. Ground and Excited State CASPT2 Geometry
  Optimizations of Small Organic Molecules. \emph{J. Comput. Chem.}
  \textbf{2003}, \emph{24}, 298--309\relax
\mciteBstWouldAddEndPuncttrue
\mciteSetBstMidEndSepPunct{\mcitedefaultmidpunct}
{\mcitedefaultendpunct}{\mcitedefaultseppunct}\relax
\EndOfBibitem
\bibitem[Bousquet \latin{et~al.}(2013)Bousquet, Fukuda, Maitarad, Jacquemin,
  Ciofini, Adamo, and Ehara]{Bou13}
Bousquet,~D.; Fukuda,~R.; Maitarad,~P.; Jacquemin,~D.; Ciofini,~I.; Adamo,~C.;
  Ehara,~M. Excited-State Geometries of Heteroaromatic Compounds: A Comparative
  TD-DFT and SAC-CI Study. \emph{J. Chem. Theory Comput.} \textbf{2013},
  \emph{9}, 2368--2379\relax
\mciteBstWouldAddEndPuncttrue
\mciteSetBstMidEndSepPunct{\mcitedefaultmidpunct}
{\mcitedefaultendpunct}{\mcitedefaultseppunct}\relax
\EndOfBibitem
\bibitem[Bousquet \latin{et~al.}(2014)Bousquet, Fukuda, Jacquemin, Ciofini,
  Adamo, and Ehara]{Bou14b}
Bousquet,~D.; Fukuda,~R.; Jacquemin,~D.; Ciofini,~I.; Adamo,~C.; Ehara,~M.
  Benchmark Study on the Triplet Excited-State Geometries and Phosphorescence
  Energies of Heterocyclic Compounds: Comparison Between TD-PBE0 and SAC-CI.
  \emph{J. Chem. Theory Comput.} \textbf{2014}, \emph{10}, 3969--3979\relax
\mciteBstWouldAddEndPuncttrue
\mciteSetBstMidEndSepPunct{\mcitedefaultmidpunct}
{\mcitedefaultendpunct}{\mcitedefaultseppunct}\relax
\EndOfBibitem
\bibitem[Jagau and Gauss(2012)Jagau, and Gauss]{Jag12}
Jagau,~T.-C.; Gauss,~J. Ground and Excited State Geometries via Mukherjee's
  Multireference Coupled-Cluster Method. \emph{Chem. Phys.} \textbf{2012},
  \emph{401}, 73--87\relax
\mciteBstWouldAddEndPuncttrue
\mciteSetBstMidEndSepPunct{\mcitedefaultmidpunct}
{\mcitedefaultendpunct}{\mcitedefaultseppunct}\relax
\EndOfBibitem
\bibitem[Guareschi and Filippi(2013)Guareschi, and Filippi]{Gua13}
Guareschi,~R.; Filippi,~C. Ground- and Excited-State Geometry Optimization of
  Small Organic Molecules with Quantum Monte Carlo. \emph{J. Chem. Theory
  Comput.} \textbf{2013}, \emph{9}, 5513--5525\relax
\mciteBstWouldAddEndPuncttrue
\mciteSetBstMidEndSepPunct{\mcitedefaultmidpunct}
{\mcitedefaultendpunct}{\mcitedefaultseppunct}\relax
\EndOfBibitem
\bibitem[Tuna \latin{et~al.}(2016)Tuna, Lu, Koslowski, and Thiel]{Tun16}
Tuna,~D.; Lu,~Y.; Koslowski,~A.; Thiel,~W. Semiempirical Quantum-Chemical
  Orthogonalization-Corrected Methods: Benchmarks of Electronically Excited
  States. \emph{J. Chem. Theory Comput.} \textbf{2016}, \emph{12},
  4400--4422\relax
\mciteBstWouldAddEndPuncttrue
\mciteSetBstMidEndSepPunct{\mcitedefaultmidpunct}
{\mcitedefaultendpunct}{\mcitedefaultseppunct}\relax
\EndOfBibitem
\bibitem[Budz{\'a}k \latin{et~al.}(2017)Budz{\'a}k, Scalmani, and
  Jacquemin]{Bud17}
Budz{\'a}k,~{\v S}.; Scalmani,~G.; Jacquemin,~D. Accurate Excited-State
  Geometries: a CASPT2 and Coupled-Cluster Reference Database for Small
  Molecules. \emph{J. Chem. Theory Comput.} \textbf{2017}, \emph{13},
  6237--6252\relax
\mciteBstWouldAddEndPuncttrue
\mciteSetBstMidEndSepPunct{\mcitedefaultmidpunct}
{\mcitedefaultendpunct}{\mcitedefaultseppunct}\relax
\EndOfBibitem
\bibitem[Jacquemin(2018)]{Jac18a}
Jacquemin,~D. What is the Key for Accurate Absorption and Emission Calculations
  ? Energy or Geometry ? \emph{J. Chem. Theory Comput.} \textbf{2018},
  \emph{14}, 1534--1543\relax
\mciteBstWouldAddEndPuncttrue
\mciteSetBstMidEndSepPunct{\mcitedefaultmidpunct}
{\mcitedefaultendpunct}{\mcitedefaultseppunct}\relax
\EndOfBibitem
\bibitem[Fihey and Jacquemin(2019)Fihey, and Jacquemin]{Fih19}
Fihey,~A.; Jacquemin,~D. Performances of Density Functional Tight-Binding
  Methods for Describing Ground and Excited State Geometries of Organic
  Molecules. \emph{J. Chem. Theory Comput.} \textbf{2019}, \emph{15},
  6267--6276\relax
\mciteBstWouldAddEndPuncttrue
\mciteSetBstMidEndSepPunct{\mcitedefaultmidpunct}
{\mcitedefaultendpunct}{\mcitedefaultseppunct}\relax
\EndOfBibitem
\bibitem[Tajti \latin{et~al.}(2018)Tajti, Stanton, Matthews, and Szalay]{Taj18}
Tajti,~A.; Stanton,~J.~F.; Matthews,~D.~A.; Szalay,~P.~G. Accuracy of Coupled
  Cluster Excited State Potential Energy Surfaces. \emph{J. Chem. Theory
  Comput.} \textbf{2018}, \emph{14}, 5859--5869\relax
\mciteBstWouldAddEndPuncttrue
\mciteSetBstMidEndSepPunct{\mcitedefaultmidpunct}
{\mcitedefaultendpunct}{\mcitedefaultseppunct}\relax
\EndOfBibitem
\bibitem[Gozem \latin{et~al.}(2012)Gozem, Schapiro, Ferr\'e, and
  Olivucci]{Goz12}
Gozem,~S.; Schapiro,~I.; Ferr\'e,~N.; Olivucci,~M. The Molecular Mechanism of
  Thermal Noise in Rod Photoreceptors. \emph{Science} \textbf{2012},
  \emph{337}, 1225--1228\relax
\mciteBstWouldAddEndPuncttrue
\mciteSetBstMidEndSepPunct{\mcitedefaultmidpunct}
{\mcitedefaultendpunct}{\mcitedefaultseppunct}\relax
\EndOfBibitem
\bibitem[Tuna \latin{et~al.}(2015)Tuna, Lefrancois, Wola{\'n}ski, Gozem,
  Schapiro, Andruni{\'o}w, Dreuw, and Olivucci]{Tun15}
Tuna,~D.; Lefrancois,~D.; Wola{\'n}ski,~{\L}.; Gozem,~S.; Schapiro,~I.;
  Andruni{\'o}w,~T.; Dreuw,~A.; Olivucci,~M. Assessment of Approximate
  Coupled-Cluster and Algebraic-Diagrammatic-Construction Methods for Ground-
  and Excited-State Reaction Paths and the Conical-Intersection Seam of a
  Retinal-Chromophore Model. \emph{J. Chem. Theory Comput.} \textbf{2015},
  \emph{11}, 5758--5781\relax
\mciteBstWouldAddEndPuncttrue
\mciteSetBstMidEndSepPunct{\mcitedefaultmidpunct}
{\mcitedefaultendpunct}{\mcitedefaultseppunct}\relax
\EndOfBibitem
\bibitem[K{\'a}nn{\'a}r and Szalay(2014)K{\'a}nn{\'a}r, and Szalay]{Kan14}
K{\'a}nn{\'a}r,~D.; Szalay,~P.~G. Benchmarking Coupled Cluster Methods on
  Valence Singlet Excited States. \emph{J. Chem. Theory Comput.} \textbf{2014},
  \emph{10}, 3757--3765\relax
\mciteBstWouldAddEndPuncttrue
\mciteSetBstMidEndSepPunct{\mcitedefaultmidpunct}
{\mcitedefaultendpunct}{\mcitedefaultseppunct}\relax
\EndOfBibitem
\bibitem[Robinson(2018)]{Rob18}
Robinson,~D. Comparison of the Transition Dipole Moments Calculated by TDDFT
  with High Level Wave Function Theory. \emph{J. Chem. Theory Comput.}
  \textbf{2018}, \emph{14}, 5303--5309\relax
\mciteBstWouldAddEndPuncttrue
\mciteSetBstMidEndSepPunct{\mcitedefaultmidpunct}
{\mcitedefaultendpunct}{\mcitedefaultseppunct}\relax
\EndOfBibitem
\bibitem[Caricato \latin{et~al.}(2010)Caricato, Trucks, Frisch, and
  Wiberg]{Car10d}
Caricato,~M.; Trucks,~G.; Frisch,~M.; Wiberg,~K. Oscillator Strength: How Does
  TDDFT Compare to EOM-CCSD? \emph{J. Chem. Theory Comput.} \textbf{2010},
  \emph{7}, 456--466\relax
\mciteBstWouldAddEndPuncttrue
\mciteSetBstMidEndSepPunct{\mcitedefaultmidpunct}
{\mcitedefaultendpunct}{\mcitedefaultseppunct}\relax
\EndOfBibitem
\bibitem[Li \latin{et~al.}(2007)Li, Wan, and Xu]{Li07b}
Li,~Y.; Wan,~J.; Xu,~X. Theoretical Study of the Vertical Excited States of
  Benzene, Pyrimidine, and Pyrazineby the Symmetry Adapted Cluster --
  Configuration Interaction Method. \emph{J. Comput. Chem.} \textbf{2007},
  \emph{28}, 1658--1667\relax
\mciteBstWouldAddEndPuncttrue
\mciteSetBstMidEndSepPunct{\mcitedefaultmidpunct}
{\mcitedefaultendpunct}{\mcitedefaultseppunct}\relax
\EndOfBibitem
\bibitem[Silva-Junior \latin{et~al.}(2010)Silva-Junior, Sauer, Schreiber, and
  Thiel]{Sil10b}
Silva-Junior,~M.~R.; Sauer,~S. P.~A.; Schreiber,~M.; Thiel,~W. Basis Set
  Effects on Coupled Cluster Benchmarks of Electronically Excited States: CC3,
  CCSDR(3) and CC2. \emph{Mol. Phys.} \textbf{2010}, \emph{108}, 453--465\relax
\mciteBstWouldAddEndPuncttrue
\mciteSetBstMidEndSepPunct{\mcitedefaultmidpunct}
{\mcitedefaultendpunct}{\mcitedefaultseppunct}\relax
\EndOfBibitem
\bibitem[Jacquemin \latin{et~al.}(2016)Jacquemin, Duchemin, Blondel, and
  Blase]{Jac16b}
Jacquemin,~D.; Duchemin,~I.; Blondel,~A.; Blase,~X. Assessment of the Accuracy
  of the Bethe-Salpeter (BSE/GW) Oscillator Strengths. \emph{J. Chem. Theory
  Comput.} \textbf{2016}, \emph{12}, 3969--3981\relax
\mciteBstWouldAddEndPuncttrue
\mciteSetBstMidEndSepPunct{\mcitedefaultmidpunct}
{\mcitedefaultendpunct}{\mcitedefaultseppunct}\relax
\EndOfBibitem
\bibitem[Hellweg(2011)]{Hel11}
Hellweg,~A. The Accuracy of Dipole Moments from Spin-Component Scaled CC2 in
  Ground and Electronically Excited States. \emph{J. Chem. Phys.}
  \textbf{2011}, \emph{134}, 064103\relax
\mciteBstWouldAddEndPuncttrue
\mciteSetBstMidEndSepPunct{\mcitedefaultmidpunct}
{\mcitedefaultendpunct}{\mcitedefaultseppunct}\relax
\EndOfBibitem
\bibitem[Jacquemin(2016)]{Jac16d}
Jacquemin,~D. Excited-State Dipole and Quadrupole Moments: TD-DFT versus CC2.
  \emph{J. Chem. Theory Comput.} \textbf{2016}, \emph{12}, 3993--4003\relax
\mciteBstWouldAddEndPuncttrue
\mciteSetBstMidEndSepPunct{\mcitedefaultmidpunct}
{\mcitedefaultendpunct}{\mcitedefaultseppunct}\relax
\EndOfBibitem
\bibitem[Hait and Head-Gordon(2018)Hait, and Head-Gordon]{Hai18}
Hait,~D.; Head-Gordon,~M. How Accurate Is Density Functional Theory at
  Predicting Dipole Moments? An Assessment Using a New Database of 200
  Benchmark Values. \emph{J. Chem. Theory Comput.} \textbf{2018}, \emph{14},
  1969--1981\relax
\mciteBstWouldAddEndPuncttrue
\mciteSetBstMidEndSepPunct{\mcitedefaultmidpunct}
{\mcitedefaultendpunct}{\mcitedefaultseppunct}\relax
\EndOfBibitem
\bibitem[Borges(2006)]{Bor06a}
Borges,~I. Influences on the Calculation of Accurate and Basis Set Extrapolated
  Oscillator Strengths: the $\tilde{A}^1B_1\leftarrow\tilde{X}^1A_1$Transition
  of H$_2$O. \emph{J. Phys. B: At. Mol. Phys.} \textbf{2006}, \emph{39},
  641--650\relax
\mciteBstWouldAddEndPuncttrue
\mciteSetBstMidEndSepPunct{\mcitedefaultmidpunct}
{\mcitedefaultendpunct}{\mcitedefaultseppunct}\relax
\EndOfBibitem
\bibitem[Nielsen \latin{et~al.}(1980)Nielsen, Jorgensen, and
  Oddershede]{Nie80b}
Nielsen,~E.~S.; Jorgensen,~P.; Oddershede,~J. Transition Moments and Dynamic
  Polarizabilities in a Second Order Polarization Propagator Approach. \emph{J.
  Chem. Phys.} \textbf{1980}, \emph{73}, 6238--6246\relax
\mciteBstWouldAddEndPuncttrue
\mciteSetBstMidEndSepPunct{\mcitedefaultmidpunct}
{\mcitedefaultendpunct}{\mcitedefaultseppunct}\relax
\EndOfBibitem
\bibitem[Schmitt and Meerts(2018)Schmitt, and Meerts]{Sch08b}
Schmitt,~M.; Meerts,~L. In \emph{Frontiers and Advances in Molecular
  Spectroscopy}; Laane,~J., Ed.; Elsevier, 2018; Chapter 5, pp 143--193\relax
\mciteBstWouldAddEndPuncttrue
\mciteSetBstMidEndSepPunct{\mcitedefaultmidpunct}
{\mcitedefaultendpunct}{\mcitedefaultseppunct}\relax
\EndOfBibitem
\bibitem[Lombardi(1987)]{Lom87}
Lombardi,~J.~R. On the Comparison of Solvatochromic Shifts with Gas Phase Stark
  Effect Measurements. \emph{Spectrochim. Acta A} \textbf{1987}, \emph{43},
  1323--1324\relax
\mciteBstWouldAddEndPuncttrue
\mciteSetBstMidEndSepPunct{\mcitedefaultmidpunct}
{\mcitedefaultendpunct}{\mcitedefaultseppunct}\relax
\EndOfBibitem
\bibitem[Eriksen and Gauss(2020)Eriksen, and Gauss]{Eri20}
Eriksen,~J.~J.; Gauss,~J. arXiv:2008.03610\relax
\mciteBstWouldAddEndPuncttrue
\mciteSetBstMidEndSepPunct{\mcitedefaultmidpunct}
{\mcitedefaultendpunct}{\mcitedefaultseppunct}\relax
\EndOfBibitem
\bibitem[Christiansen \latin{et~al.}(1995)Christiansen, Koch, and
  J{\o}rgensen]{Chr95b}
Christiansen,~O.; Koch,~H.; J{\o}rgensen,~P. Response Functions in the CC3
  Iterative Triple Excitation Model. \emph{J. Chem. Phys.} \textbf{1995},
  \emph{103}, 7429--7441\relax
\mciteBstWouldAddEndPuncttrue
\mciteSetBstMidEndSepPunct{\mcitedefaultmidpunct}
{\mcitedefaultendpunct}{\mcitedefaultseppunct}\relax
\EndOfBibitem
\bibitem[Koch \latin{et~al.}(1995)Koch, Christiansen, Jorgensen, and
  Olsen]{Koc95}
Koch,~H.; Christiansen,~O.; Jorgensen,~P.; Olsen,~J. Excitation Energies of BH,
  CH$_2$ and Ne in Full Configuration Interaction and the Hierarchy CCS, CC2,
  CCSD and CC3 of Coupled Cluster Models. \emph{Chem. Phys. Lett.}
  \textbf{1995}, \emph{244}, 75--82\relax
\mciteBstWouldAddEndPuncttrue
\mciteSetBstMidEndSepPunct{\mcitedefaultmidpunct}
{\mcitedefaultendpunct}{\mcitedefaultseppunct}\relax
\EndOfBibitem
\bibitem[Aidas \latin{et~al.}(2014)Aidas, Angeli, Bak, Bakken, Bast, Boman,
  Christiansen, Cimiraglia, Coriani, Dahle, Dalskov, Ekstr{\"o}m, Enevoldsen,
  Eriksen, Ettenhuber, Fern{\'a}ndez, Ferrighi, Fliegl, Frediani, Hald,
  Halkier, H\"attig, Heiberg, Helgaker, Hennum, Hettema, Hjerten{\ae}s,
  H{\o}st, Hoyvik, Iozzi, Jansik, Jensen, Jonsson, J{\o}rgensen, Kauczor,
  Kirpekar, Kj{\ae}rgaard, Klopper, Knecht, Kobayashi, Koch, Kongsted, Krapp,
  Kristensen, Ligabue, Lutn{\ae}s, Melo, Mikkelsen, Myhre, Neiss, Nielsen,
  Norman, Olsen, Olsen, Osted, Packer, Pawlowski, Pedersen, Provasi, Reine,
  Rinkevicius, Ruden, Ruud, Rybkin, Sa{\l}ek, Samson, de~Mer{\'a}s, Saue,
  Sauer, Schimmelpfennig, Sneskov, Steindal, Sylvester-Hvid, Taylor, Teale,
  Tellgren, Tew, Thorvaldsen, Thogersen, Vahtras, Watson, Wilson, Ziolkowski,
  and {\AA}gren]{dalton}
Aidas,~K.; Angeli,~C.; Bak,~K.~L.; Bakken,~V.; Bast,~R.; Boman,~L.;
  Christiansen,~O.; Cimiraglia,~R.; Coriani,~S.; Dahle,~P.; Dalskov,~E.~K.;
  Ekstr{\"o}m,~U.; Enevoldsen,~T.; Eriksen,~J.~J.; Ettenhuber,~P.;
  Fern{\'a}ndez,~B.; Ferrighi,~L.; Fliegl,~H.; Frediani,~L.; Hald,~K.;
  Halkier,~A.; H\"attig,~C.; Heiberg,~H.; Helgaker,~T.; Hennum,~A.~C.;
  Hettema,~H.; Hjerten{\ae}s,~E.; H{\o}st,~S.; Hoyvik,~I.-M.; Iozzi,~M.~F.;
  Jansik,~B.; Jensen,~H. J.~A.; Jonsson,~D.; J{\o}rgensen,~P.; Kauczor,~J.;
  Kirpekar,~S.; Kj{\ae}rgaard,~T.; Klopper,~W.; Knecht,~S.; Kobayashi,~R.;
  Koch,~H.; Kongsted,~J.; Krapp,~A.; Kristensen,~K.; Ligabue,~A.;
  Lutn{\ae}s,~O.~B.; Melo,~J.~I.; Mikkelsen,~K.~V.; Myhre,~R.~H.; Neiss,~C.;
  Nielsen,~C.~B.; Norman,~P.; Olsen,~J.; Olsen,~J. M.~H.; Osted,~A.;
  Packer,~M.~J.; Pawlowski,~F.; Pedersen,~T.~B.; Provasi,~P.~F.; Reine,~S.;
  Rinkevicius,~Z.; Ruden,~T.~A.; Ruud,~K.; Rybkin,~V.~V.; Sa{\l}ek,~P.;
  Samson,~C. C.~M.; de~Mer{\'a}s,~A.~S.; Saue,~T.; Sauer,~S. P.~A.;
  Schimmelpfennig,~B.; Sneskov,~K.; Steindal,~A.~H.; Sylvester-Hvid,~K.~O.;
  Taylor,~P.~R.; Teale,~A.~M.; Tellgren,~E.~I.; Tew,~D.~P.; Thorvaldsen,~A.~J.;
  Thogersen,~L.; Vahtras,~O.; Watson,~M.~A.; Wilson,~D. J.~D.; Ziolkowski,~M.;
  {\AA}gren,~H. The Dalton Quantum Chemistry Program System. \emph{WIREs
  Comput. Mol. Sci.} \textbf{2014}, \emph{4}, 269--284\relax
\mciteBstWouldAddEndPuncttrue
\mciteSetBstMidEndSepPunct{\mcitedefaultmidpunct}
{\mcitedefaultendpunct}{\mcitedefaultseppunct}\relax
\EndOfBibitem
\bibitem[cfo()]{cfour}
CFOUR, Coupled-Cluster techniques for Computational Chemistry, a
  quantum-chemical program package by J.F. Stanton, J. Gauss, L. Cheng, M.E.
  Harding, D.A. Matthews, P.G. Szalay with contributions from A.A. Auer, R.J.
  Bartlett, U. Benedikt, C. Berger, D.E. Bernholdt, Y.J. Bomble, O.
  Christiansen, F. Engel, R. Faber, M. Heckert, O. Heun, M. Hilgenberg, C.
  Huber, T.-C. Jagau, D. Jonsson, J. Jus{\'e}lius, T. Kirsch, K. Klein, W.J.
  Lauderdale, F. Lipparini, T. Metzroth, L.A. M{\"u}ck, D.P. O'Neill, D.R.
  Price, E. Prochnow, C. Puzzarini, K. Ruud, F. Schiffmann, W. Schwalbach, C.
  Simmons, S. Stopkowicz, A. Tajti, J. V{\'a}zquez, F. Wang, J.D. Watts and the
  integral packages MOLECULE (J. Alml\"of and P.R. Taylor), PROPS (P.R.
  Taylor), ABACUS (T. Helgaker, H.J. Aa. Jensen, P. J{\o}rgensen, and J.
  Olsen), and ECP routines by A. V. Mitin and C. van W{\"u}llen. For the
  current version, see http://www.cfour.de.\relax
\mciteBstWouldAddEndPunctfalse
\mciteSetBstMidEndSepPunct{\mcitedefaultmidpunct}
{}{\mcitedefaultseppunct}\relax
\EndOfBibitem
\bibitem[Helgaker \latin{et~al.}(1997)Helgaker, Klopper, Koch, and
  Noga]{Hel97b}
Helgaker,~T.; Klopper,~W.; Koch,~H.; Noga,~J. Basis-Set Convergence of
  Correlated Calculations on Water. \emph{J. Chem. Phys.} \textbf{1997},
  \emph{106}, 9639--9646\relax
\mciteBstWouldAddEndPuncttrue
\mciteSetBstMidEndSepPunct{\mcitedefaultmidpunct}
{\mcitedefaultendpunct}{\mcitedefaultseppunct}\relax
\EndOfBibitem
\bibitem[Halkier \latin{et~al.}(1999)Halkier, Klopper, Helgaker, and
  Jorgensen]{Hal99b}
Halkier,~A.; Klopper,~W.; Helgaker,~T.; Jorgensen,~P. Basis-set Convergence of
  the Molecular Electric Dipole Moment. \emph{J. Chem. Phys.} \textbf{1999},
  \emph{111}, 4424--4430\relax
\mciteBstWouldAddEndPuncttrue
\mciteSetBstMidEndSepPunct{\mcitedefaultmidpunct}
{\mcitedefaultendpunct}{\mcitedefaultseppunct}\relax
\EndOfBibitem
\bibitem[Kallay \latin{et~al.}(2020)Kallay, Nagy, Mester, Rolik, Samu, Csoka,
  Szabo, Gyevi-Nagu, Hegely, Ladjanski, Szegedy, Ladoczki, Petrov, Farkas,
  Mezei, and Ganyecz]{Kal20}
Kallay,~M.; Nagy,~P.~E.; Mester,~D.; Rolik,~Z.; Samu,~J.,~G. Ans~Csontos;
  Csoka,~J.; Szabo,~P.~B.; Gyevi-Nagu,~L.; Hegely,~B.; Ladjanski,~I.;
  Szegedy,~L.; Ladoczki,~B.; Petrov,~K.; Farkas,~M.; Mezei,~P.~D.; Ganyecz,~A.
  The MRCC Program System: Accurate Quantum Chemistry from Water to Proteins.
  \emph{J. Chem. Phys.} \textbf{2020}, \emph{152}, 074107\relax
\mciteBstWouldAddEndPuncttrue
\mciteSetBstMidEndSepPunct{\mcitedefaultmidpunct}
{\mcitedefaultendpunct}{\mcitedefaultseppunct}\relax
\EndOfBibitem
\bibitem[K{\'a}llay \latin{et~al.}(2017)K{\'a}llay, Rolik, Csontos, Nagy, Samu,
  Mester, Cs{\'o}ka, Szab{\'o}, Ladj{\'a}nszki, Szegedy, Lad{\'o}czki, Petrov,
  Farkas, Mezei, and H{\'e}gely.]{mrcc}
K{\'a}llay,~M.; Rolik,~Z.; Csontos,~J.; Nagy,~P.; Samu,~G.; Mester,~D.;
  Cs{\'o}ka,~J.; Szab{\'o},~B.; Ladj{\'a}nszki,~I.; Szegedy,~L.;
  Lad{\'o}czki,~B.; Petrov,~K.; Farkas,~M.; Mezei,~P.~D.; H{\'e}gely.,~B. MRCC,
  Quantum Chemical Program. 2017; See: www.mrcc.hu.\relax
\mciteBstWouldAddEndPunctfalse
\mciteSetBstMidEndSepPunct{\mcitedefaultmidpunct}
{}{\mcitedefaultseppunct}\relax
\EndOfBibitem
\bibitem[Purvis~III and Bartlett(1982)Purvis~III, and Bartlett]{Pur82}
Purvis~III,~G.~P.; Bartlett,~R.~J. A Full Coupled-Cluster Singles and Doubles
  Model: The Inclusion of Disconnected Triples. \emph{J. Chem. Phys.}
  \textbf{1982}, \emph{76}, 1910--1918\relax
\mciteBstWouldAddEndPuncttrue
\mciteSetBstMidEndSepPunct{\mcitedefaultmidpunct}
{\mcitedefaultendpunct}{\mcitedefaultseppunct}\relax
\EndOfBibitem
\bibitem[Scuseria \latin{et~al.}(1987)Scuseria, Scheiner, Lee, Rice, and
  Schaefer]{Scu87}
Scuseria,~G.~E.; Scheiner,~A.~C.; Lee,~T.~J.; Rice,~J.~E.; Schaefer,~H.~F. The
  Closed‐Shell Coupled Cluster Single and Double Excitation (CCSD) Model for
  the Description of Electron Correlation. A Comparison with Configuration
  Interaction (CISD) Results. \emph{J. Chem. Phys.} \textbf{1987}, \emph{86},
  2881--2890\relax
\mciteBstWouldAddEndPuncttrue
\mciteSetBstMidEndSepPunct{\mcitedefaultmidpunct}
{\mcitedefaultendpunct}{\mcitedefaultseppunct}\relax
\EndOfBibitem
\bibitem[Koch \latin{et~al.}(1990)Koch, Jensen, Jorgensen, and
  Helgaker]{Koc90b}
Koch,~H.; Jensen,~H. J.~A.; Jorgensen,~P.; Helgaker,~T. Excitation Energies
  from the Coupled Cluster Singles and Doubles Linear Response Function
  (CCSDLR). Applications to Be, CH$^+$, CO, and H$_2$O. \emph{J. Chem. Phys.}
  \textbf{1990}, \emph{93}, 3345--3350\relax
\mciteBstWouldAddEndPuncttrue
\mciteSetBstMidEndSepPunct{\mcitedefaultmidpunct}
{\mcitedefaultendpunct}{\mcitedefaultseppunct}\relax
\EndOfBibitem
\bibitem[Stanton and Bartlett(1993)Stanton, and Bartlett]{Sta93}
Stanton,~J.~F.; Bartlett,~R.~J. The Equation of Motion Coupled-Cluster Method -
  A Systematic Biorthogonal Approach to Molecular Excitation Energies,
  Transition-Probabilities, and Excited-State Properties. \emph{J. Chem. Phys.}
  \textbf{1993}, \emph{98}, 7029--7039\relax
\mciteBstWouldAddEndPuncttrue
\mciteSetBstMidEndSepPunct{\mcitedefaultmidpunct}
{\mcitedefaultendpunct}{\mcitedefaultseppunct}\relax
\EndOfBibitem
\bibitem[Stanton(1993)]{Sta93b}
Stanton,~J.~F. Many-Body Methods for Excited State Potential Energy Surfaces.
  I: General Theory of Energy Gradients for the Equation-of-Motion
  Coupled-Cluster Method. \emph{J. Chem. Phys.} \textbf{1993}, \emph{99},
  8840--8847\relax
\mciteBstWouldAddEndPuncttrue
\mciteSetBstMidEndSepPunct{\mcitedefaultmidpunct}
{\mcitedefaultendpunct}{\mcitedefaultseppunct}\relax
\EndOfBibitem
\bibitem[Noga and Bartlett(1987)Noga, and Bartlett]{Nog87}
Noga,~J.; Bartlett,~R.~J. The Full CCSDT Model for Molecular Electronic
  Structure. \emph{J. Chem. Phys.} \textbf{1987}, \emph{86}, 7041--7050\relax
\mciteBstWouldAddEndPuncttrue
\mciteSetBstMidEndSepPunct{\mcitedefaultmidpunct}
{\mcitedefaultendpunct}{\mcitedefaultseppunct}\relax
\EndOfBibitem
\bibitem[Scuseria and Schaefer(1988)Scuseria, and Schaefer]{Scu88}
Scuseria,~G.~E.; Schaefer,~H.~F. A New Implementation of the Full CCSDT Model
  for Molecular Electronic Structure. \emph{Chem. Phys. Lett.} \textbf{1988},
  \emph{152}, 382--386\relax
\mciteBstWouldAddEndPuncttrue
\mciteSetBstMidEndSepPunct{\mcitedefaultmidpunct}
{\mcitedefaultendpunct}{\mcitedefaultseppunct}\relax
\EndOfBibitem
\bibitem[Kucharski \latin{et~al.}(2001)Kucharski, W{\l}och, Musia{\l}, and
  Bartlett]{Kuc01}
Kucharski,~S.~A.; W{\l}och,~M.; Musia{\l},~M.; Bartlett,~R.~J. Coupled-Cluster
  Theory for Excited Electronic States: The Full Equation-Of-Motion
  Coupled-Cluster Single, Double, and Triple Excitation Method. \emph{J. Chem.
  Phys.} \textbf{2001}, \emph{115}, 8263--8266\relax
\mciteBstWouldAddEndPuncttrue
\mciteSetBstMidEndSepPunct{\mcitedefaultmidpunct}
{\mcitedefaultendpunct}{\mcitedefaultseppunct}\relax
\EndOfBibitem
\bibitem[Kowalski and Piecuch(2001)Kowalski, and Piecuch]{Kow01}
Kowalski,~K.; Piecuch,~P. The Active-Space Equation-of-Motion Coupled-Cluster
  Methods for Excited Electronic States: Full EOMCCSDt. \emph{J. Chem. Phys.}
  \textbf{2001}, \emph{115}, 643--651\relax
\mciteBstWouldAddEndPuncttrue
\mciteSetBstMidEndSepPunct{\mcitedefaultmidpunct}
{\mcitedefaultendpunct}{\mcitedefaultseppunct}\relax
\EndOfBibitem
\bibitem[Kowalski and Piecuch(2001)Kowalski, and Piecuch]{Kow01b}
Kowalski,~K.; Piecuch,~P. Excited-State Potential Energy Curves of CH$^+$: a
  Comparison of the EOMCCSDt And Full EOMCCSDT Results. \emph{Chem. Phys.
  Lett.} \textbf{2001}, \emph{347}, 237--246\relax
\mciteBstWouldAddEndPuncttrue
\mciteSetBstMidEndSepPunct{\mcitedefaultmidpunct}
{\mcitedefaultendpunct}{\mcitedefaultseppunct}\relax
\EndOfBibitem
\bibitem[Kucharski and Bartlett(1991)Kucharski, and Bartlett]{Kuc91}
Kucharski,~S.~A.; Bartlett,~R.~J. Recursive Intermediate Factorization and
  Complete Computational Linearization of the Coupled-Cluster Single, Double,
  Triple, and Quadruple Excitation Equations. \emph{Theor. Chim. Acta}
  \textbf{1991}, \emph{80}, 387--405\relax
\mciteBstWouldAddEndPuncttrue
\mciteSetBstMidEndSepPunct{\mcitedefaultmidpunct}
{\mcitedefaultendpunct}{\mcitedefaultseppunct}\relax
\EndOfBibitem
\bibitem[K{\'a}llay \latin{et~al.}(2003)K{\'a}llay, Gauss, and Szalay]{Kal03}
K{\'a}llay,~M.; Gauss,~J.; Szalay,~P.~G. Analytic First Derivatives for General
  Coupled-Cluster and Configuration Interaction Models. \emph{J. Chem. Phys.}
  \textbf{2003}, \emph{119}, 2991--3004\relax
\mciteBstWouldAddEndPuncttrue
\mciteSetBstMidEndSepPunct{\mcitedefaultmidpunct}
{\mcitedefaultendpunct}{\mcitedefaultseppunct}\relax
\EndOfBibitem
\bibitem[K{\'a}llay and Gauss(2004)K{\'a}llay, and Gauss]{Kal04}
K{\'a}llay,~M.; Gauss,~J. Calculation of Excited-State Properties Using General
  Coupled-Cluster and Configuration-Interaction Models. \emph{J. Chem. Phys.}
  \textbf{2004}, \emph{121}, 9257--9269\relax
\mciteBstWouldAddEndPuncttrue
\mciteSetBstMidEndSepPunct{\mcitedefaultmidpunct}
{\mcitedefaultendpunct}{\mcitedefaultseppunct}\relax
\EndOfBibitem
\bibitem[Hirata(2004)]{Hir04}
Hirata,~S. Higher-Order Equation-of-Motion Coupled-Cluster Methods. \emph{J.
  Chem. Phys.} \textbf{2004}, \emph{121}, 51--59\relax
\mciteBstWouldAddEndPuncttrue
\mciteSetBstMidEndSepPunct{\mcitedefaultmidpunct}
{\mcitedefaultendpunct}{\mcitedefaultseppunct}\relax
\EndOfBibitem
\bibitem[Suellen \latin{et~al.}(2019)Suellen, Garcia~Freitas, Loos, and
  Jacquemin]{Sue19}
Suellen,~C.; Garcia~Freitas,~R.; Loos,~P.-F.; Jacquemin,~D. Cross Comparisons
  Between Experiment, TD-DFT, CC, and ADC for Transition Energies. \emph{J.
  Chem. Theory Comput.} \textbf{2019}, \emph{15}, 4581--4590\relax
\mciteBstWouldAddEndPuncttrue
\mciteSetBstMidEndSepPunct{\mcitedefaultmidpunct}
{\mcitedefaultendpunct}{\mcitedefaultseppunct}\relax
\EndOfBibitem
\bibitem[Frisch \latin{et~al.}(2016)Frisch, Trucks, Schlegel, Scuseria, Robb,
  Cheeseman, Scalmani, Barone, Petersson, Nakatsuji, Li, Caricato, Marenich,
  Bloino, Janesko, Gomperts, Mennucci, Hratchian, Ortiz, Izmaylov, Sonnenberg,
  Williams-Young, Ding, Lipparini, Egidi, Goings, Peng, Petrone, Henderson,
  Ranasinghe, Zakrzewski, Gao, Rega, Zheng, Liang, Hada, Ehara, Toyota, Fukuda,
  Hasegawa, Ishida, Nakajima, Honda, Kitao, Nakai, Vreven, Throssell,
  Montgomery, Peralta, Ogliaro, Bearpark, Heyd, Brothers, Kudin, Staroverov,
  Keith, Kobayashi, Normand, Raghavachari, Rendell, Burant, Iyengar, Tomasi,
  Cossi, Millam, Klene, Adamo, Cammi, Ochterski, Martin, Morokuma, Farkas,
  Foresman, and Fox]{Gaussian16}
Frisch,~M.~J.; Trucks,~G.~W.; Schlegel,~H.~B.; Scuseria,~G.~E.; Robb,~M.~A.;
  Cheeseman,~J.~R.; Scalmani,~G.; Barone,~V.; Petersson,~G.~A.; Nakatsuji,~H.;
  Li,~X.; Caricato,~M.; Marenich,~A.~V.; Bloino,~J.; Janesko,~B.~G.;
  Gomperts,~R.; Mennucci,~B.; Hratchian,~H.~P.; Ortiz,~J.~V.; Izmaylov,~A.~F.;
  Sonnenberg,~J.~L.; Williams-Young,~D.; Ding,~F.; Lipparini,~F.; Egidi,~F.;
  Goings,~J.; Peng,~B.; Petrone,~A.; Henderson,~T.; Ranasinghe,~D.;
  Zakrzewski,~V.~G.; Gao,~J.; Rega,~N.; Zheng,~G.; Liang,~W.; Hada,~M.;
  Ehara,~M.; Toyota,~K.; Fukuda,~R.; Hasegawa,~J.; Ishida,~M.; Nakajima,~T.;
  Honda,~Y.; Kitao,~O.; Nakai,~H.; Vreven,~T.; Throssell,~K.;
  Montgomery,~J.~A.,~{Jr.}; Peralta,~J.~E.; Ogliaro,~F.; Bearpark,~M.~J.;
  Heyd,~J.~J.; Brothers,~E.~N.; Kudin,~K.~N.; Staroverov,~V.~N.; Keith,~T.~A.;
  Kobayashi,~R.; Normand,~J.; Raghavachari,~K.; Rendell,~A.~P.; Burant,~J.~C.;
  Iyengar,~S.~S.; Tomasi,~J.; Cossi,~M.; Millam,~J.~M.; Klene,~M.; Adamo,~C.;
  Cammi,~R.; Ochterski,~J.~W.; Martin,~R.~L.; Morokuma,~K.; Farkas,~O.;
  Foresman,~J.~B.; Fox,~D.~J. Gaussian 16 {R}evision {A}.03. 2016; Gaussian
  Inc. Wallingford CT\relax
\mciteBstWouldAddEndPuncttrue
\mciteSetBstMidEndSepPunct{\mcitedefaultmidpunct}
{\mcitedefaultendpunct}{\mcitedefaultseppunct}\relax
\EndOfBibitem
\bibitem[Krylov and Gill(2013)Krylov, and Gill]{Kry13}
Krylov,~A.~I.; Gill,~P.~M. Q-Chem: an Engine for Innovation. \emph{WIREs
  Comput. Mol. Sci.} \textbf{2013}, \emph{3}, 317--326\relax
\mciteBstWouldAddEndPuncttrue
\mciteSetBstMidEndSepPunct{\mcitedefaultmidpunct}
{\mcitedefaultendpunct}{\mcitedefaultseppunct}\relax
\EndOfBibitem
\bibitem[Folkestad \latin{et~al.}(2020)Folkestad, Kj{\o}nstad, Myhre, Andersen,
  Balbi, Coriani, Giovannini, Goletto, Haugland, Hutcheson, H{\o}yvik, Moitra,
  Paul, Scavino, Skeidsvoll, Tveten, and Koch]{eT}
Folkestad,~S.~D.; Kj{\o}nstad,~E.~F.; Myhre,~R.~H.; Andersen,~J.~H.; Balbi,~A.;
  Coriani,~S.; Giovannini,~T.; Goletto,~L.; Haugland,~T.~S.; Hutcheson,~A.;
  H{\o}yvik,~I.-M.; Moitra,~T.; Paul,~A.~C.; Scavino,~M.; Skeidsvoll,~A.~S.;
  Tveten,~{\AA}.~H.; Koch,~H. eT 1.0: An Open Source Electronic Structure
  Program with Emphasis on Coupled lCuster and Multilevel Methods. \emph{J.
  Chem. Phys.} \textbf{2020}, \emph{152}, 184103\relax
\mciteBstWouldAddEndPuncttrue
\mciteSetBstMidEndSepPunct{\mcitedefaultmidpunct}
{\mcitedefaultendpunct}{\mcitedefaultseppunct}\relax
\EndOfBibitem
\bibitem[Koch and J{\o}rgensen(1990)Koch, and J{\o}rgensen]{Koc90}
Koch,~H.; J{\o}rgensen,~P. Coupled Cluster Response Functions. \emph{J. Chem.
  Phys.} \textbf{1990}, \emph{93}, 3333--3344\relax
\mciteBstWouldAddEndPuncttrue
\mciteSetBstMidEndSepPunct{\mcitedefaultmidpunct}
{\mcitedefaultendpunct}{\mcitedefaultseppunct}\relax
\EndOfBibitem
\bibitem[Christiansen \latin{et~al.}(1998)Christiansen, J{\o}rgensen, and
  H\"attig]{Chr98d}
Christiansen,~O.; J{\o}rgensen,~P.; H\"attig,~C. Response Functions from
  Fourier Component Variational Perturbation Theory Applied to a Time-Averaged
  Quasienergy. \emph{Int. J. Quantum Chem.} \textbf{1998}, \emph{68},
  1--52\relax
\mciteBstWouldAddEndPuncttrue
\mciteSetBstMidEndSepPunct{\mcitedefaultmidpunct}
{\mcitedefaultendpunct}{\mcitedefaultseppunct}\relax
\EndOfBibitem
\bibitem[Caricato \latin{et~al.}(2009)Caricato, Trucks, and Frisch]{Car09b}
Caricato,~M.; Trucks,~G.~W.; Frisch,~M.~J. On the Difference Between the
  Transition Properties Calculated with Linear Response- and Equation of
  Motion-CCSD Approaches. \emph{J. Chem. Phys.} \textbf{2009}, \emph{131},
  174104\relax
\mciteBstWouldAddEndPuncttrue
\mciteSetBstMidEndSepPunct{\mcitedefaultmidpunct}
{\mcitedefaultendpunct}{\mcitedefaultseppunct}\relax
\EndOfBibitem
\bibitem[Paw{\l}owski \latin{et~al.}(2004)Paw{\l}owski, J{\o}rgensen, and
  H{\"a}ttig]{Paw04}
Paw{\l}owski,~F.; J{\o}rgensen,~P.; H{\"a}ttig,~C. Gauge Invariance of
  Oscillator Strengths in the Approximate Coupled Cluster Triples Model CC3.
  \emph{Chem. Phys. Lett.} \textbf{2004}, \emph{389}, 413--420\relax
\mciteBstWouldAddEndPuncttrue
\mciteSetBstMidEndSepPunct{\mcitedefaultmidpunct}
{\mcitedefaultendpunct}{\mcitedefaultseppunct}\relax
\EndOfBibitem
\bibitem[Trucks \latin{et~al.}(1988)Trucks, Salter, Sosa, and Bartlett]{Tru88}
Trucks,~G.~W.; Salter,~E.; Sosa,~C.; Bartlett,~R.~J. Theory and Implementation
  of the MBPT Density Matrix. An Application to One-Electron Properties.
  \emph{Chem. Phys. Lett.} \textbf{1988}, \emph{147}, 359--366\relax
\mciteBstWouldAddEndPuncttrue
\mciteSetBstMidEndSepPunct{\mcitedefaultmidpunct}
{\mcitedefaultendpunct}{\mcitedefaultseppunct}\relax
\EndOfBibitem
\bibitem[Christiansen \latin{et~al.}(1995)Christiansen, Koch, and
  J{\o}rgensen]{Chr95}
Christiansen,~O.; Koch,~H.; J{\o}rgensen,~P. The Second-Order Approximate
  Coupled Cluster Singles and Doubles Model CC2. \emph{Chem. Phys. Lett.}
  \textbf{1995}, \emph{243}, 409--418\relax
\mciteBstWouldAddEndPuncttrue
\mciteSetBstMidEndSepPunct{\mcitedefaultmidpunct}
{\mcitedefaultendpunct}{\mcitedefaultseppunct}\relax
\EndOfBibitem
\bibitem[Balabanov and Peterson(2006)Balabanov, and Peterson]{Bal06}
Balabanov,~N.~B.; Peterson,~K.~A. Basis set Limit Electronic Excitation
  Energies, Ionization Potentials, and Electron Affinities for the $3d$
  Transition Metal Atoms: Coupled Cluster and Multireference Methods. \emph{J.
  Chem. Phys.} \textbf{2006}, \emph{125}, 074110\relax
\mciteBstWouldAddEndPuncttrue
\mciteSetBstMidEndSepPunct{\mcitedefaultmidpunct}
{\mcitedefaultendpunct}{\mcitedefaultseppunct}\relax
\EndOfBibitem
\bibitem[Kamiya and Hirata(2006)Kamiya, and Hirata]{Kam06b}
Kamiya,~M.; Hirata,~S. Higher-Order Equation-of-Motion Coupled-Cluster Methods
  for Ionization Processes. \emph{J. Chem. Phys.} \textbf{2006}, \emph{125},
  074111\relax
\mciteBstWouldAddEndPuncttrue
\mciteSetBstMidEndSepPunct{\mcitedefaultmidpunct}
{\mcitedefaultendpunct}{\mcitedefaultseppunct}\relax
\EndOfBibitem
\bibitem[Watson and Chan(2012)Watson, and Chan]{Wat12}
Watson,~M.~A.; Chan,~G. K.-L. Excited States of Butadiene to Chemical Accuracy:
  Reconciling Theory and Experiment. \emph{J. Chem. Theory Comput.}
  \textbf{2012}, \emph{8}, 4013--4018\relax
\mciteBstWouldAddEndPuncttrue
\mciteSetBstMidEndSepPunct{\mcitedefaultmidpunct}
{\mcitedefaultendpunct}{\mcitedefaultseppunct}\relax
\EndOfBibitem
\bibitem[Feller \latin{et~al.}(2014)Feller, Peterson, and Davidson]{Fel14}
Feller,~D.; Peterson,~K.~A.; Davidson,~E.~R. A Systematic Approach to
  Vertically Excited States of Ethylene Using Configuration Interaction and
  Coupled Cluster Techniques. \emph{J. Chem. Phys.} \textbf{2014}, \emph{141},
  104302\relax
\mciteBstWouldAddEndPuncttrue
\mciteSetBstMidEndSepPunct{\mcitedefaultmidpunct}
{\mcitedefaultendpunct}{\mcitedefaultseppunct}\relax
\EndOfBibitem
\bibitem[Franke \latin{et~al.}(2019)Franke, Moore, Schaefer, and
  Douberly]{Fra19}
Franke,~P.~R.; Moore,~K.~B.; Schaefer,~H.~F.; Douberly,~G.~E. tert-Butyl Peroxy
  Radical: Ground and First Excited State Energetics and Fundamental
  Frequencies. \emph{Phys. Chem. Chem. Phys.} \textbf{2019}, \emph{21},
  9747--9758\relax
\mciteBstWouldAddEndPuncttrue
\mciteSetBstMidEndSepPunct{\mcitedefaultmidpunct}
{\mcitedefaultendpunct}{\mcitedefaultseppunct}\relax
\EndOfBibitem
\bibitem[Koput(2015)]{Kop15}
Koput,~J. Ab Initio Spectroscopic Characterization of Borane, BH, in its
  $X^1\Sigma^+$ Electronic State. \emph{J. Comput. Chem.} \textbf{2015},
  \emph{36}, 2219--2227\relax
\mciteBstWouldAddEndPuncttrue
\mciteSetBstMidEndSepPunct{\mcitedefaultmidpunct}
{\mcitedefaultendpunct}{\mcitedefaultseppunct}\relax
\EndOfBibitem
\bibitem[Engin \latin{et~al.}(2012)Engin, Sisourat, and Carniato]{Eng12}
Engin,~S.; Sisourat,~N.; Carniato,~S. Ab initio Study of Low-Lying Excited
  States of HCl: Accurate Calculations of Optical Valence-Shell Excitations.
  \emph{J. Chem. Phys.} \textbf{2012}, \emph{137}, 154304\relax
\mciteBstWouldAddEndPuncttrue
\mciteSetBstMidEndSepPunct{\mcitedefaultmidpunct}
{\mcitedefaultendpunct}{\mcitedefaultseppunct}\relax
\EndOfBibitem
\bibitem[Thomson and Dalby(1969)Thomson, and Dalby]{Tho69}
Thomson,~R.; Dalby,~F.~W. An Experimental Determination of the Dipole Moments
  of the $X(^1\Sigma)$ and $A(^1\Pi)$ States of the BH Molecule. \emph{Can. J.
  Phys.} \textbf{1969}, \emph{47}, 1155--1158\relax
\mciteBstWouldAddEndPuncttrue
\mciteSetBstMidEndSepPunct{\mcitedefaultmidpunct}
{\mcitedefaultendpunct}{\mcitedefaultseppunct}\relax
\EndOfBibitem
\bibitem[Douglass \latin{et~al.}(1989)Douglass, Nelson, and Rice]{Dou89}
Douglass,~C.~H.; Nelson,~H.~H.; Rice,~J.~K. Spectra, Radiative Lifetimes, and
  Band Oscillator Strengths of the $A ^1\Pi - X ^1\Sigma^+$ Transition of BH.
  \emph{J. Chem. Phys.} \textbf{1989}, \emph{90}, 6940--6948\relax
\mciteBstWouldAddEndPuncttrue
\mciteSetBstMidEndSepPunct{\mcitedefaultmidpunct}
{\mcitedefaultendpunct}{\mcitedefaultseppunct}\relax
\EndOfBibitem
\bibitem[Dufayard and Nedelec(1978)Dufayard, and Nedelec]{Duf78}
Dufayard,~J.; Nedelec,~O. Lifetime of the BH $A^1\Pi$ State Excited by a Pulsed
  Dye Laser. \emph{J. Chem. Phys.} \textbf{1978}, \emph{69}, 4708--4709\relax
\mciteBstWouldAddEndPuncttrue
\mciteSetBstMidEndSepPunct{\mcitedefaultmidpunct}
{\mcitedefaultendpunct}{\mcitedefaultseppunct}\relax
\EndOfBibitem
\bibitem[Cheng \latin{et~al.}(2002)Cheng, Chung, Bahou, Lee, and Lee]{Che02}
Cheng,~B.-M.; Chung,~C.-Y.; Bahou,~M.; Lee,~Y.-P.; Lee,~L.~C. Quantitative
  Spectral Analysis of HCl and DCl in 120--220 nm: Effects of Singlet--Triplet
  Mixing. \emph{J. Chem. Phys.} \textbf{2002}, \emph{117}, 4293--4298\relax
\mciteBstWouldAddEndPuncttrue
\mciteSetBstMidEndSepPunct{\mcitedefaultmidpunct}
{\mcitedefaultendpunct}{\mcitedefaultseppunct}\relax
\EndOfBibitem
\bibitem[Li \latin{et~al.}(2006)Li, Zhu, Yuan, Liu, and Xu]{Li06c}
Li,~W.-B.; Zhu,~L.-F.; Yuan,~Z.-S.; Liu,~X.-J.; Xu,~K.-Z. Optical Oscillator
  Strengths for Valence-Shell and Br-$3d$ Inner-Shell Excitations of HCl and
  HBr. \emph{J. Chem. Phys.} \textbf{2006}, \emph{125}, 154310\relax
\mciteBstWouldAddEndPuncttrue
\mciteSetBstMidEndSepPunct{\mcitedefaultmidpunct}
{\mcitedefaultendpunct}{\mcitedefaultseppunct}\relax
\EndOfBibitem
\bibitem[Xu \latin{et~al.}(2019)Xu, Liu, Du, Xu, and Zhu]{Xu19}
Xu,~Y.-C.; Liu,~Y.-W.; Du,~X.-J.; Xu,~L.-Q.; Zhu,~L.-F. Oscillator Strengths
  and Integral Cross Sections of the Valence-Shell Excitations of HCl Studied
  by Fast Electron Scattering. \emph{Phys. Chem. Chem. Phys.} \textbf{2019},
  \emph{21}, 17433--17440\relax
\mciteBstWouldAddEndPuncttrue
\mciteSetBstMidEndSepPunct{\mcitedefaultmidpunct}
{\mcitedefaultendpunct}{\mcitedefaultseppunct}\relax
\EndOfBibitem
\bibitem[Lu(2003)]{Lu03}
Lu,~S.-I. Performance of Ornstein--Uhlenbeck Diffusion Quantum Monte Carlo for
  First-Row Diatomic Dissociation Energies and Dipole Moments. \emph{J. Chem.
  Phys.} \textbf{2003}, \emph{118}, 6152--6156\relax
\mciteBstWouldAddEndPuncttrue
\mciteSetBstMidEndSepPunct{\mcitedefaultmidpunct}
{\mcitedefaultendpunct}{\mcitedefaultseppunct}\relax
\EndOfBibitem
\bibitem[P{\'a}len{\'\i}kov{\'a} \latin{et~al.}(2008)P{\'a}len{\'\i}kov{\'a},
  Kraus, Neogr{\'a}dy, Kell{\"o}, and Urban]{Pal08}
P{\'a}len{\'\i}kov{\'a},~J.; Kraus,~M.; Neogr{\'a}dy,~P.; Kell{\"o},~V.;
  Urban,~M. Theoretical Study of Molecular Properties of Low-Lying Electronic
  Excited States of H$_2$O and H$_2$S. \emph{Mol. Phys.} \textbf{2008},
  \emph{106}, 2333--2344\relax
\mciteBstWouldAddEndPuncttrue
\mciteSetBstMidEndSepPunct{\mcitedefaultmidpunct}
{\mcitedefaultendpunct}{\mcitedefaultseppunct}\relax
\EndOfBibitem
\bibitem[Borges(2006)]{Bor06b}
Borges,~I. Configuration Interaction Oscillator Strengths of the H$_2$O
  Molecule: Transitions from the Ground to the $\tilde{B} ^1A_1, \tilde{C}
  ^1B_1, \tilde{D} ^1A_1$, and $1 ^1B_2$ Excited States. \emph{Chem. Phys.}
  \textbf{2006}, \emph{328}, 284--290\relax
\mciteBstWouldAddEndPuncttrue
\mciteSetBstMidEndSepPunct{\mcitedefaultmidpunct}
{\mcitedefaultendpunct}{\mcitedefaultseppunct}\relax
\EndOfBibitem
\bibitem[John \latin{et~al.}(1980)John, Bacskay, and Hush]{Joh80}
John,~I.; Bacskay,~G.; Hush,~N. Finite Field Method Calculations. Vi. Raman
  Scatering Activities, Infrared Absorption Intensities and Higher-Order
  Moments: SCF and CI Calculations for the Isotopic Derivatives of H$_2$O and
  SCF Calculations for CH$_4$. \emph{Chem. Phys.} \textbf{1980}, \emph{51},
  49--60\relax
\mciteBstWouldAddEndPuncttrue
\mciteSetBstMidEndSepPunct{\mcitedefaultmidpunct}
{\mcitedefaultendpunct}{\mcitedefaultseppunct}\relax
\EndOfBibitem
\bibitem[Ralphs \latin{et~al.}(2013)Ralphs, Serna, Hargreaves, Khakoo,
  Winstead, and McKoy]{Ral13}
Ralphs,~K.; Serna,~G.; Hargreaves,~L.~R.; Khakoo,~M.~A.; Winstead,~C.;
  McKoy,~V. Excitation of the Six Lowest Electronic Transitions in Water by
  9--20 eV Electrons. \emph{J. Phys. B} \textbf{2013}, \emph{46}, 125201\relax
\mciteBstWouldAddEndPuncttrue
\mciteSetBstMidEndSepPunct{\mcitedefaultmidpunct}
{\mcitedefaultendpunct}{\mcitedefaultseppunct}\relax
\EndOfBibitem
\bibitem[Thorn \latin{et~al.}(2007)Thorn, Brunger, Teubner, Diakomichalis,
  Maddern, Bolorizadeh, Newell, Kato, Hoshino, Tanaka, Cho, and Kim]{Tho07}
Thorn,~P.~A.; Brunger,~M.~J.; Teubner,~P. J.~O.; Diakomichalis,~N.;
  Maddern,~T.; Bolorizadeh,~M.~A.; Newell,~W.~R.; Kato,~H.; Hoshino,~M.;
  Tanaka,~H.; Cho,~H.; Kim,~Y.-K. Cross Sections and Oscillator Strengths for
  Electron-Impact Excitation of the $\tilde{A} ^1B_1$ Electronic State of
  Water. \emph{J. Chem. Phys.} \textbf{2007}, \emph{126}, 064306\relax
\mciteBstWouldAddEndPuncttrue
\mciteSetBstMidEndSepPunct{\mcitedefaultmidpunct}
{\mcitedefaultendpunct}{\mcitedefaultseppunct}\relax
\EndOfBibitem
\bibitem[Mota \latin{et~al.}(2005)Mota, Parafita, Giuliani, Hubin-Franskin,
  Louren{\c c}o, Garcia, Hoffmann, Mason, Ribeiro, Raposo, and
  Lim{\~a}o-Vieira]{Mot05}
Mota,~R.; Parafita,~R.; Giuliani,~A.; Hubin-Franskin,~M.-J.; Louren{\c c}o,~J.;
  Garcia,~G.; Hoffmann,~S.; Mason,~N.; Ribeiro,~P.; Raposo,~M.;
  Lim{\~a}o-Vieira,~P. Water VUV Electronic State Spectroscopy by Synchrotron
  Radiation. \emph{Chem. Phys. Lett.} \textbf{2005}, \emph{416}, 152--159\relax
\mciteBstWouldAddEndPuncttrue
\mciteSetBstMidEndSepPunct{\mcitedefaultmidpunct}
{\mcitedefaultendpunct}{\mcitedefaultseppunct}\relax
\EndOfBibitem
\bibitem[Ertan \latin{et~al.}(2020)Ertan, Lundberg, S{\o}rensen, and
  Odelius]{Ert20}
Ertan,~E.; Lundberg,~M.; S{\o}rensen,~L.~K.; Odelius,~M. Setting the Stage for
  Theoretical X-Ray Spectra of the H$_2$S Molecule with Multi-Configurational
  Quantum Chemical Calculations of the Energy Landscape. \emph{J. Chem. Phys.}
  \textbf{2020}, \emph{152}, 094305\relax
\mciteBstWouldAddEndPuncttrue
\mciteSetBstMidEndSepPunct{\mcitedefaultmidpunct}
{\mcitedefaultendpunct}{\mcitedefaultseppunct}\relax
\EndOfBibitem
\bibitem[Huiszoon and Dymanus(1965)Huiszoon, and Dymanus]{Hui65}
Huiszoon,~C.; Dymanus,~A. Stark Effect of Millimeter Wave Transitions: I.
  Hydrogen Sulfide. \emph{Physica} \textbf{1965}, \emph{31}, 1049--1052\relax
\mciteBstWouldAddEndPuncttrue
\mciteSetBstMidEndSepPunct{\mcitedefaultmidpunct}
{\mcitedefaultendpunct}{\mcitedefaultseppunct}\relax
\EndOfBibitem
\bibitem[Masuko \latin{et~al.}(1979)Masuko, Morioka, Nakamura, Ishiguro, and
  Sasanuma]{Mas79}
Masuko,~H.; Morioka,~Y.; Nakamura,~M.; Ishiguro,~E.; Sasanuma,~M. Absorption
  Spectrum of the H$_2$S Molecule In The Vacuum Ultraviolet Region. \emph{Can.
  J. Phys.} \textbf{1979}, \emph{57}, 745--760\relax
\mciteBstWouldAddEndPuncttrue
\mciteSetBstMidEndSepPunct{\mcitedefaultmidpunct}
{\mcitedefaultendpunct}{\mcitedefaultseppunct}\relax
\EndOfBibitem
\bibitem[Lee \latin{et~al.}(1987)Lee, Wang, and Suto]{Lee87}
Lee,~L.~C.; Wang,~X.; Suto,~M. Quantitative Photoabsorption and Fluorescence
  Spectroscopy of H$_2$S and D$_2$S at 49--240 nm. \emph{J. Chem. Phys.}
  \textbf{1987}, \emph{86}, 4353--4361\relax
\mciteBstWouldAddEndPuncttrue
\mciteSetBstMidEndSepPunct{\mcitedefaultmidpunct}
{\mcitedefaultendpunct}{\mcitedefaultseppunct}\relax
\EndOfBibitem
\bibitem[Feng \latin{et~al.}(1999)Feng, Cooper, and Brion]{Fen99}
Feng,~R.; Cooper,~G.; Brion,~C. Absolute Oscillator Strengths for Hydrogen
  Sulphide: I. Photoabsorption in the Valence-Shell and the $S 2p$ and $2s$
  Inner-Shell Regions (4--260 eV). \emph{Chem. Phys.} \textbf{1999},
  \emph{244}, 127--142\relax
\mciteBstWouldAddEndPuncttrue
\mciteSetBstMidEndSepPunct{\mcitedefaultmidpunct}
{\mcitedefaultendpunct}{\mcitedefaultseppunct}\relax
\EndOfBibitem
\bibitem[Pericou-Cayere \latin{et~al.}(1997)Pericou-Cayere, Gelize, and
  Dargelos]{Per97d}
Pericou-Cayere,~M.; Gelize,~M.; Dargelos,~A. Ab initio Calculations of
  Electronic Spectra of H$_2$S and H$_2$S$_2$. \emph{Chem. Phys.}
  \textbf{1997}, \emph{214}, 81--89\relax
\mciteBstWouldAddEndPuncttrue
\mciteSetBstMidEndSepPunct{\mcitedefaultmidpunct}
{\mcitedefaultendpunct}{\mcitedefaultseppunct}\relax
\EndOfBibitem
\bibitem[Rauk and Collins(1984)Rauk, and Collins]{Rau84}
Rauk,~A.; Collins,~S. The Ground and Excited States of Hydrogen Sulfide,
  Methanethiol, and Hydrogen Selenide. \emph{J. Mol. Spectrosc.} \textbf{1984},
  \emph{105}, 438--452\relax
\mciteBstWouldAddEndPuncttrue
\mciteSetBstMidEndSepPunct{\mcitedefaultmidpunct}
{\mcitedefaultendpunct}{\mcitedefaultseppunct}\relax
\EndOfBibitem
\bibitem[Lovas and Johnson(1971)Lovas, and Johnson]{Lov71}
Lovas,~F.~J.; Johnson,~D.~R. Microwave Spectrum of BF. \emph{J. Chem. Phys.}
  \textbf{1971}, \emph{55}, 41--44\relax
\mciteBstWouldAddEndPuncttrue
\mciteSetBstMidEndSepPunct{\mcitedefaultmidpunct}
{\mcitedefaultendpunct}{\mcitedefaultseppunct}\relax
\EndOfBibitem
\bibitem[Honigmann \latin{et~al.}(1993)Honigmann, Hirsch, and Buenker]{Hon93}
Honigmann,~M.; Hirsch,~G.; Buenker,~R.~J. Theoretical Study of the Optical and
  Generalized Oscillator Strengths for Transitions Between Low-Lying Electronic
  States of the BF Molecule. \emph{Chem. Phys.} \textbf{1993}, \emph{172},
  59--71\relax
\mciteBstWouldAddEndPuncttrue
\mciteSetBstMidEndSepPunct{\mcitedefaultmidpunct}
{\mcitedefaultendpunct}{\mcitedefaultseppunct}\relax
\EndOfBibitem
\bibitem[Magoulas \latin{et~al.}(2013)Magoulas, Kalemos, and Mavridis]{Mag13}
Magoulas,~I.; Kalemos,~A.; Mavridis,~A. An \emph{ab initio} Study of the
  Electronic Structure of BF and BF$^+$. \emph{J. Chem. Phys.} \textbf{2013},
  \emph{138}, 104312\relax
\mciteBstWouldAddEndPuncttrue
\mciteSetBstMidEndSepPunct{\mcitedefaultmidpunct}
{\mcitedefaultendpunct}{\mcitedefaultseppunct}\relax
\EndOfBibitem
\bibitem[M{\'e}rawa \latin{et~al.}(1997)M{\'e}rawa, B{\'e}gu{\'e}, R{\'e}rat,
  and Pouchan]{Mer97}
M{\'e}rawa,~M.; B{\'e}gu{\'e},~D.; R{\'e}rat,~M.; Pouchan,~C. Dynamic
  Polarizability and Hyperpolarizability for the 14 Electron Molecules CO and
  BF. \emph{Chem. Phys. Lett.} \textbf{1997}, \emph{280}, 203--211\relax
\mciteBstWouldAddEndPuncttrue
\mciteSetBstMidEndSepPunct{\mcitedefaultmidpunct}
{\mcitedefaultendpunct}{\mcitedefaultseppunct}\relax
\EndOfBibitem
\bibitem[Huber and Herzberg(1979)Huber, and Herzberg]{Hub79}
Huber,~K.~P.; Herzberg,~G. \emph{Constants of Diatomic Molecules}; Molecular
  Spectra and Molecular Structure; Van Nostrand: Princeton, 1979; Vol.~4\relax
\mciteBstWouldAddEndPuncttrue
\mciteSetBstMidEndSepPunct{\mcitedefaultmidpunct}
{\mcitedefaultendpunct}{\mcitedefaultseppunct}\relax
\EndOfBibitem
\bibitem[Coe \latin{et~al.}(2013)Coe, Taylor, and Paterson]{Coe13b}
Coe,~J.~P.; Taylor,~D.~J.; Paterson,~M.~J. Monte Carlo Configuration
  Interaction Applied to Multipole Moments, Ionization Energies, and Electron
  Affinities. \emph{J. Comput. Chem.} \textbf{2013}, \emph{34},
  1083--1093\relax
\mciteBstWouldAddEndPuncttrue
\mciteSetBstMidEndSepPunct{\mcitedefaultmidpunct}
{\mcitedefaultendpunct}{\mcitedefaultseppunct}\relax
\EndOfBibitem
\bibitem[Cooper and Kirby(1987)Cooper, and Kirby]{Coo87}
Cooper,~D.~L.; Kirby,~K. Theoretical Study of Low Lying $^1\Sigma+$ and $^1\Pi$
  States of CO. I. Potential Energy Curves and Dipole Moments. \emph{J. Chem.
  Phys.} \textbf{1987}, \emph{87}, 424--432\relax
\mciteBstWouldAddEndPuncttrue
\mciteSetBstMidEndSepPunct{\mcitedefaultmidpunct}
{\mcitedefaultendpunct}{\mcitedefaultseppunct}\relax
\EndOfBibitem
\bibitem[Chan \latin{et~al.}(1993)Chan, Cooper, and Brion]{Cha93b}
Chan,~W.; Cooper,~G.; Brion,~C. Absolute Optical Oscillator Strengths for
  Discrete and Continuum Photoabsorption of Carbon Monoxide (7--200 eV) and
  Transition Moments for the $X ^1\Sigma^+ \rightarrow A ^1\Pi$ System.
  \emph{Chem. Phys.} \textbf{1993}, \emph{170}, 123--138\relax
\mciteBstWouldAddEndPuncttrue
\mciteSetBstMidEndSepPunct{\mcitedefaultmidpunct}
{\mcitedefaultendpunct}{\mcitedefaultseppunct}\relax
\EndOfBibitem
\bibitem[Fisher and Dalby(1976)Fisher, and Dalby]{Fis76}
Fisher,~N.~J.; Dalby,~F.~W. On the Dipole Moments of Excited Singlet States of
  Carbon Monoxide. \emph{Can. J. Phys.} \textbf{1976}, \emph{54},
  258--261\relax
\mciteBstWouldAddEndPuncttrue
\mciteSetBstMidEndSepPunct{\mcitedefaultmidpunct}
{\mcitedefaultendpunct}{\mcitedefaultseppunct}\relax
\EndOfBibitem
\bibitem[Kang \latin{et~al.}(2015)Kang, Liu, Xu, Ni, Yang, Hiraoka, Tsuei, and
  Zhu]{Kan15}
Kang,~X.; Liu,~Y.~W.; Xu,~L.~Q.; Ni,~D.~D.; Yang,~K.; Hiraoka,~N.;
  Tsuei,~K.~D.; Zhu,~L.~F. Oscillator Strength Measurement for the
  $A(0-6)-X(0)$, $C(0)-X(0)$, and $E(0)-X(0)$ Transitions of CO by the Dipole
  ($\gamma$,$\gamma$) Method. \emph{Astrophys. J.} \textbf{2015}, \emph{807},
  96\relax
\mciteBstWouldAddEndPuncttrue
\mciteSetBstMidEndSepPunct{\mcitedefaultmidpunct}
{\mcitedefaultendpunct}{\mcitedefaultseppunct}\relax
\EndOfBibitem
\bibitem[Drabbels \latin{et~al.}(1993)Drabbels, Meerts, and ter Meulen]{Dra93}
Drabbels,~M.; Meerts,~W.~L.; ter Meulen,~J.~J. Determination of Electric Dipole
  Moments and Transition Probabilities of Low-Lying Singlet States of CO.
  \emph{J. Chem. Phys.} \textbf{1993}, \emph{99}, 2352--2358\relax
\mciteBstWouldAddEndPuncttrue
\mciteSetBstMidEndSepPunct{\mcitedefaultmidpunct}
{\mcitedefaultendpunct}{\mcitedefaultseppunct}\relax
\EndOfBibitem
\bibitem[Scuseria \latin{et~al.}(1991)Scuseria, Miller, Jensen, and
  Geertsen]{Scu91}
Scuseria,~G.~E.; Miller,~M.~D.; Jensen,~F.; Geertsen,~J. The Dipole Moment of
  Carbon Monoxide. \emph{J. Chem. Phys.} \textbf{1991}, \emph{94},
  6660--6663\relax
\mciteBstWouldAddEndPuncttrue
\mciteSetBstMidEndSepPunct{\mcitedefaultmidpunct}
{\mcitedefaultendpunct}{\mcitedefaultseppunct}\relax
\EndOfBibitem
\bibitem[Nielsen \latin{et~al.}(1980)Nielsen, Jorgensen, and Oddershede]{Nie80}
Nielsen,~E.~S.; Jorgensen,~P.; Oddershede,~J. \emph{J. Chem. Phys.}
  \textbf{1980}, \emph{22}, 1539--1548\relax
\mciteBstWouldAddEndPuncttrue
\mciteSetBstMidEndSepPunct{\mcitedefaultmidpunct}
{\mcitedefaultendpunct}{\mcitedefaultseppunct}\relax
\EndOfBibitem
\bibitem[Carlson \latin{et~al.}(1978)Carlson, Đuri{\'c}, Erman, and
  Larsson]{Car78}
Carlson,~T.~A.; Đuri{\'c},~N.; Erman,~P.; Larsson,~M. Correlation Between
  Perturbation and Collisional Transfers In The $A$, $B$, $C$ and $b$ States of
  CO as Revealed by High Resolution Lifetime Measurements. \emph{Z. Phys. A}
  \textbf{1978}, \emph{287}, 123--136\relax
\mciteBstWouldAddEndPuncttrue
\mciteSetBstMidEndSepPunct{\mcitedefaultmidpunct}
{\mcitedefaultendpunct}{\mcitedefaultseppunct}\relax
\EndOfBibitem
\bibitem[Rocha \latin{et~al.}(1998)Rocha, Borges, and Bielschowsky]{Roc98}
Rocha,~A.~B.; Borges,~I.; Bielschowsky,~C.~E. Optical and Generalized
  Oscillator Strengths for the ${B}^{1}{\ensuremath{\Sigma}}^{+}$,
  ${C}^{1}{\ensuremath{\Sigma}}^{+}$, and ${E}^{1}\ensuremath{\Pi}$ Vibronic
  Bands in the CO Molecule. \emph{Phys. Rev. A} \textbf{1998}, \emph{57},
  4394--4400\relax
\mciteBstWouldAddEndPuncttrue
\mciteSetBstMidEndSepPunct{\mcitedefaultmidpunct}
{\mcitedefaultendpunct}{\mcitedefaultseppunct}\relax
\EndOfBibitem
\bibitem[Oddershede \latin{et~al.}(1985)Oddershede, Gr{\=u}ner, and
  Diercksen]{Odd85}
Oddershede,~J.; Gr{\=u}ner,~N.~E.; Diercksen,~G.~H. Comparison Between Equation
  of Motion and Polarization Propagator Calculations. \emph{Chem. Phys.}
  \textbf{1985}, \emph{97}, 303--310\relax
\mciteBstWouldAddEndPuncttrue
\mciteSetBstMidEndSepPunct{\mcitedefaultmidpunct}
{\mcitedefaultendpunct}{\mcitedefaultseppunct}\relax
\EndOfBibitem
\bibitem[Neugebauer \latin{et~al.}(2004)Neugebauer, Baerends, and
  Nooijen]{Neu04}
Neugebauer,~J.; Baerends,~E.~J.; Nooijen,~M. Vibronic Coupling and Double
  Excitations in Linear Response Time-Dependent Density Functional
  Calculations: Dipole-Allowed States of N$_2$. \emph{J. Chem. Phys.}
  \textbf{2004}, \emph{121}, 6155--6166\relax
\mciteBstWouldAddEndPuncttrue
\mciteSetBstMidEndSepPunct{\mcitedefaultmidpunct}
{\mcitedefaultendpunct}{\mcitedefaultseppunct}\relax
\EndOfBibitem
\bibitem[Chan \latin{et~al.}(1993)Chan, Cooper, Sodhi, and Brion]{Cha93}
Chan,~W.; Cooper,~G.; Sodhi,~R.; Brion,~C. Absolute Optical Oscillator
  Strengths for Discrete and Continuum Photoabsorption of Molecular Nitrogen
  (11--200 eV). \emph{Chem. Phys.} \textbf{1993}, \emph{170}, 81--97\relax
\mciteBstWouldAddEndPuncttrue
\mciteSetBstMidEndSepPunct{\mcitedefaultmidpunct}
{\mcitedefaultendpunct}{\mcitedefaultseppunct}\relax
\EndOfBibitem
\bibitem[Ben-Shlomo and Kaldor(1990)Ben-Shlomo, and Kaldor]{Ben90}
Ben-Shlomo,~S.~B.; Kaldor,~U. N$_2$ Excitations Below 15 eV by the
  Multireference Coupled-Cluster Method. \emph{J. Chem. Phys.} \textbf{1990},
  \emph{92}, 3680--3682\relax
\mciteBstWouldAddEndPuncttrue
\mciteSetBstMidEndSepPunct{\mcitedefaultmidpunct}
{\mcitedefaultendpunct}{\mcitedefaultseppunct}\relax
\EndOfBibitem
\bibitem[Ajello \latin{et~al.}(1989)Ajello, James, Franklin, and
  Shemansky]{Aje89}
Ajello,~J.~M.; James,~G.~K.; Franklin,~B.~O.; Shemansky,~D.~E.
  Medium-Resolution Studies of Extreme Ultraviolet Emission from
  ${\mathrm{N}}_{2}$ by Electron Impact: Vibrational Perturbations and Cross
  Sections of the ${c}_{4}^{\mathcal{'}}$ $^{1}\ensuremath{\Sigma}_{u}^{+}$ and
  b' $^{1}\ensuremath{\Sigma}_{u}^{+}$ States. \emph{Phys. Rev. A}
  \textbf{1989}, \emph{40}, 3524--3556\relax
\mciteBstWouldAddEndPuncttrue
\mciteSetBstMidEndSepPunct{\mcitedefaultmidpunct}
{\mcitedefaultendpunct}{\mcitedefaultseppunct}\relax
\EndOfBibitem
\bibitem[zzz()]{zzz-mountain}
The first, second, and third $\Pi_u$ transitions have as largest MO
  contribution, $A_g \rightarrow B_{2u}$, $B_{1u} \rightarrow B_{3g}$, and
  $B_{2u} \rightarrow A_{g}$ excitation, respectively, within the commonly used
  $D_{2h}$ point group. There is nevertheless significant mixing, and we used
  the largest coefficient to discriminate the various $\Pi_u$.\relax
\mciteBstWouldAddEndPunctfalse
\mciteSetBstMidEndSepPunct{\mcitedefaultmidpunct}
{}{\mcitedefaultseppunct}\relax
\EndOfBibitem
\bibitem[Liu \latin{et~al.}(2016)Liu, Kang, Xu, Ni, Yang, Hiraoka, Tsuei, and
  Zhu]{Liu16c}
Liu,~Y.-W.; Kang,~X.; Xu,~L.-Q.; Ni,~D.-D.; Yang,~K.; Hiraoka,~N.;
  Tsuei,~K.-D.; Zhu,~L.-F. Oscillator Strengths of Vibronic Excitations of
  Nitrogen Determined by the Dipole ($\gamma$,$\gamma$) Method.
  \emph{Astrophys. J.} \textbf{2016}, \emph{819}, 142\relax
\mciteBstWouldAddEndPuncttrue
\mciteSetBstMidEndSepPunct{\mcitedefaultmidpunct}
{\mcitedefaultendpunct}{\mcitedefaultseppunct}\relax
\EndOfBibitem
\bibitem[Lavin \latin{et~al.}(2004)Lavin, Velasco, Martin, and Bustos]{Lav04}
Lavin,~C.; Velasco,~A.; Martin,~I.; Bustos,~E. MQDO Oscillator Strengths and
  Emission Coefficients for Electronic Transitions in N$_2$ and NO. \emph{Chem.
  Phys. Lett.} \textbf{2004}, \emph{394}, 114--119\relax
\mciteBstWouldAddEndPuncttrue
\mciteSetBstMidEndSepPunct{\mcitedefaultmidpunct}
{\mcitedefaultendpunct}{\mcitedefaultseppunct}\relax
\EndOfBibitem
\bibitem[Robin(1985)]{Rob85b}
Robin,~M.~B. In \emph{Higher Excited States of Polyatomic Molecules};
  Robin,~M.~B., Ed.; Academic Press, 1985; Vol. III\relax
\mciteBstWouldAddEndPuncttrue
\mciteSetBstMidEndSepPunct{\mcitedefaultmidpunct}
{\mcitedefaultendpunct}{\mcitedefaultseppunct}\relax
\EndOfBibitem
\bibitem[Serrano-Andr\'es \latin{et~al.}(1993)Serrano-Andr\'es, Mech\'an,
  Nebot-Gil, Lindh, and Roos]{Ser93}
Serrano-Andr\'es,~L.; Mech\'an,~M.; Nebot-Gil,~I.; Lindh,~R.; Roos,~B.~O.
  Towards an Accurate Molecular Orbital Theory for Excited States: Ethene,
  Butadiene, and Hexatriene. \emph{J. Chem. Phys.} \textbf{1993}, \emph{98},
  3151--3162\relax
\mciteBstWouldAddEndPuncttrue
\mciteSetBstMidEndSepPunct{\mcitedefaultmidpunct}
{\mcitedefaultendpunct}{\mcitedefaultseppunct}\relax
\EndOfBibitem
\bibitem[Merer and Mulliken(1969)Merer, and Mulliken]{Mer69}
Merer,~A.~J.; Mulliken,~R.~S. Ultraviolet Spectra and Excited States of
  Ethylene and its Alkyl Derivatives. \emph{Chem. Rev.} \textbf{1969},
  \emph{69}, 639--656\relax
\mciteBstWouldAddEndPuncttrue
\mciteSetBstMidEndSepPunct{\mcitedefaultmidpunct}
{\mcitedefaultendpunct}{\mcitedefaultseppunct}\relax
\EndOfBibitem
\bibitem[Zelikoff and Watanabe(1953)Zelikoff, and Watanabe]{Zel53}
Zelikoff,~M.; Watanabe,~K. Absorption Coefficients of Ethylene in the Vacuum
  Ultraviolet. \emph{J. Opt. Soc. Am.} \textbf{1953}, \emph{43}, 756--759\relax
\mciteBstWouldAddEndPuncttrue
\mciteSetBstMidEndSepPunct{\mcitedefaultmidpunct}
{\mcitedefaultendpunct}{\mcitedefaultseppunct}\relax
\EndOfBibitem
\bibitem[Watts \latin{et~al.}(1996)Watts, Gwaltney, and Bartlett]{Wat96b}
Watts,~J.~D.; Gwaltney,~S.~R.; Bartlett,~R.~J. Coupled-Cluster Calculations of
  the Excitation Energies of Ethylene, Butadiene, and Cyclopentadiene. \emph{J.
  Chem. Phys.} \textbf{1996}, \emph{105}, 6979--6988\relax
\mciteBstWouldAddEndPuncttrue
\mciteSetBstMidEndSepPunct{\mcitedefaultmidpunct}
{\mcitedefaultendpunct}{\mcitedefaultseppunct}\relax
\EndOfBibitem
\bibitem[Cooper \latin{et~al.}(1995)Cooper, Olney, and Brion]{Coo95}
Cooper,~G.; Olney,~T.~N.; Brion,~C. Absolute UV and Soft X-ray Photoabsorption
  of Ethylene by High Resolution Dipole (e,e) Spectroscopy. \emph{Chem. Phys.}
  \textbf{1995}, \emph{194}, 175--184\relax
\mciteBstWouldAddEndPuncttrue
\mciteSetBstMidEndSepPunct{\mcitedefaultmidpunct}
{\mcitedefaultendpunct}{\mcitedefaultseppunct}\relax
\EndOfBibitem
\bibitem[Hammond and Price(1955)Hammond, and Price]{Ham55}
Hammond,~V.~J.; Price,~W.~C. Oscillator Strengths of the Vacuum Ultra-Violet
  Absorption Bands of Benzene and Ethylene. \emph{Trans. Faraday Soc.}
  \textbf{1955}, \emph{51}, 605--610\relax
\mciteBstWouldAddEndPuncttrue
\mciteSetBstMidEndSepPunct{\mcitedefaultmidpunct}
{\mcitedefaultendpunct}{\mcitedefaultseppunct}\relax
\EndOfBibitem
\bibitem[Foresman \latin{et~al.}(1992)Foresman, Head-Gordon, Pople, and
  Frisch]{For92b}
Foresman,~J.~B.; Head-Gordon,~M.; Pople,~J.~A.; Frisch,~M.~J. Toward a
  Systematic Molecular Orbital Theory for Excited States. \emph{J. Phys. Chem.}
  \textbf{1992}, \emph{96}, 135--149\relax
\mciteBstWouldAddEndPuncttrue
\mciteSetBstMidEndSepPunct{\mcitedefaultmidpunct}
{\mcitedefaultendpunct}{\mcitedefaultseppunct}\relax
\EndOfBibitem
\bibitem[Hadad \latin{et~al.}(1993)Hadad, Foresman, and Wiberg]{Had93}
Hadad,~C.~M.; Foresman,~J.~B.; Wiberg,~K.~B. Excited States of Carbonyl
  Compounds. 1. Formaldehyde and Acetaldehyde. \emph{J. Phys. Chem.}
  \textbf{1993}, \emph{97}, 4293--4312\relax
\mciteBstWouldAddEndPuncttrue
\mciteSetBstMidEndSepPunct{\mcitedefaultmidpunct}
{\mcitedefaultendpunct}{\mcitedefaultseppunct}\relax
\EndOfBibitem
\bibitem[Head-Gordon \latin{et~al.}(1994)Head-Gordon, Rico, Oumi, and
  Lee]{Hea94}
Head-Gordon,~M.; Rico,~R.~J.; Oumi,~M.; Lee,~T.~J. A Doubles Correction to
  Electronic Excited States From Configuration Interaction in the Space of
  Single Substitutions. \emph{Chem. Phys. Lett.} \textbf{1994}, \emph{219},
  21--29\relax
\mciteBstWouldAddEndPuncttrue
\mciteSetBstMidEndSepPunct{\mcitedefaultmidpunct}
{\mcitedefaultendpunct}{\mcitedefaultseppunct}\relax
\EndOfBibitem
\bibitem[Head-Gordon \latin{et~al.}(1995)Head-Gordon, Maurice, and Oumi]{Hea95}
Head-Gordon,~M.; Maurice,~D.; Oumi,~M. A Perturbative Correction to Restricted
  Open-Shell Configuration-Interaction with Single Substitutions for
  Excited-States of Radicals. \emph{Chem. Phys. Lett.} \textbf{1995},
  \emph{246}, 114--121\relax
\mciteBstWouldAddEndPuncttrue
\mciteSetBstMidEndSepPunct{\mcitedefaultmidpunct}
{\mcitedefaultendpunct}{\mcitedefaultseppunct}\relax
\EndOfBibitem
\bibitem[Gwaltney and Bartlett(1995)Gwaltney, and Bartlett]{Gwa95}
Gwaltney,~S.~R.; Bartlett,~R.~J. An Application of the Equation-Of-Motion
  Coupled Cluster Method to the Excited States of Formaldehyde, Acetaldehyde,
  and Acetone. \emph{Chem. Phys. Lett.} \textbf{1995}, \emph{241}, 26--32\relax
\mciteBstWouldAddEndPuncttrue
\mciteSetBstMidEndSepPunct{\mcitedefaultmidpunct}
{\mcitedefaultendpunct}{\mcitedefaultseppunct}\relax
\EndOfBibitem
\bibitem[Wiberg \latin{et~al.}(1998)Wiberg, Stratmann, and Frisch]{Wib98}
Wiberg,~K.~B.; Stratmann,~R.~E.; Frisch,~M.~J. A Time-Dependent Density
  Functional Theory Study of the Electronically Excited States of Formaldehyde,
  Acetaldehyde and Acetone. \emph{Chem. Phys. Lett.} \textbf{1998}, \emph{297},
  60--64\relax
\mciteBstWouldAddEndPuncttrue
\mciteSetBstMidEndSepPunct{\mcitedefaultmidpunct}
{\mcitedefaultendpunct}{\mcitedefaultseppunct}\relax
\EndOfBibitem
\bibitem[Wilberg \latin{et~al.}(2002)Wilberg, de~Oliveria, and Trucks]{Wib02}
Wilberg,~K.~B.; de~Oliveria,~A.~E.; Trucks,~G. A Comparison of the Electronic
  Transition Energies for Ethene, Isobutene, Formaldehyde, and Acetone
  Calculated Using RPA, TDDFT, and EOM-CCSD. Effect of Basis Sets. \emph{J.
  Phys. Chem. A} \textbf{2002}, \emph{106}, 4192--4199\relax
\mciteBstWouldAddEndPuncttrue
\mciteSetBstMidEndSepPunct{\mcitedefaultmidpunct}
{\mcitedefaultendpunct}{\mcitedefaultseppunct}\relax
\EndOfBibitem
\bibitem[Peach \latin{et~al.}(2008)Peach, Benfield, Helgaker, and Tozer]{Pea08}
Peach,~M. J.~G.; Benfield,~P.; Helgaker,~T.; Tozer,~D.~J. Excitation Energies
  in Density Functional Theory: an Evaluation and a Diagnostic Test. \emph{J.
  Chem. Phys.} \textbf{2008}, \emph{128}, 044118\relax
\mciteBstWouldAddEndPuncttrue
\mciteSetBstMidEndSepPunct{\mcitedefaultmidpunct}
{\mcitedefaultendpunct}{\mcitedefaultseppunct}\relax
\EndOfBibitem
\bibitem[Shen and Li(2009)Shen, and Li]{She09b}
Shen,~J.; Li,~S. Block Correlated Coupled Cluster Method with the Complete
  Active-Space Self-Consistent-Field Reference Function: Applications for
  Low-Lying Electronic Excited States. \emph{J. Chem. Phys.} \textbf{2009},
  \emph{131}, 174101\relax
\mciteBstWouldAddEndPuncttrue
\mciteSetBstMidEndSepPunct{\mcitedefaultmidpunct}
{\mcitedefaultendpunct}{\mcitedefaultseppunct}\relax
\EndOfBibitem
\bibitem[Caricato \latin{et~al.}(2010)Caricato, Trucks, Frisch, and
  Wiberg]{Car10}
Caricato,~M.; Trucks,~G.~W.; Frisch,~M.~J.; Wiberg,~K.~B. Electronic Transition
  Energies: A Study of the Performance of a Large Range of Single Reference
  Density Functional and Wave Function Methods on Valence and Rydberg States
  Compared to Experiment. \emph{J. Chem. Theory Comput.} \textbf{2010},
  \emph{6}, 370--383\relax
\mciteBstWouldAddEndPuncttrue
\mciteSetBstMidEndSepPunct{\mcitedefaultmidpunct}
{\mcitedefaultendpunct}{\mcitedefaultseppunct}\relax
\EndOfBibitem
\bibitem[Li and Paldus(2011)Li, and Paldus]{Li11}
Li,~X.; Paldus,~J. Multi-Reference State-Universal Coupled-Cluster Approaches
  to Electronically Excited States. \emph{J. Chem. Phys.} \textbf{2011},
  \emph{134}, 214118\relax
\mciteBstWouldAddEndPuncttrue
\mciteSetBstMidEndSepPunct{\mcitedefaultmidpunct}
{\mcitedefaultendpunct}{\mcitedefaultseppunct}\relax
\EndOfBibitem
\bibitem[Leang \latin{et~al.}(2012)Leang, Zahariev, and Gordon]{Lea12}
Leang,~S.~S.; Zahariev,~F.; Gordon,~M.~S. Benchmarking the Performance of
  Time-Dependent Density Functional Methods. \emph{J. Chem. Phys.}
  \textbf{2012}, \emph{136}, 104101\relax
\mciteBstWouldAddEndPuncttrue
\mciteSetBstMidEndSepPunct{\mcitedefaultmidpunct}
{\mcitedefaultendpunct}{\mcitedefaultseppunct}\relax
\EndOfBibitem
\bibitem[Hoyer \latin{et~al.}(2016)Hoyer, Ghosh, Truhlar, and Gagliardi]{Hoy16}
Hoyer,~C.~E.; Ghosh,~S.; Truhlar,~D.~G.; Gagliardi,~L. Multiconfiguration
  Pair-Density Functional Theory Is as Accurate as CASPT2 for Electronic
  Excitation. \emph{J. Phys. Chem. Lett.} \textbf{2016}, \emph{7},
  586--591\relax
\mciteBstWouldAddEndPuncttrue
\mciteSetBstMidEndSepPunct{\mcitedefaultmidpunct}
{\mcitedefaultendpunct}{\mcitedefaultseppunct}\relax
\EndOfBibitem
\bibitem[K{\'a}nn{\'a}r \latin{et~al.}(2017)K{\'a}nn{\'a}r, Tajti, and
  Szalay]{Kan17}
K{\'a}nn{\'a}r,~D.; Tajti,~A.; Szalay,~P.~G. Accuracy of Coupled Cluster
  Excitation Energies in Diffuse Basis Sets. \emph{J. Chem. Theory Comput.}
  \textbf{2017}, \emph{13}, 202--209\relax
\mciteBstWouldAddEndPuncttrue
\mciteSetBstMidEndSepPunct{\mcitedefaultmidpunct}
{\mcitedefaultendpunct}{\mcitedefaultseppunct}\relax
\EndOfBibitem
\bibitem[M{\"u}ller and Lischka(2001)M{\"u}ller, and Lischka]{Mul01}
M{\"u}ller,~T.; Lischka,~H. Simultaneous Calculation of Rydberg and Valence
  Excited States of Formaldehyde. \emph{Theor. Chem. Acc.} \textbf{2001},
  \emph{106}, 369--378\relax
\mciteBstWouldAddEndPuncttrue
\mciteSetBstMidEndSepPunct{\mcitedefaultmidpunct}
{\mcitedefaultendpunct}{\mcitedefaultseppunct}\relax
\EndOfBibitem
\bibitem[Hachey \latin{et~al.}(1994)Hachey, Bruna, and Grein]{Hac94}
Hachey,~M.; Bruna,~P.~J.; Grein,~F. Configuration Interaction Studies on the
  $S_2$ Surface of H$_2$CO: $2 ^1A'(\sigma,\pi^\star/\pi,\pi^\star)$ as
  Perturber of $1 ^1B_2(n,3s)$. \emph{J. Chem. Soc.{,} Faraday Trans.}
  \textbf{1994}, \emph{90}, 683--688\relax
\mciteBstWouldAddEndPuncttrue
\mciteSetBstMidEndSepPunct{\mcitedefaultmidpunct}
{\mcitedefaultendpunct}{\mcitedefaultseppunct}\relax
\EndOfBibitem
\bibitem[G{\'o}mez-Carrasco \latin{et~al.}(2010)G{\'o}mez-Carrasco, M{\"u}ller,
  and K\"oppel]{Gom10b}
G{\'o}mez-Carrasco,~S.; M{\"u}ller,~T.; K\"oppel,~H. Ab Initio Study of the
  VUV-Induced Multistate Photodynamics of Formaldehyde. \emph{J. Phys. Chem. A}
  \textbf{2010}, \emph{114}, 11436--11449\relax
\mciteBstWouldAddEndPuncttrue
\mciteSetBstMidEndSepPunct{\mcitedefaultmidpunct}
{\mcitedefaultendpunct}{\mcitedefaultseppunct}\relax
\EndOfBibitem
\bibitem[Fabricant \latin{et~al.}(1977)Fabricant, Krieger, and Muenter]{Fab77}
Fabricant,~B.; Krieger,~D.; Muenter,~J.~S. Molecular Beam Electric Resonance
  Study of Formaldehyde, Thioformaldehyde, and Ketene. \emph{J. Chem. Phys.}
  \textbf{1977}, \emph{67}, 1576--1586\relax
\mciteBstWouldAddEndPuncttrue
\mciteSetBstMidEndSepPunct{\mcitedefaultmidpunct}
{\mcitedefaultendpunct}{\mcitedefaultseppunct}\relax
\EndOfBibitem
\bibitem[Haner and Dows(1968)Haner, and Dows]{Han68b}
Haner,~D.~A.; Dows,~D.~A. Electric-Field-Induced Spectra: Excited-State Dipole
  Moment from Line-Shape Analysis. \emph{J. Chem. Phys.} \textbf{1968},
  \emph{49}, 601--605\relax
\mciteBstWouldAddEndPuncttrue
\mciteSetBstMidEndSepPunct{\mcitedefaultmidpunct}
{\mcitedefaultendpunct}{\mcitedefaultseppunct}\relax
\EndOfBibitem
\bibitem[Weiss \latin{et~al.}(1971)Weiss, Kuyatt, and Mielczarek]{Wei71}
Weiss,~M.~J.; Kuyatt,~C.~E.; Mielczarek,~S. Inelastic Electron Scattering from
  Formaldehyde. \emph{J. Chem. Phys.} \textbf{1971}, \emph{54},
  4147--4150\relax
\mciteBstWouldAddEndPuncttrue
\mciteSetBstMidEndSepPunct{\mcitedefaultmidpunct}
{\mcitedefaultendpunct}{\mcitedefaultseppunct}\relax
\EndOfBibitem
\bibitem[Causley and Russell(1978)Causley, and Russell]{Cau78}
Causley,~G.~C.; Russell,~B.~R. Electric Dipole Moment and Polarizability of the
  1749 \AA, $^1B_2$ Excited State of Formaldehyde. \emph{J. Chem. Phys.}
  \textbf{1978}, \emph{68}, 3797--3800\relax
\mciteBstWouldAddEndPuncttrue
\mciteSetBstMidEndSepPunct{\mcitedefaultmidpunct}
{\mcitedefaultendpunct}{\mcitedefaultseppunct}\relax
\EndOfBibitem
\bibitem[Causley and Russell(1979)Causley, and Russell]{Cau79}
Causley,~G.~C.; Russell,~B.~R. Electric Dichroism Spectroscopy in the Vacuum
  Ultraviolet. 2. Formaldehyde, Acetaldehyde, adn Acetone. \emph{J. Am. Chem.
  Soc.} \textbf{1979}, \emph{101}, 5573--5578\relax
\mciteBstWouldAddEndPuncttrue
\mciteSetBstMidEndSepPunct{\mcitedefaultmidpunct}
{\mcitedefaultendpunct}{\mcitedefaultseppunct}\relax
\EndOfBibitem
\bibitem[Vaccaro \latin{et~al.}(1989)Vaccaro, Zabludoff, Carrera-Pati{\~n}o,
  Kinsey, and Field]{Vac89}
Vaccaro,~P.~H.; Zabludoff,~A.; Carrera-Pati{\~n}o,~M.~E.; Kinsey,~J.~L.;
  Field,~R.~W. High Precision Dipole Moments in $\tilde{A}$ $^1A_2$
  Formaldehyde Determined via Stark Quantum Beat Spectroscopy. \emph{J. Chem.
  Phys.} \textbf{1989}, \emph{90}, 4150--4167\relax
\mciteBstWouldAddEndPuncttrue
\mciteSetBstMidEndSepPunct{\mcitedefaultmidpunct}
{\mcitedefaultendpunct}{\mcitedefaultseppunct}\relax
\EndOfBibitem
\bibitem[Cooper \latin{et~al.}(1996)Cooper, Anderson, and Brion]{Coo96}
Cooper,~G.; Anderson,~J.~E.; Brion,~C. Absolute Photoabsorption and
  Photoionization of Formaldehyde in the VUV and Soft X-ray Regions (3--200
  eV). \emph{Chem. Phys.} \textbf{1996}, \emph{209}, 61--77\relax
\mciteBstWouldAddEndPuncttrue
\mciteSetBstMidEndSepPunct{\mcitedefaultmidpunct}
{\mcitedefaultendpunct}{\mcitedefaultseppunct}\relax
\EndOfBibitem
\bibitem[Benkov{\'a} \latin{et~al.}(2007)Benkov{\'a}, {\v C}ernu{\v s}{\'a}k,
  and Zahradn{\'\i}k]{Ben07b}
Benkov{\'a},~Z.; {\v C}ernu{\v s}{\'a}k,~I.; Zahradn{\'\i}k,~P. Electric
  Properties of Formaldehyde, Thioformaldehyde, Urea, Formamide, and
  Thioformamide -- Post-HF and DFT Study. \emph{Int. J. Quantum Chem.}
  \textbf{2007}, \emph{107}, 2133--2152\relax
\mciteBstWouldAddEndPuncttrue
\mciteSetBstMidEndSepPunct{\mcitedefaultmidpunct}
{\mcitedefaultendpunct}{\mcitedefaultseppunct}\relax
\EndOfBibitem
\bibitem[Clouthier and Ramsay(1983)Clouthier, and Ramsay]{Clo83}
Clouthier,~D.~J.; Ramsay,~D.~A. The Spectroscopy of Formaldehyde and
  Thioformaldehyde. \emph{Annu. Rev. Phys. Chem.} \textbf{1983}, \emph{34},
  31--58\relax
\mciteBstWouldAddEndPuncttrue
\mciteSetBstMidEndSepPunct{\mcitedefaultmidpunct}
{\mcitedefaultendpunct}{\mcitedefaultseppunct}\relax
\EndOfBibitem
\bibitem[Suzuki \latin{et~al.}(1985)Suzuki, Saito, and Hirota]{Suz85}
Suzuki,~T.; Saito,~S.; Hirota,~E. Dipole Moments of H$_2$CS in the
  $\tilde{A}^1A_2 (\nu=0)$ and $\tilde{a}^3A_2 (\nu_3=1)$ States by MODR
  Spectroscopy. \emph{J. Mol. Spectrosc.} \textbf{1985}, \emph{111},
  54--61\relax
\mciteBstWouldAddEndPuncttrue
\mciteSetBstMidEndSepPunct{\mcitedefaultmidpunct}
{\mcitedefaultendpunct}{\mcitedefaultseppunct}\relax
\EndOfBibitem
\bibitem[Goetz \latin{et~al.}(1981)Goetz, Moule, and Ramsay]{Goe81}
Goetz,~W.; Moule,~D.~C.; Ramsay,~D.~A. The Electric Dipole Moment of the
  $\tilde{C}^1B_2 (3s \leftarrow n)$ State of Thioformaldehyde. \emph{Can. J.
  Phys.} \textbf{1981}, \emph{59}, 1635--1639\relax
\mciteBstWouldAddEndPuncttrue
\mciteSetBstMidEndSepPunct{\mcitedefaultmidpunct}
{\mcitedefaultendpunct}{\mcitedefaultseppunct}\relax
\EndOfBibitem
\bibitem[Johnson \latin{et~al.}(1971)Johnson, Powell, and Kirchhoff]{Joh71}
Johnson,~D.~R.; Powell,~F.~X.; Kirchhoff,~W.~H. Microwave Spectrum, Ground
  State Structure, and Fipole Moment of Thioformaldehyde. \emph{J. Mol.
  Spectrosc.} \textbf{1971}, \emph{39}, 136--145\relax
\mciteBstWouldAddEndPuncttrue
\mciteSetBstMidEndSepPunct{\mcitedefaultmidpunct}
{\mcitedefaultendpunct}{\mcitedefaultseppunct}\relax
\EndOfBibitem
\bibitem[Fung and Ramsay(1996)Fung, and Ramsay]{Fun96}
Fung,~K.~H.; Ramsay,~D.~A. Doppler-free Stark Spectra of the $4_0^1$ Band of
  the $\tilde{A}^1A_2 - \tilde{X}^1A_1$ System of Thioformaldehyde. \emph{Mol.
  Phys.} \textbf{1996}, \emph{88}, 997--1004\relax
\mciteBstWouldAddEndPuncttrue
\mciteSetBstMidEndSepPunct{\mcitedefaultmidpunct}
{\mcitedefaultendpunct}{\mcitedefaultseppunct}\relax
\EndOfBibitem
\bibitem[Freeman and Klemperer(1964)Freeman, and Klemperer]{Fre64}
Freeman,~D.~E.; Klemperer,~W. Dipole Moments of Excited Electronic States of
  Molecules: The $^1A_2$ State of Formaldehyde. \emph{J. Chem. Phys.}
  \textbf{1964}, \emph{40}, 604--605\relax
\mciteBstWouldAddEndPuncttrue
\mciteSetBstMidEndSepPunct{\mcitedefaultmidpunct}
{\mcitedefaultendpunct}{\mcitedefaultseppunct}\relax
\EndOfBibitem
\bibitem[Freeman and Klemperer(1966)Freeman, and Klemperer]{Fre66}
Freeman,~D.~E.; Klemperer,~W. Electric Dipole Moment of the $^1A_2$ Electronic
  State of Formaldehyde. \emph{J. Chem. Phys.} \textbf{1966}, \emph{45},
  52--57\relax
\mciteBstWouldAddEndPuncttrue
\mciteSetBstMidEndSepPunct{\mcitedefaultmidpunct}
{\mcitedefaultendpunct}{\mcitedefaultseppunct}\relax
\EndOfBibitem
\bibitem[Bridge \latin{et~al.}(1968)Bridge, Haner, and Dows]{Bri68b}
Bridge,~N.~J.; Haner,~D.~A.; Dows,~D.~A. Electric-Field-Induced Spectra: Theory
  and Experimental Study of Formaldehyde. \emph{J. Chem. Phys.} \textbf{1968},
  \emph{48}, 4196--4210\relax
\mciteBstWouldAddEndPuncttrue
\mciteSetBstMidEndSepPunct{\mcitedefaultmidpunct}
{\mcitedefaultendpunct}{\mcitedefaultseppunct}\relax
\EndOfBibitem
\bibitem[Mentall \latin{et~al.}(1971)Mentall, Gentieu, Krauss, and
  Neumann]{Men71}
Mentall,~J.~E.; Gentieu,~E.~P.; Krauss,~M.; Neumann,~D. Photoionization and
  Absorption Spectrum of Formaldehyde in the Vacuum Ultraviolet. \emph{J. Chem.
  Phys.} \textbf{1971}, \emph{55}, 5471--5479\relax
\mciteBstWouldAddEndPuncttrue
\mciteSetBstMidEndSepPunct{\mcitedefaultmidpunct}
{\mcitedefaultendpunct}{\mcitedefaultseppunct}\relax
\EndOfBibitem
\bibitem[Suto \latin{et~al.}(1986)Suto, Wang, and Lee]{Sut86}
Suto,~M.; Wang,~X.; Lee,~L.~C. Fluorescence from VUV excitation of
  formaldehyde. \emph{J. Chem. Phys.} \textbf{1986}, \emph{85},
  4228--4233\relax
\mciteBstWouldAddEndPuncttrue
\mciteSetBstMidEndSepPunct{\mcitedefaultmidpunct}
{\mcitedefaultendpunct}{\mcitedefaultseppunct}\relax
\EndOfBibitem
\bibitem[Matsuzawa \latin{et~al.}(2001)Matsuzawa, Ishitani, Dixon, and
  Uda]{Mat01c}
Matsuzawa,~N.~N.; Ishitani,~A.; Dixon,~D.~A.; Uda,~T. Time-Dependent Density
  Functional Theory Calculations of Photoabsorption Spectra in the Vacuum
  Ultraviolet Region. \emph{J. Phys. Chem. A} \textbf{2001}, \emph{105},
  4953--4962\relax
\mciteBstWouldAddEndPuncttrue
\mciteSetBstMidEndSepPunct{\mcitedefaultmidpunct}
{\mcitedefaultendpunct}{\mcitedefaultseppunct}\relax
\EndOfBibitem
\bibitem[Jensen and Bunker(1982)Jensen, and Bunker]{Jen82b}
Jensen,~P.; Bunker,~P. The Geometry and the Out-Of-Plane Bending Potential
  Function of Thioformaldehyde in the {\~A}$^1A_2$ and {\~a}$^3A_2$ Electronic
  States. \emph{J. Mol. Spectrosc.} \textbf{1982}, \emph{95}, 92--100\relax
\mciteBstWouldAddEndPuncttrue
\mciteSetBstMidEndSepPunct{\mcitedefaultmidpunct}
{\mcitedefaultendpunct}{\mcitedefaultseppunct}\relax
\EndOfBibitem
\bibitem[Dunlop \latin{et~al.}(1991)Dunlop, Karolczak, Clouthier, and
  Ross]{Dun91}
Dunlop,~J.~R.; Karolczak,~J.; Clouthier,~D.~J.; Ross,~S.~C. Pyrolysis Jet
  Spectroscopy: The $S_1-S_0$ Band System of Thioformaldehyde and the
  Excited-State Bending Potential. \emph{J. Phys. Chem.} \textbf{1991},
  \emph{95}, 3045--3062\relax
\mciteBstWouldAddEndPuncttrue
\mciteSetBstMidEndSepPunct{\mcitedefaultmidpunct}
{\mcitedefaultendpunct}{\mcitedefaultseppunct}\relax
\EndOfBibitem
\bibitem[Dixon and Webster(1978)Dixon, and Webster]{Dix78}
Dixon,~R.; Webster,~C. The Determination of the Electric Dipole Moment of
  H$_2$CS in the $\tilde{A}^1A_2$ Excited State Using Dye-Laser Electric Field
  Spectroscopy. \emph{J. Mol. Spectrosc.} \textbf{1978}, \emph{70},
  314--322\relax
\mciteBstWouldAddEndPuncttrue
\mciteSetBstMidEndSepPunct{\mcitedefaultmidpunct}
{\mcitedefaultendpunct}{\mcitedefaultseppunct}\relax
\EndOfBibitem
\bibitem[Dixon and Gunson(1983)Dixon, and Gunson]{Dix83}
Dixon,~R.; Gunson,~M. The Dipole Moment of Thioformaldehyde in its Singlet and
  Triplet $\pi^\star - n$ Excited States. \emph{J. Mol. Spectrosc.}
  \textbf{1983}, \emph{101}, 369--378\relax
\mciteBstWouldAddEndPuncttrue
\mciteSetBstMidEndSepPunct{\mcitedefaultmidpunct}
{\mcitedefaultendpunct}{\mcitedefaultseppunct}\relax
\EndOfBibitem
\bibitem[Judge \latin{et~al.}(1978)Judge, Drury-Lessard, and Moule]{Jud78}
Judge,~R.; Drury-Lessard,~C.; Moule,~D. The Far Ultraviolet Spectrum of
  Thioformaldehyde. \emph{Chem. Phys. Lett.} \textbf{1978}, \emph{53},
  82--83\relax
\mciteBstWouldAddEndPuncttrue
\mciteSetBstMidEndSepPunct{\mcitedefaultmidpunct}
{\mcitedefaultendpunct}{\mcitedefaultseppunct}\relax
\EndOfBibitem
\bibitem[Saito and Takagi(1973)Saito, and Takagi]{Sai73}
Saito,~S.; Takagi,~K. Microwave Spectrum of Nitroxyl. \emph{J. Mol. Spectrosc.}
  \textbf{1973}, \emph{47}, 99--106\relax
\mciteBstWouldAddEndPuncttrue
\mciteSetBstMidEndSepPunct{\mcitedefaultmidpunct}
{\mcitedefaultendpunct}{\mcitedefaultseppunct}\relax
\EndOfBibitem
\bibitem[Dixon and Noble(1980)Dixon, and Noble]{Dix80}
Dixon,~R.; Noble,~M. The Dipole Moment of HNO in its $\tilde{A}^1A''$ Excited
  State Determined using Optical-Optical Double Resonance Stark Spectroscopy.
  \emph{Chem. Phys.} \textbf{1980}, \emph{50}, 331--339\relax
\mciteBstWouldAddEndPuncttrue
\mciteSetBstMidEndSepPunct{\mcitedefaultmidpunct}
{\mcitedefaultendpunct}{\mcitedefaultseppunct}\relax
\EndOfBibitem
\bibitem[Takagi \latin{et~al.}(1985)Takagi, Suzuki, Saito, and Hirota]{Tak85}
Takagi,~K.; Suzuki,~T.; Saito,~S.; Hirota,~E. Microwave Optical Double
  Resonance of HNO: Dipole Moment of HNO in $\tilde{A}^1A"(100)$. \emph{J.
  Chem. Phys.} \textbf{1985}, \emph{83}, 535--538\relax
\mciteBstWouldAddEndPuncttrue
\mciteSetBstMidEndSepPunct{\mcitedefaultmidpunct}
{\mcitedefaultendpunct}{\mcitedefaultseppunct}\relax
\EndOfBibitem
\bibitem[Johns and McKellar(1977)Johns, and McKellar]{Joh77}
Johns,~J. W.~C.; McKellar,~A. R.~W. Laser Stark Spectroscopy of the Fundamental
  Bands of HNO ($\nu_2$ and $\nu_3$) and DNO ($\nu_1$ and $\nu_2$). \emph{J.
  Chem. Phys.} \textbf{1977}, \emph{66}, 1217--1224\relax
\mciteBstWouldAddEndPuncttrue
\mciteSetBstMidEndSepPunct{\mcitedefaultmidpunct}
{\mcitedefaultendpunct}{\mcitedefaultseppunct}\relax
\EndOfBibitem
\bibitem[Scuseria \latin{et~al.}(1986)Scuseria, Duran, Maclagan, and
  Schaefer]{Scu86}
Scuseria,~G.~E.; Duran,~M.; Maclagan,~R. G. A.~R.; Schaefer,~H.~F. Halocarbenes
  CHF, CHCl, and CHBr: Geometries, Singlet-Triplet Separations, and Vibrational
  Frequencies. \emph{J. Am. Chem. Soc.} \textbf{1986}, \emph{108},
  3248--3253\relax
\mciteBstWouldAddEndPuncttrue
\mciteSetBstMidEndSepPunct{\mcitedefaultmidpunct}
{\mcitedefaultendpunct}{\mcitedefaultseppunct}\relax
\EndOfBibitem
\bibitem[Kakimoto \latin{et~al.}(1981)Kakimoto, Saito, and Hirota]{Kak81}
Kakimoto,~M.; Saito,~S.; Hirota,~E. Doppler-Limited Dye Laser Excitation
  Spectroscopy of HCF. \emph{J. Mol. Spectrosc.} \textbf{1981}, \emph{88},
  300--310\relax
\mciteBstWouldAddEndPuncttrue
\mciteSetBstMidEndSepPunct{\mcitedefaultmidpunct}
{\mcitedefaultendpunct}{\mcitedefaultseppunct}\relax
\EndOfBibitem
\bibitem[Wagner \latin{et~al.}(2000)Wagner, Gamperling, Braun, Prohaska, and
  H{\"u}ttner]{Wag00}
Wagner,~M.; Gamperling,~M.; Braun,~D.; Prohaska,~M.; H{\"u}ttner,~W. Rotational
  Transitions and Electric Dipole Moment of Fluorocarbene, HCF. \emph{J. Mol.
  Struct.} \textbf{2000}, \emph{517-518}, 327--334\relax
\mciteBstWouldAddEndPuncttrue
\mciteSetBstMidEndSepPunct{\mcitedefaultmidpunct}
{\mcitedefaultendpunct}{\mcitedefaultseppunct}\relax
\EndOfBibitem
\bibitem[Lu \latin{et~al.}(2012)Lu, Hao, Wilke, Yamaguchi, Fang, and
  Schaefer]{Lu12}
Lu,~T.; Hao,~Q.; Wilke,~J.~J.; Yamaguchi,~Y.; Fang,~D.-C.; Schaefer,~H.~F.
  Silylidene (SiCH$_2$) and its Isomers: Anharmonic Rovibrational Analyses for
  Silylidene, Silaacetylene, and Silavinylidene. \emph{J. Mol. Struct.}
  \textbf{2012}, \emph{1009}, 103--110\relax
\mciteBstWouldAddEndPuncttrue
\mciteSetBstMidEndSepPunct{\mcitedefaultmidpunct}
{\mcitedefaultendpunct}{\mcitedefaultseppunct}\relax
\EndOfBibitem
\bibitem[Smith \latin{et~al.}(2003)Smith, Evans, and Clouthier]{Smi03}
Smith,~T.~C.; Evans,~C.~J.; Clouthier,~D.~J. Discovery of the Optically
  Forbidden $S_1-S_0$ Transition of Silylidene (H$_2$C=Si). \emph{J. Chem.
  Phys.} \textbf{2003}, \emph{118}, 1642--1648\relax
\mciteBstWouldAddEndPuncttrue
\mciteSetBstMidEndSepPunct{\mcitedefaultmidpunct}
{\mcitedefaultendpunct}{\mcitedefaultseppunct}\relax
\EndOfBibitem
\bibitem[Harper \latin{et~al.}(1997)Harper, Waddell, and Clouthier]{Har97b}
Harper,~W.~W.; Waddell,~K.~W.; Clouthier,~D.~J. Jet Spectroscopy, Structure,
  Anomalous Fluorescence, and Molecular Quantum Beats of Silylidene
  (H$_2$C=Si), the Simplest Unsaturated Silylene. \emph{J. Chem. Phys.}
  \textbf{1997}, \emph{107}, 8829--8839\relax
\mciteBstWouldAddEndPuncttrue
\mciteSetBstMidEndSepPunct{\mcitedefaultmidpunct}
{\mcitedefaultendpunct}{\mcitedefaultseppunct}\relax
\EndOfBibitem
\bibitem[Hilliard and Grev(1997)Hilliard, and Grev]{Hil97}
Hilliard,~R.~K.; Grev,~R.~S. The Excited Electronic States of H$_2$CSi.
  \emph{J. Chem. Phys.} \textbf{1997}, \emph{107}, 8823--8828\relax
\mciteBstWouldAddEndPuncttrue
\mciteSetBstMidEndSepPunct{\mcitedefaultmidpunct}
{\mcitedefaultendpunct}{\mcitedefaultseppunct}\relax
\EndOfBibitem
\end{mcitethebibliography}

\end{document}